\RequirePackage{rotating}

\documentclass[fleqn,usenatbib]{mnras}

\usepackage{newtxtext,newtxmath}
\usepackage{supertabular}

\usepackage[T1]{fontenc}

\DeclareRobustCommand{\VAN}[3]{#2}
\let\VANthebibliography\thebibliography
\def\thebibliography{\DeclareRobustCommand{\VAN}[3]{##3}\VANthebibliography}

\usepackage{graphicx}	
\usepackage{amsmath}	

\usepackage{rotating}
\usepackage{bm}
\usepackage{longtable,tabu}
\usepackage{xcolor}
\usepackage{pifont}
\usepackage{soul}
\usepackage{ulem}
\usepackage{caption}

\usepackage{graphicx,pifont}
\let\oldding\ding
\renewcommand{\ding}[2][1]{\scalebox{#1}{\oldding{#2}}}

\usepackage{array}
\newcolumntype{H}{>{\setbox0=\hbox\bgroup}c<{\egroup}@{}}

\title[Spectroscopy of distant Galactic halo RR Lyrae]{RR Lyrae stars as probes of the outer Galactic halo: Chemical and kinematic analysis of a pilot sample}

\author[Medina et al.]{
Gustavo E. Medina,$^{1}$\thanks{E-mail: gustavo.medina@uni-heidelberg.de}
Camilla J. Hansen,$^{2}$
Ricardo R. Mu\~{n}oz,$^{3}$
Eva K. Grebel,$^{1}$
A. Katherina Vivas,$^{4}$
\newauthor
Jeffrey L. Carlin$^{5}$ and 
Clara E. Mart\'inez-V\'azquez$^{6}$
\\
$^{1}$Astronomisches Rechen-Institut, Zentrum f{\"u}r Astronomie der Universit{\"a}t Heidelberg, M{\"o}nchhofstr. 12-14, 69120 Heidelberg, Germany\\
$^{2}$Goethe University Frankfurt, Institute for Applied Physics, Max-von-Laue Str. 11, 60438 Frankfurt am Main, Germany\\
$^{3}$Departamento de Astronom\'ia, Universidad de Chile, Camino El Observatorio 1515, Las Condes, Santiago, Chile\\
$^{4}$Cerro Tololo Inter-American Observatory/NSF's NOIRLab, Casilla 603, La Serena, Chile\\
$^{5}$AURA/Rubin Observatory Project Office, 950 North Cherry Avenue, Tucson, AZ 85719, USA\\
$^{6}$Gemini Observatory/NSF's NOIRLab, 670 N. A'ohoku Place, Hilo, HI 96720, USA\\
}

\date{Accepted XXX. Received YYY; in original form ZZZ}

\pubyear{2022}

\begin{document}
\label{firstpage}
\pagerange{\pageref{firstpage}--\pageref{lastpage}}
\maketitle

\begin{abstract}

We report the spectroscopic analysis of 20 halo ab-type RR Lyrae stars with heliocentric distances between 15 and 165\,kpc, conducted using medium-resolution spectra from the Magellan Inamori Kyocera Echelle (MIKE) spectrograph. 
We obtain the systemic line-of-sight velocities of our targets with typical uncertainties of $5$--$10$\,km\,s$^{-1}$, and compute orbital parameters for a subsample out to 50\,kpc from the Galactic centre, including proper motion data from {\it Gaia} DR3.
The orientation of our stars' orbits, determined for an isolated Milky Way and for a model perturbed by the Large Magellanic Cloud, appears to suggest an accreted origin for at least half of the sample. 
In addition, we derive atmospheric parameters and chemical abundance ratios for seven stars beyond 20\,kpc.
The derived $\alpha$-abundances of five of these stars follow a Milky Way halo-like trend, while the other two display an underabundance of $\alpha$-elements for their [Fe/H], indicating an association with accretion events.
Furthermore, based on the [Sr/Ba] ratio, we can speculate about the conditions for the formation of a potential chemically peculiar carbon-enhanced metal-poor (CEMP) RR Lyrae star.
By analysing the stars' orbital parameters and abundance ratios, we find hints of association of two of our stars with two massive satellites, namely the Large Magellanic Cloud and Sagittarius. 
Overall, our results are in line with the suggestion that the accretion of sub-haloes largely contributes to the outer halo stellar populations.

\end{abstract}

\begin{keywords}
Galaxy: halo -- Galaxy: kinematics and dynamics -- stars: variables: RR Lyrae -- stars: abundances 
\end{keywords}




\section{Introduction}
Under the $\Lambda$ cold dark matter cosmological paradigm, the galaxies that we observe today are formed through the merger of smaller structures \citep[e.g.,][]{Seargle1978,White1991,Kauffmann1993,Cole1994,BJ05,Fattahi2020}. 
In recent years, growing evidence of these interactions, and in particular of the accretion of massive satellites, has been found by studying the dynamics and chemical patterns of Milky Way (MW) stars.  
This is the case of the Sagittarius merger event \citep{Ibata1994}, {\it{Gaia}}-Sausage-Enceladus \citep[GSE;][]{Belokurov2018b,Haywood2018,Helmi2018}, Kraken \citep{Massari2019,Kruijssen2019}, and
Sequoia \citep{Myeong2019},
among others.

The stellar halo notably contains valuable probes of the MW's history and is thought to be composed mainly of accreted substructures \cite[e.g.,][]{Rodriguez2016,Malhan2018b,Naidu2020}, making it a natural laboratory to study the evolution of the entire Galaxy. 
In order to reconstruct the assembly history of our Galaxy, it has become customary to study the observed six-dimensional phase-space (positions and velocities) of present-day stellar populations and overdensities as signatures of tidal stripping \citep[e.g.,][]{Li2021,Cook2022}, especially in the halo, given the relatively long dynamical time-scales resulting from partial phase mixing.
Thus, dynamical information of halo stars can be used to look for associations and even to determine the parent populations of single stars.
Furthermore, halo stars contain evidence of the chemical composition of the environment in which old stars were formed, hence they are good tracers of the early chemical evolution of the Galaxy.
However, the phase-space and detailed abundance patterns of outer halo stars remain vastly unexplored, mostly due to distance determination limitations.

RR Lyrae stars (RRLs, or RRL for single stars) have served as powerful probes of the chemical and dynamical evolution of the disc and halo of our Galaxy \citep[e.g.,][]{Vivas2005,Keller2008,For2011a,For2011b,Hansen2011,Hansen2016,Belokurov2018a,Li2022}, given that the small scatter in their mean absolute magnitudes make them precise distance indicators with which distances can be easily determined at a 5\% level precision \citep[see e.g.][]{Christy1966,Catelan2015,Beaton2018}.
These stars are old low-mass horizontal branch (HB) stars that pulsate in the instability strip, with periods typically shorter than one day.
Depending on the pulsation mode, and consequently their periods and light curve shapes, they have been mainly classified in three different types; the radial fundamental mode pulsators (RRab stars), the radial first overtone RRLs (RRc stars), and those that pulsate in both modes at the same time (RRd stars).
Given the relatively high luminosity and characteristic pulsation properties of RRLs, they are easily identifiable in time domain surveys, and have been widely used as tracers of Galactic substructures \citep[][]{Vivas2006,Watkins2009,Dekany2018,Mateu2018,Martinez2019,Prudil2019,Torrealba2019,Cook2022}.

Radial velocity measurements and spectroscopic metallicity derivations are more challenging to conduct in RRLs, owing to their variability on short time-scales.
In spite of these difficulties, different authors have measured these quantities using both low- and high-resolution spectra. 
The usage of the former began with introduction of the $\Delta S$ method by \citet{Preston1959} which relates the absorption line strengths of hydrogen and calcium K-lines to the metallicity, and was recently revised by \citet{Crestani21a}.
Other methods rely on the use of a known correlation between the Ca II triplet and RRLs metallicities \citep[][]{Wallerstein2012,Kunder2016,Martinez2016}, or the determination of photometric and luminosity-based metallicities \citep[e.g.,][]{Jurcsik1996,Smolec2005,Hajdu2018,Dekany2021,Mullen2021,Mullen2022,Garofalo2022}. 
The most precise methods to estimate RRLs metallicities are based on high-resolution spectra. 
However, the number of RRLs analysed with this method is still relatively low \citep[see e.g.][]{Clementini1995,Kolenberg10, For2011b, Hansen2011, Pancino2015,Chadid2017,Sneden2017,Magurno2018,Magurno2019,Gilligan2021,Crestani21b}.
For distant RRLs, in particular, this is 
mainly due to the need for large telescopes and long exposure times, which is in clear conflict with their short-term pulsations.
The determination of centre-of-mass radial velocities has also proven to be a challenging task given the pulsation phase of the RRLs that needs to be considered. 
To take this effect into account, observations can be conveniently scheduled so they take place at pulsation phases where the pulsation contributes the least to the observed radial velocity, or the measured velocities can be corrected by assuming a pulsation model.

Because of these observational challenges, only a handful of spectroscopic studies have been performed specifically on halo RRLs \citep[e.g.,][]{Liu2020,Fabrizio2021}, and even fewer have focused on RRLs at large heliocentric distances ($d_{\rm H}$), mostly due to the low number of RRLs detected in these regions, and due to instrumental limitations. 
At these limits one would expect, for instance, RRLs as faint as $g\sim21.0$ or $V\sim20.6$ at distances of $d_{\rm H}\sim100$\,kpc, and correspondingly fainter magnitudes with increasing distance.
However, the increasing amount of data from deep photometric surveys has allowed astronomers to detect significant samples of distant RRLs, as the ones found using the Catalina surveys \citep{Drake14,Drake17,Torrealba15}, 
the Panoramic Survey Telescope And Rapid Response System survey \citep[Pan-STARRS-1;][]{Chambers2016, Sesar2017}, the High cadence Transient Survey \citep[HiTS;][]{Forster2016, Medina2018}, the Dark Energy Survey \citep[DES;][]{DES2016, Stringer21},
or the Zwicky Transient Facility \citep[ZTF;][]{Chen2020,Huang2022}. 
\citet{Medina2018} found 16 RRL candidates beyond 100\,kpc in a survey area of $\sim$ 120\,deg$^2$, whereas more recently \citet{Stringer21} identified 800 RRLs candidates further than 100\,kpc using the footprint of the DES ($>$ 5,000\,deg$^2$) with a limiting magnitude of $g\sim$ 23.5, in line with the predicted number of RRLs from current accretion models \citep{BJ05, Sanderson2017}.
These stars have thus become intrinsically alluring targets for spectroscopic follow-up observations. 

In this work we describe our effort to spectroscopically analyse a selection of distant RRLs taken from publicly available catalogs, including our own previous work.  
Section~\ref{sec:sample} briefly describes the sample of RRLs selected for our analysis, and the data acquired for it.
In Section~\ref{sec:spectral} we describe the derivation of our sample's systemic velocities, stellar parameters, element abundance, and integrated orbits. 
We present the results of our spectroscopic and kinematic analysis in Section~\ref{sec:results} and use them to identify potential parent population for our RRL sample in Section~\ref{sec:accreted}. 
Finally, in Section~\ref{sec:conclusions} we summarize the outcomes of this study, and put them into a broader Galactic context to draw our conclusions.

\begin{figure*}

\includegraphics[angle=0,scale=.193]{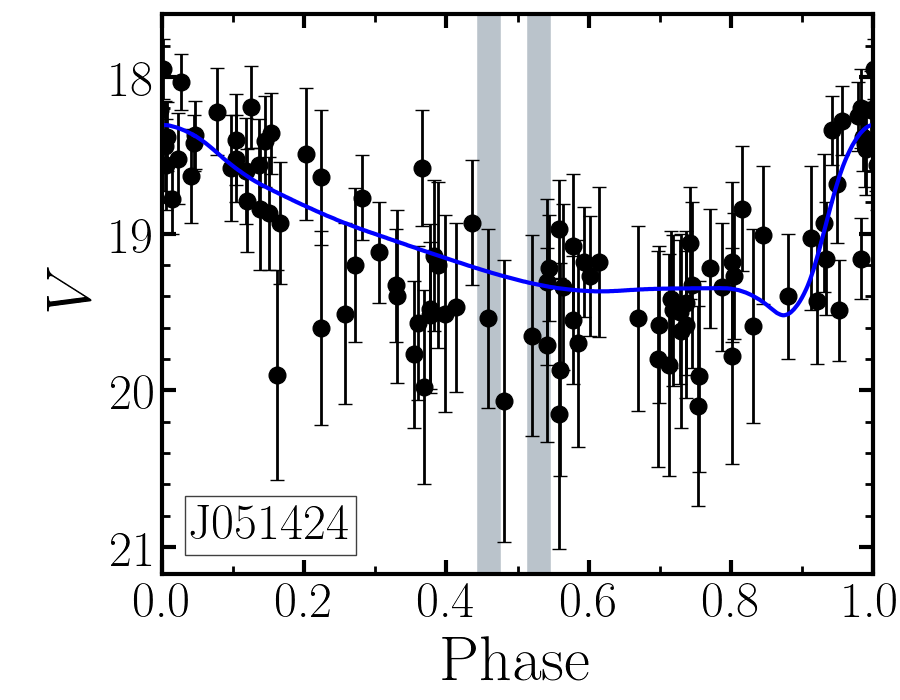} 
\includegraphics[angle=0,scale=.193]{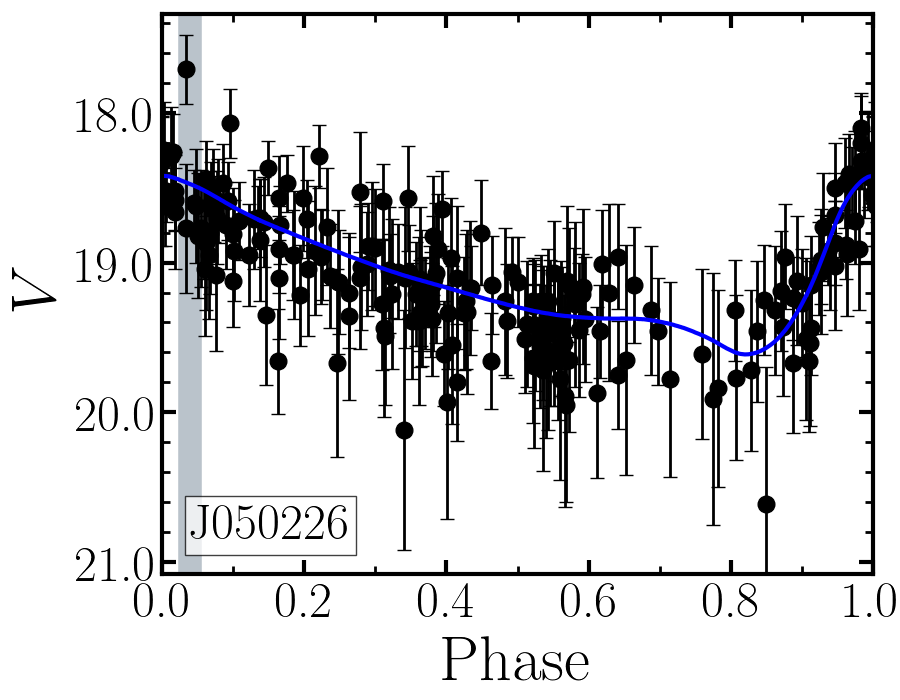} 
\includegraphics[angle=0,scale=.193]{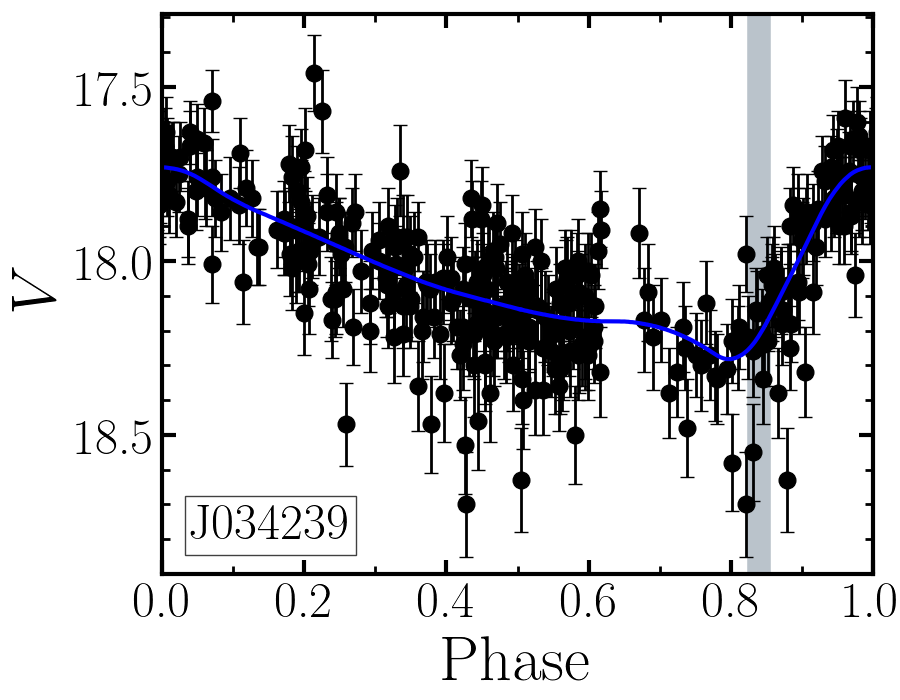} 
\includegraphics[angle=0,scale=.193]{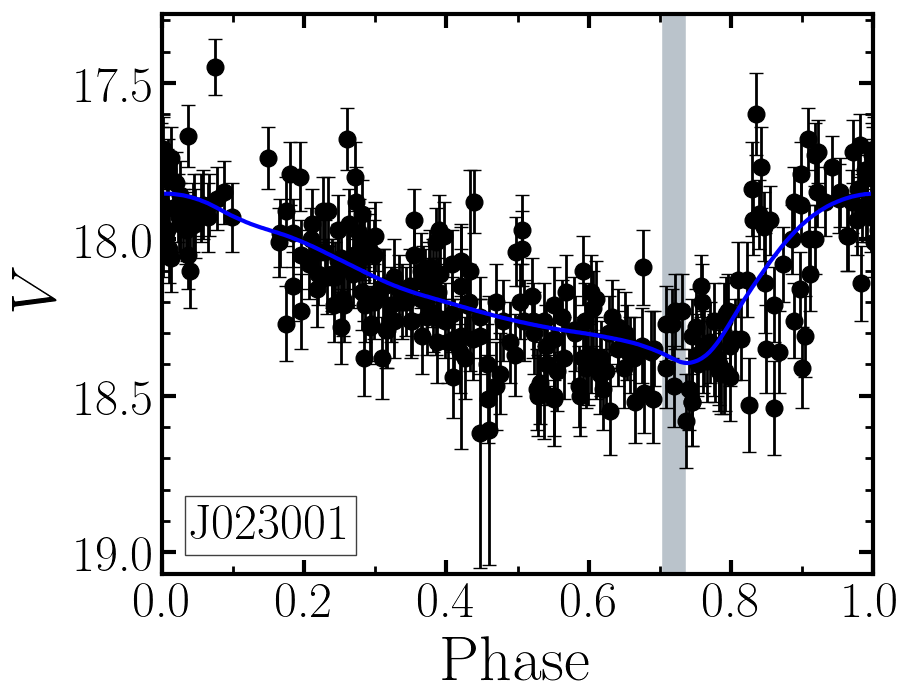}\\ 

\includegraphics[angle=0,scale=.193]{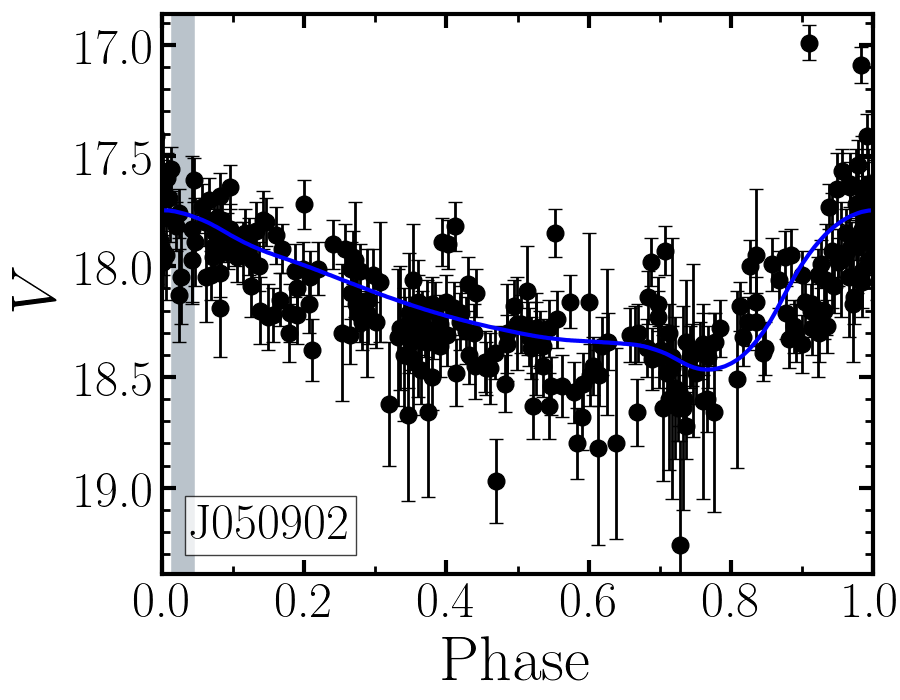} 
\includegraphics[angle=0,scale=.193]{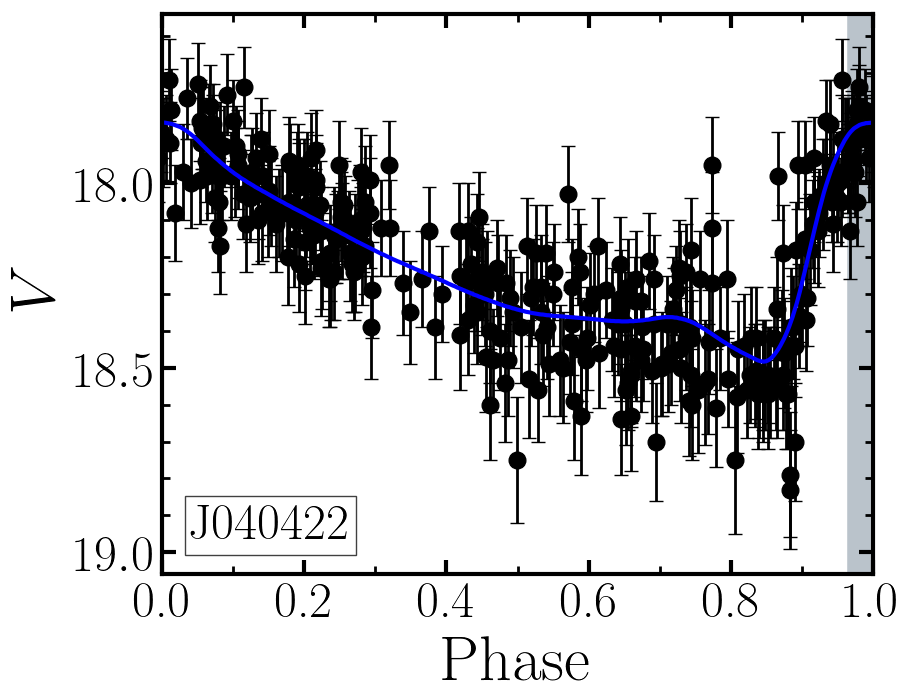} 
\includegraphics[angle=0,scale=.193]{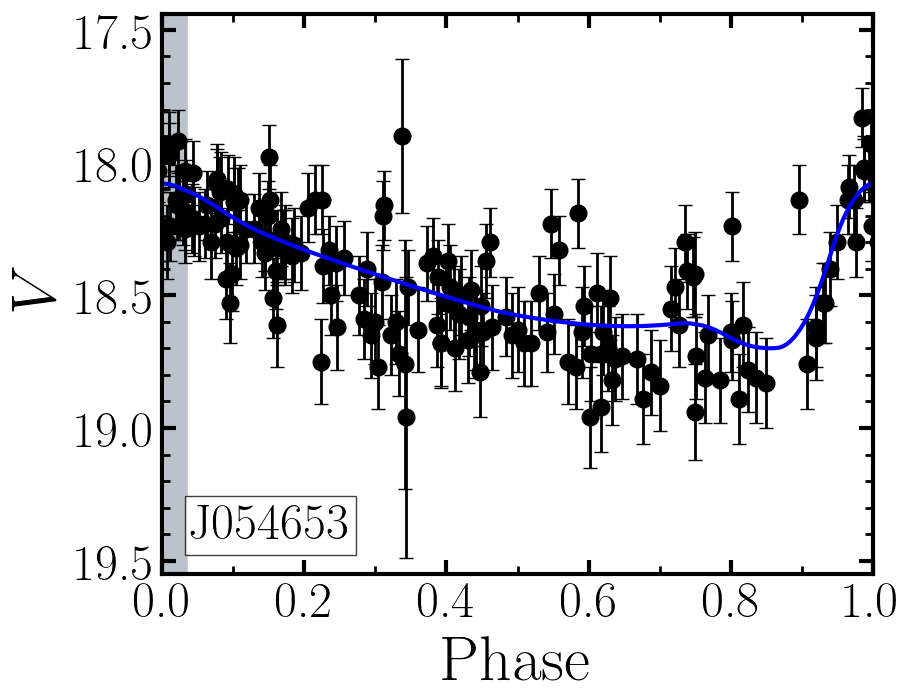} 
\includegraphics[angle=0,scale=.193]{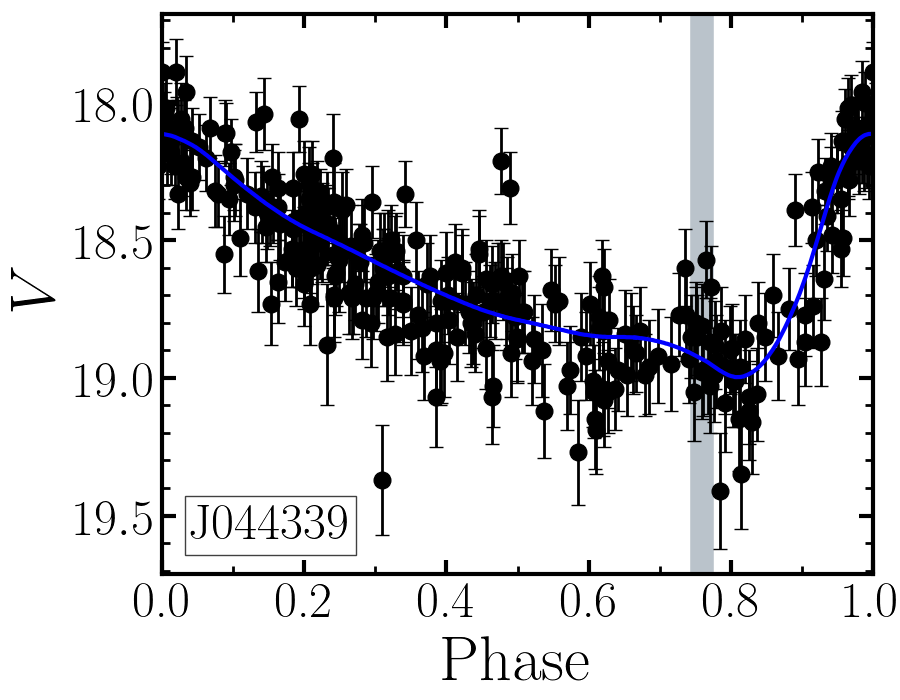}\\ 

\includegraphics[angle=0,scale=.193]{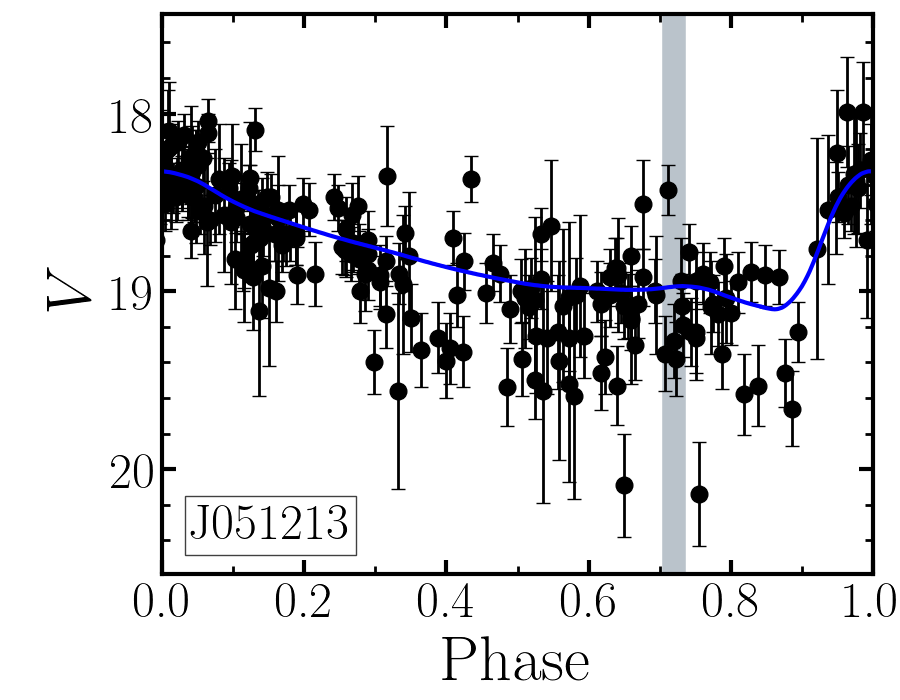} 
\includegraphics[angle=0,scale=.193]{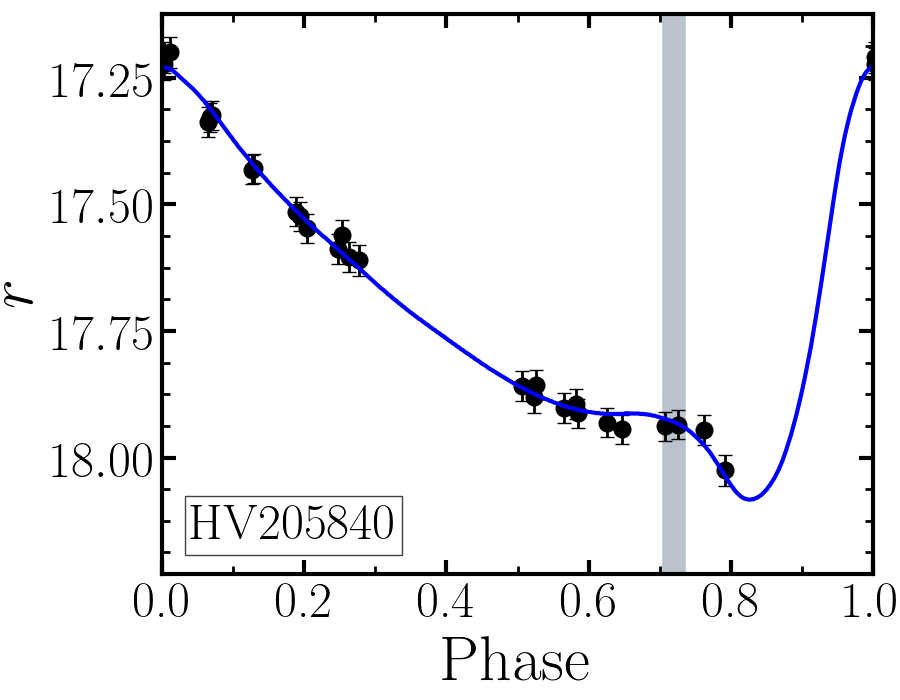} 
\includegraphics[angle=0,scale=.193]{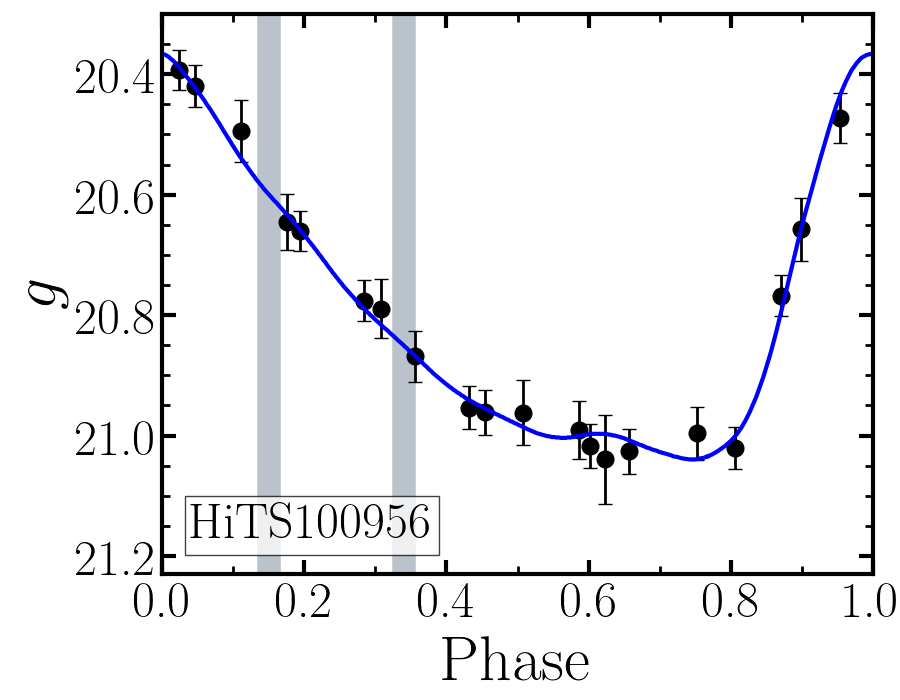} 
\includegraphics[angle=0,scale=.193]{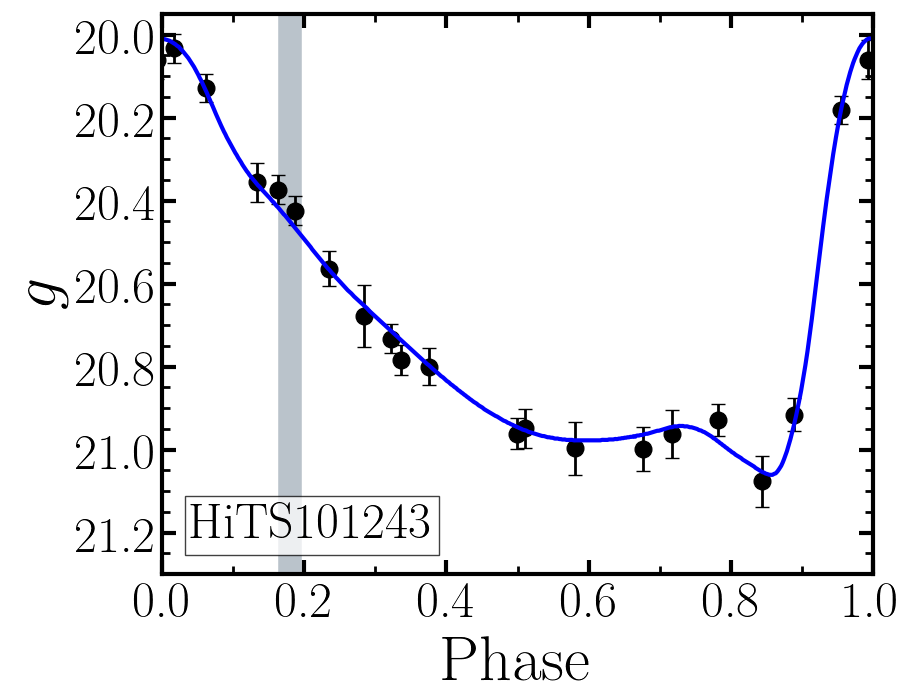}\\ 

\includegraphics[angle=0,scale=.193]{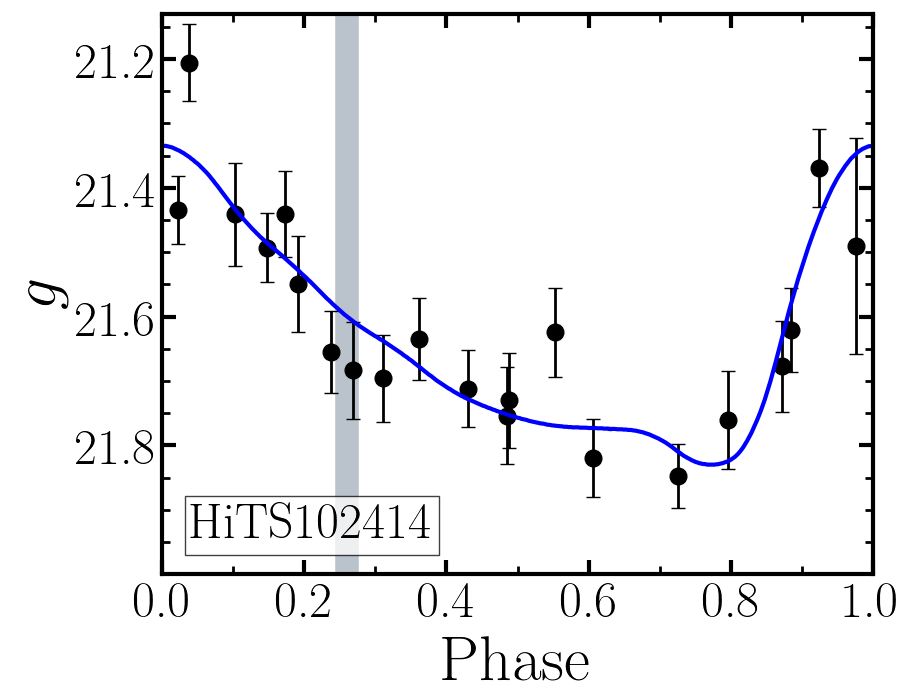} 
\includegraphics[angle=0,scale=.193]{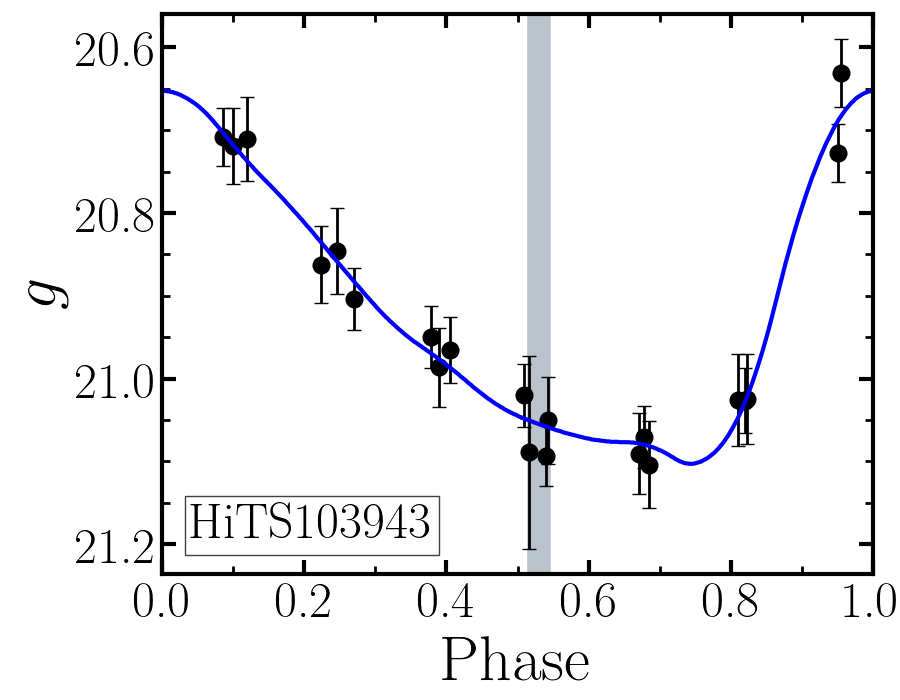} 
\includegraphics[angle=0,scale=.193]{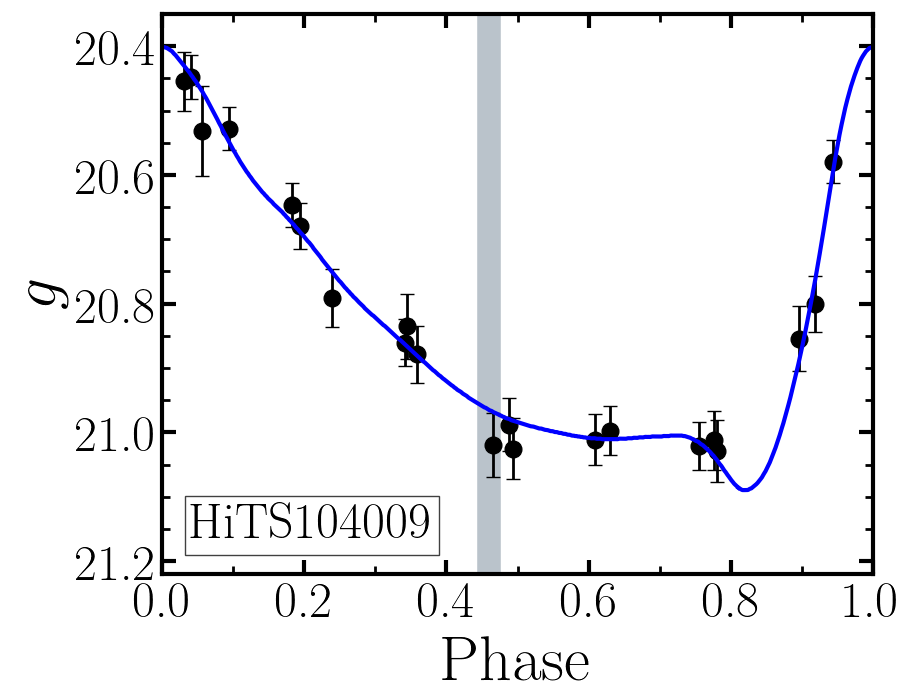} 
\includegraphics[angle=0,scale=.193]{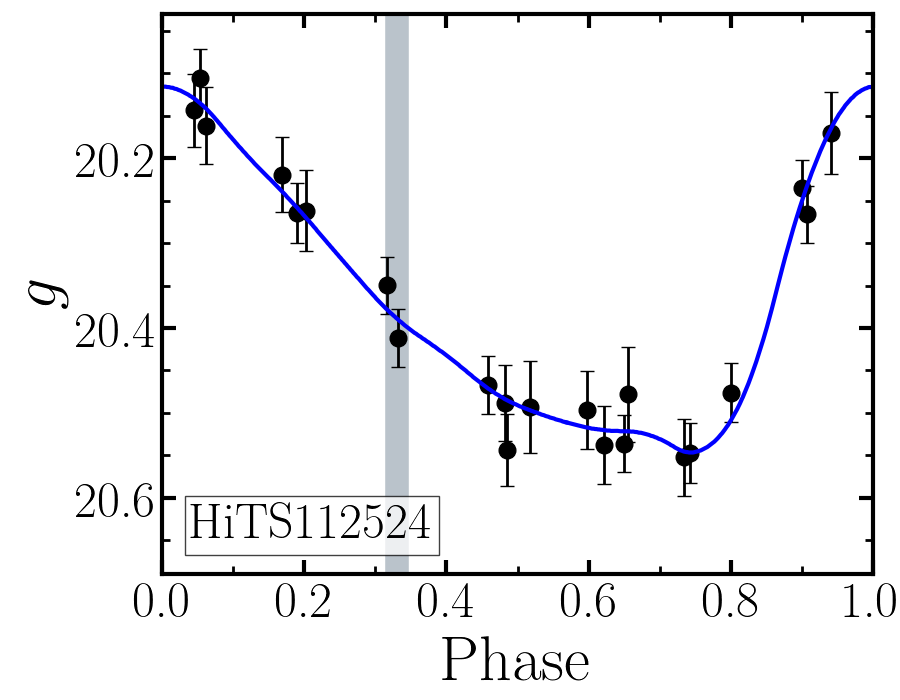}\\ 

\includegraphics[angle=0,scale=.193]{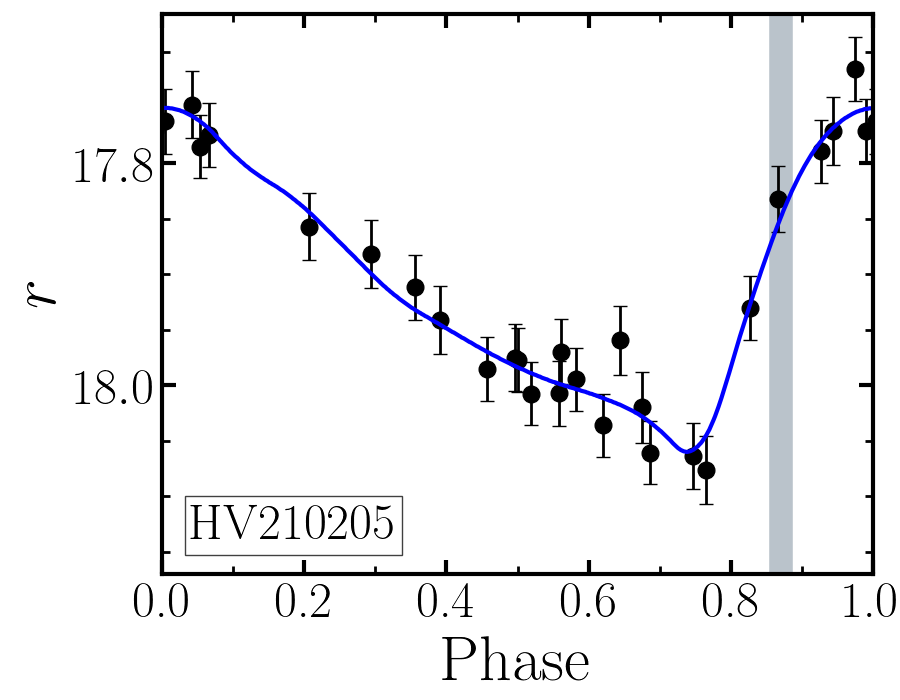} 
\includegraphics[angle=0,scale=.193]{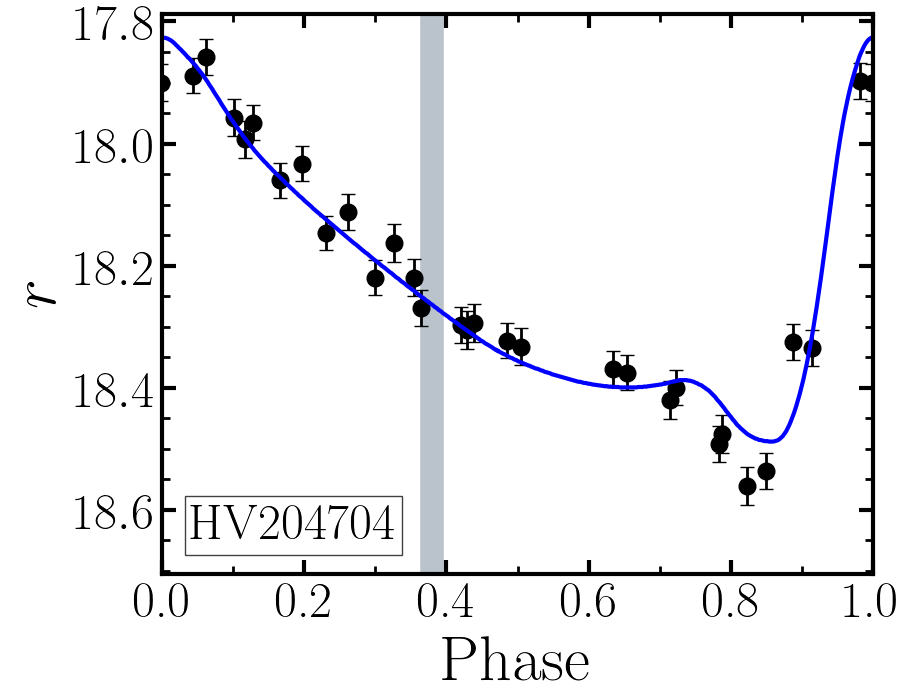} 
\includegraphics[angle=0,scale=.193]{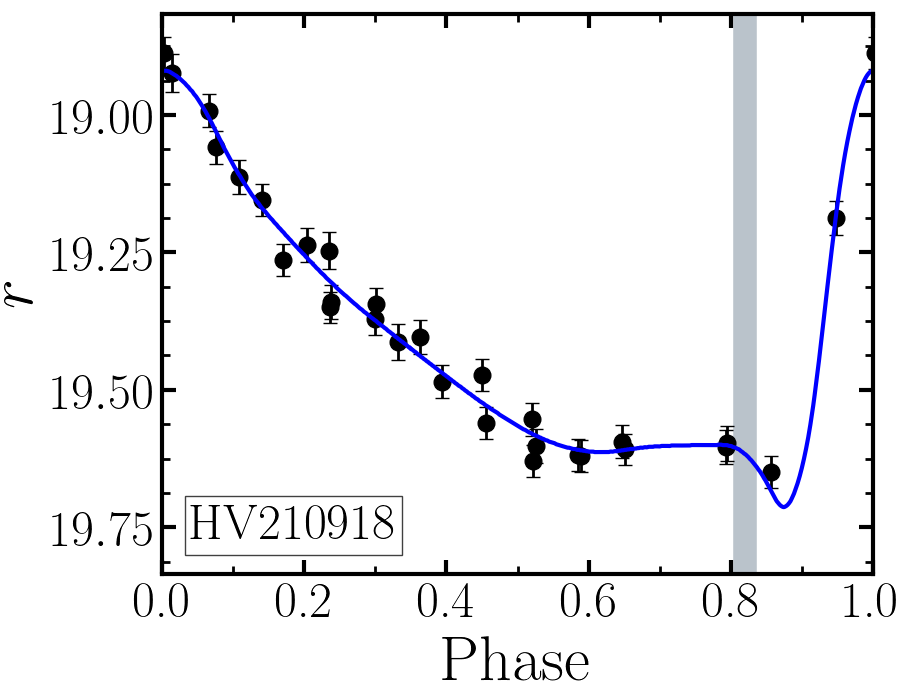} 
\includegraphics[angle=0,scale=.193]{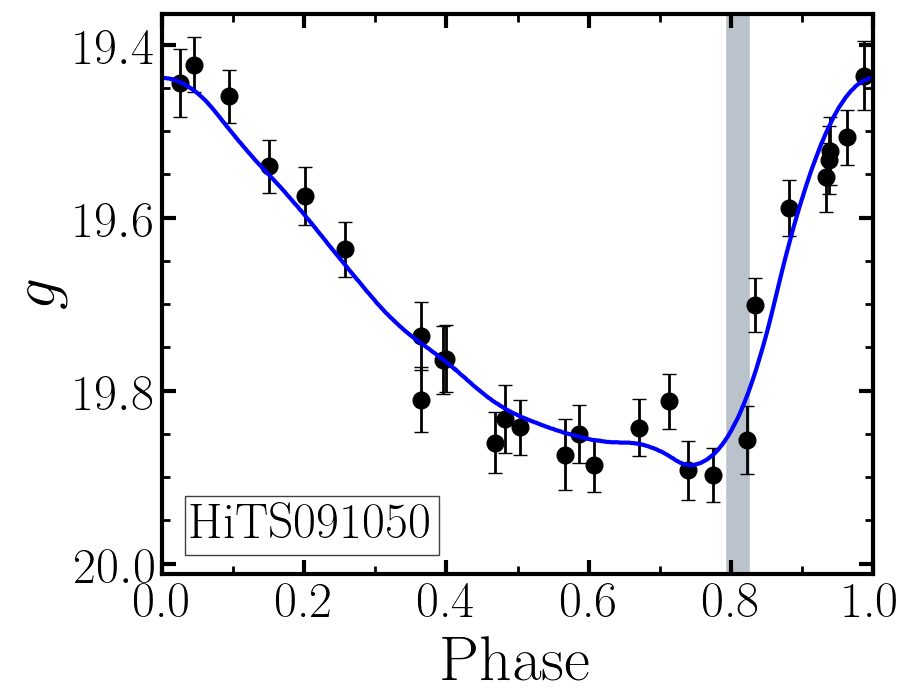}\\ 

\caption{Folded light curves of our program stars, based on time series from the Catalina survey, HiTS, and HOWVAST.  
On each panel, a solid blue line displays a model generated with \textsc{gatspy} \citep{VanderPlas2015}, which is based on the light curve templates from the Sloan Digital Sky Survey (SDSS) Stripe 82 RRLs \citep{Sesar10}. The grey vertical regions represent the estimated phase of the pulsation period of the RRLs in which the spectra were obtained, and their widths approximate the duration of the co-added observations.   
}
\label{fig:lcs}
\end{figure*}

\begin{figure}
\includegraphics[angle=0,scale=.30]{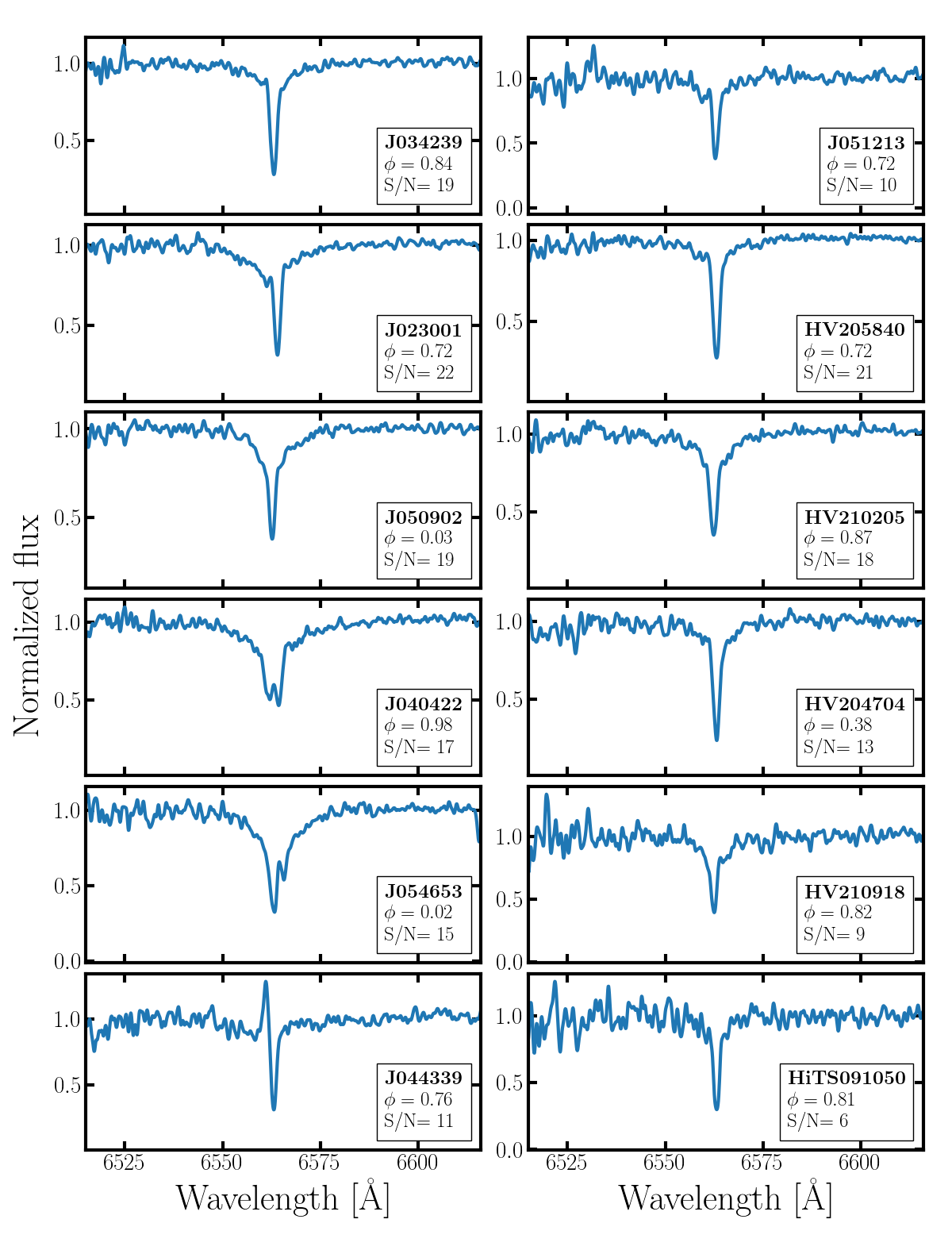} 
\caption{Spectral region surrounding the H$_\alpha$ line for the stars observed during our second campaign. A Gaussian convolution with $\sigma = 3$ was applied to smooth the spectra in order to help visualize the emission lines affecting the  H$_\alpha$ profile in certain phases during the RRLs pulsation cycle. 
}
\label{fig:Halphas}
\end{figure}

\begin{sidewaystable}

\scriptsize

\caption{
Summary of the stars observed in our program, including the estimated time of maximum light T$_0$, the time of observation T$_{\rm obs}$, and the phase of observation $\phi$ of our RRLs. 
}
\label{tab:Targets}
\begin{center}
\begin{tabular}{|c|c|c|c|c|c|c|c|c|c|c|c|c|c|c|}
\hline
    \multicolumn{1}{|c|}{ID} &
  \multicolumn{1}{c|}{RA} &
  \multicolumn{1}{c|}{DEC} &
  \multicolumn{1}{c|}{$<$mag$>$} &
    \multicolumn{1}{c|}{Filter} &
  \multicolumn{1}{c|}{Period} &
    \multicolumn{1}{c|}{$d_{\rm H}$} &
    \multicolumn{1}{c|}{T$_0$ (MJD)} &
    \multicolumn{1}{c|}{T$_{\rm obs}$ (MJD)} &
    \multicolumn{1}{c|}{${\rm \phi}$} &
  \multicolumn{1}{c|}{Run} &
  \multicolumn{1}{c|}{Slit size} &
  \multicolumn{1}{c|}{Binning} &
  \multicolumn{1}{c|}{Coadded spectra} &
  \multicolumn{1}{c|}{S/N$^1$} \\
 & (deg) & (deg) &  & & (days)  & (kpc) & & & &  &  (\,arcsec $\times$ \,arcsec)  & (spatial $\times$ spectral)  &   &      \\
\hline
      J051424.2-595954 (J051424) &    $78.59417$ &  $-59.99483$ &  19.1 &        $V$ &  0.4838 &   $47.1\pm2.4$ &       53647.1685 & 58498.1524  &        0.46 &       1 &  1.0 $\times$ 5.0 &  3 $\times$ 8 &   5 ($\times$ 900\,s) &              14 \\  
      &  &   &  &   & &    &       &   58499.1558  &       0.53 &       1 &  1.0 $\times$ 5.0 &  3 $\times$ 8 &   2 ($\times$ 900\,s) &              14 \\
    J050226.9-395429 (J050226) &     $75.61042$ &  $-39.91011$ &  19.2 &        $V$ &  0.5200 &   $47.9\pm2.4$ &       53597.4055 &  58498.0516   &      0.04 &       1 &  1.0 $\times$ 5.0 &  3 $\times$ 8 &   3 ($\times$ 900\,s) &              12 \\
   HiTS112524-024348 (HiTS112524) &   $171.35042$ &   $-2.73200$ &  20.2 &        $g$ &  0.6373 &   $85.7\pm3.7$ &       56717.4255 & 58498.2525  &        0.33 &       1 &  1.0 $\times$ 5.0 &  3 $\times$ 8 &   4 ($\times$ 900\,s) &               8 \\
  HiTS100956+013212 (HiTS100956) &   $152.48583$ &    $1.53700$ &  20.3 &        $g$ &  0.6220 &   $88.6\pm3.8$ &       56717.2142 &  58498.2087 &        0.34 &       1 &  1.0 $\times$ 5.0 &  3 $\times$ 8 &   4 ($\times$ 900\,s) &               8 \\
  &   &   &  &      &   &   &     & 58499.3346 &          0.15 &       1 &  1.0 $\times$ 5.0 &  3 $\times$ 8 &   3 ($\times$ 900\,s) &               8 \\
   HiTS101243+022118 (HiTS101243) &    $153.17958$ &    $2.35514$ &  20.5 &        $g$ &  0.5346 &   $90.6\pm3.5$ &       56717.4971 &  58498.3472  &       0.18 &       1 &  1.0 $\times$ 5.0 &  3 $\times$ 8 &   2 ($\times$ 900\,s) &               5 \\
  HiTS104009-063304 (HiTS104009) &    $160.04083$ &   $-6.55275$ &  20.7 &        $g$ &  0.6376 &  $104.7\pm4.5$ &       56717.1908 &  58498.3032  &       0.46 &       1 &  1.0 $\times$ 5.0 &  3 $\times$ 8 &   5 ($\times$ 900\,s) &               3 \\
   HiTS103943-021726 (HiTS103943) &    $159.93375$ &   $-2.29472$ &  20.7 &        $g$ &  0.6956 &  $110.8\pm5.1$ &       56717.4797 &  58499.2813 &        0.53 &       1 &  1.0 $\times$ 5.0 &  3 $\times$ 8 &   6 ($\times$ 900\,s) &               6 \\
   HiTS102414-095518 (HiTS102414) &    $156.06000$ &   $-9.92625$ &  21.5 &        $g$ &  0.7641 &  $166.8\pm8.3$ &       56717.1339 & 58499.2119  &        0.26 &       1 &  1.0 $\times$ 5.0 &  3 $\times$ 8 &   6 ($\times$ 900\,s) &               5 \\

 CS 22874-042 (CS 22874) &  $219.50710$ & $-24.97972$ & 14.0 & $V$ & -- & -- & -- & 58499.3659 & -- & 1  &  1.0 $\times$ 5.0 & 3 $\times$ 8  & 2 ($\times$ 240s) & 114 \\
  
  \hline
      J054653.1-020350 (J054653) &    $86.72131$ &   $-2.06410$ &  18.6 &        $V$ &  0.6144 &   $14.4\pm0.7$ &       53650.0886 & 59164.2922  &          0.02 &       2 &  1.5 $\times$ 5.0 &  2 $\times$ 2 &  4 ($\times$ 1200\,s) &              15 \\
      J040422.4-012011 (J040422) &     $61.09358$ &   $-1.33642$ &  18.3 &        $V$ &  0.6354 &   $25.2\pm1.3$ &       53627.1991 & 59163.2132  &         0.98 &       2 &  1.5 $\times$ 5.0 &  2 $\times$ 2 &  3 ($\times$ 1200\,s) &              17 \\
       J050902.1-123926 (J050902) &    $77.25909$ &  $-12.65738$ &  18.2 &        $V$ &  0.6180 &   $25.6\pm1.3$ &       53620.2295 & 59164.1659 &         0.03 &       2 &  1.5 $\times$ 5.0 &  2 $\times$ 2 &  4 ($\times$ 1200\,s) &              19 \\
       J034239.9-000009 (J034239) &    $55.66652$ &   $-0.00260$ &  18.1 &        $V$ &  0.6118 &   $26.9\pm1.3$ &       53626.4146 & 59163.2975  &        0.84 &       2 &  1.5 $\times$ 5.0 &  2 $\times$ 2 &  4 ($\times$ 1200\,s) &              19 \\
     HV205840-342000 (HV205840) &    $314.66548$ &  $-34.33322$ &  17.6 &        $r$ &  0.6664 &   $27.1\pm0.9$ &       57992.3662 & 59163.0433 &         0.72 &       2 &  1.5 $\times$ 5.0 &  2 $\times$ 2 &  4 ($\times$ 1200\,s) &              21 \\
       J023001.9-011146 (J023001) &    $37.50817$ &   $-1.19623$ &  18.2 &        $V$ &  0.6491 &   $30.5\pm1.5$ &       53627.2430 & 59164.2295 &         0.72 &       2 &  1.5 $\times$ 5.0 &  2 $\times$ 2 &  4 ($\times$ 1200\,s) &              22 \\
     HV210205-341427 (HV210205) &    $315.52120$ &  $-34.24092$ &  17.9 &        $r$ &  0.6454 &   $32.2\pm1.1$ &       57992.1996 & 59164.1085  &        0.87 &       2 &  1.5 $\times$ 5.0 &  2 $\times$ 2 &  3 ($\times$ 1200\,s) &              18 \\
   HV204704-382019 (HV204704) &    $311.76850$ &  $-38.33850$ &  18.1 &        $r$ &  0.6104 &   $35.5\pm1.2$ &       57992.1393 &  59163.0983  &        0.38 &       2 &  1.5 $\times$ 5.0 &  2 $\times$ 2 &  3 ($\times$ 1200\,s) &              13 \\
      J044339.2-005841 (J044339) &    $70.91353$ &   $-0.97819$ &  18.7 &        $V$ &  0.6433 &   $38.2\pm1.9$ &       53651.2801 & 59163.3507 &         0.76 &       2 &  1.5 $\times$ 5.0 &  2 $\times$ 2 &  3 ($\times$ 1200\,s) &              11 \\
    J051213.6-151041 (J051213) &     $78.05674$ &  $-15.17828$ &  18.9 &        $V$ &  0.6660 &   $39.4\pm2.0$ &       53598.0793 & 59163.1549 &         0.72 &       2 &  1.5 $\times$ 5.0 &  2 $\times$ 2 &  4 ($\times$ 1200\,s) &              10 \\
 HiTS091050-055917 (HiTS091050) &    $137.70940$ &   $-5.98800$ &  19.4 &        $g$ &  0.6468 &   $61.4\pm2.7$ &       57070.1314 &  59164.3469  &        0.81 &       2 &  1.5 $\times$ 5.0 &  2 $\times$ 2 &  3 ($\times$ 1200\,s) &               6 \\
   HV210918-335828 (HV210918) &    $317.32527$ &  $-33.97447$ &  19.4 &        $r$ &  0.6340 &   $62.3\pm2.1$ &       57992.5420 &  59164.0519  &         0.82 &       2 &  1.5 $\times$ 5.0 &  2 $\times$ 2 &  4 ($\times$ 1200\,s) &               9 \\

 HD 76483 &  $133.88150$ & $-27.68186$ & 4.9 & $V$ & -- & -- & -- &  59164.3710 & -- & 2 &  1.5 $\times$ 5.0 & 2 $\times$ 2  & 1 ($\times$ 3s) & 154 \\

\hline

\end{tabular} 
\end{center}
$^1$The S/N ratio is estimated from the continuum surrounding H$_\alpha$. For J050226 it is measured around H$_\beta$ instead. 

\end{sidewaystable}

\begin{sidewaystable}\scriptsize

\caption{
Radial velocity measurements for our targets after heliocentric correction. In addition to line-of-sight velocities ($v_{\rm los}$), we include the number of MIKE orders used to estimate such velocities in both the blue and the red side of the detector (N$_{\rm ap}$), their propagated uncertainties (e $v_{\rm los}$), and the scatter in the measurements from different orders ($\sigma_{v_{\rm los}}$).
In the case of the systemic velocities ($v_{\rm sys}$), we provide the uncertainties propagated from using the Balmer and metallic line-based RV templates (e $v_{\rm sys}$), and the scatter in the best fits ($\sigma v_{\rm sys}$).
}
\label{tab:RVsTable}
\begin{center}
\begin{tabular}{|c|c|c|c|c|c|c|c|c|}
\hline
  \multicolumn{1}{|c|}{ID} &
  \multicolumn{1}{c|}{N$_{\rm ap}$ red} &
  \multicolumn{1}{c|}{N$_{\rm ap}$ blue} &
  \multicolumn{1}{c|}{$v_{\rm los}$} &
  \multicolumn{1}{c|}{e $v_{\rm los}$} &
  \multicolumn{1}{c|}{$\sigma {v_{\rm los}}$} &
  \multicolumn{1}{c|}{$v_{\rm sys}$} &
  \multicolumn{1}{c|}{e $v_{\rm sys}$} &
  \multicolumn{1}{c|}{$\sigma v_{\rm sys}$} \\

     &            &         &  (km\,s$^{-1}$)   &   (km\,s$^{-1}$) &  (km\,s$^{-1}$) &(km\,s$^{-1}$) &  (km\,s$^{-1}$) &  (km\,s$^{-1}$) \\

\hline
J051424 &           3 &          3 &   221.2 &        1.7 &    9.6 &   216.9 &    7.7 &      4.0 \\
     
    &           3 &          3 &   247.7 &        0.8 &    7.6 &   240.3 &    8.0 &      0.2 \\
J050226 &           3 &         -- &    79.3 &        1.6 &    3.9 &   132.3 &    7.9 &      8.4 \\
    HiTS112524 &           3 &          3 &   188.4 &        0.7 &    7.3 &   205.4 &    4.3 &      2.8 \\
    HiTS100956 &           1 &          1 &    58.6 &        0.9 &    0.0 &    75.4 &   12.7 &      0.0 \\
   
    &           3 &          3 &    51.5 &        1.5 &    4.8 &    66.0 &    4.3 &      9.3 \\
    HiTS101243 &           1 &          1 &   293.8 &        1.4 &    6.4 &   305.6 &    3.8 &      1.0 \\
    HiTS104009 &           1 &          1 &   146.5 &        1.9 &    3.3 &   140.5 &    7.6 &      1.9 \\
    HiTS103943 &          -- &          2 &   171.2 &        1.8 &    0.6 &   167.0 &   12.7 &      0.0 \\
    HiTS102414 &          -- &          2 &    50.0 &        0.8 &    1.9 &    72.3 &   12.7 &      0.0 \\
    
\hline
    J054653 &           2 &          1 &   $-$66.0 &        1.6 &    7.4 &     3.4 &    6.8 &     12.5 \\
    J040422 &           3 &          3 &  $-$122.3 &        0.5 &    4.5 &  $-$138.4 &    6.1 &      2.6 \\
    J050902 &           3 &          2 &   102.7 &        0.7 &    5.2 &   145.7 &    6.2 &      5.1 \\
    J034239 &           3 &          4 &  $-$101.2 &        0.8 &    3.4 &  $-$117.6 &    5.9 &      9.5 \\
  HV205840 &           3 &          4 &   105.5 &        1.1 &    7.8 &    93.7 &    7.0 &      1.0 \\
     J023001 &           2 &         9 &  $-$142.6$^*$ &        0.7 &    3.8 &  $-$137.1 &    5.9 &      8.0 \\
       HV210205 &           3 &          3 &    97.3 &        0.3 &   10.5 &    72.0 &    5.5 &     26.0 \\ 
 HV204704 &           3 &          4 &    97.6 &        0.2 &    4.7 &   108.6 &    6.6 &      4.0 \\
    J044339 &           3 &          2 &   $-$27.5 &        0.2 &    2.8 &   $-$44.5 &    6.7 &      5.2 \\
    J051213 &           3 &          3 &    78.6 &        0.6 &    2.6 &    66.1 &    6.4 &      4.1 \\
   HiTS091050 &           2 &          2 &    60.7 &        0.3 &    0.7 &    46.2 &    5.4 &      9.0 \\
 HV210918 &           3 &          2 &   $-$78.1 &        0.7 &   10.7 &  $-$108.7 &    6.8 &     25.0 \\

\hline

\end{tabular} 
\end{center}
$^*$The $v_{\rm los}$
used for J023001's radial velocity correction
was computed using orders with metallic lines only.  

\end{sidewaystable}

\begin{table*}\scriptsize
\caption{
Summary of the atmospheric parameters determined for the RRLs with higher S/N in our sample using empirical estimations (emp) and spectrum synthesis (synth).  
}
\label{tab:atms}
\begin{center}

\begin{tabular}{|c|c|c|c|c|c|c|c|c|c|c|c|c|c|c|}

\hline

ID &
[Fe/H] &
e\_[Fe/H] &
$\sigma_{ \rm [Fe/H]}$ &
[Fe/H] &
e\_[Fe/H] &
$\sigma_{ \rm [Fe/H]}$   &

$T_{\rm eff}$ &
$\sigma_{T_{\rm eff}}$ &

$T_{\rm eff}$ &
e\_{$T_{\rm eff}$}   &
$\sigma_{T_{\rm eff}}$ &

log g &
e\_{log g}   &
$\sigma_{\rm log g }$ 

\\

     &   (dex) &   (dex) &   (dex) &  (dex) &  (dex) &  (dex) & (K) & 
  
     (K) & (K) &  (K) &  (K) & 
     (cgs) & (cgs)  & (cgs)   \\

\hline
    &   \multicolumn{3}{c|}{(emp)}  &  \multicolumn{3}{c|}{(synth)} & 
    \multicolumn{2}{c|}{(emp)} & \multicolumn{3}{c|}{(synth)} & 
    \multicolumn{3}{c|}{(synth)}   \\

\hline

\multicolumn{15}{|c|}{ Primary sample } \\

\hline

 J040422 &           $-2.24$ &                $0.09$ &                $0.32$ &   $-1.56$ &      $0.17$ &          -- &  $8057$ &       $416$ &     $8003$ &        $473$ &           $12$ &      $4.5$ &        $0.9$ &          $0.3$ \\
  J050902 &           $-2.83$ &                $0.07$ &                $0.19$ &   $-1.43$ &      $0.29$ &        $0.11$ &  $8375$ &       $509$ &     $7548$ &        $389$ &          $147$ &      $4.6$ &        $0.9$ &          $0.3$ \\
   J034239 &           $-2.54$ &                $0.08$ &                $0.42$ &   $-1.34$ &      $0.42$ &        $0.06$ &  $6973$ &       $185$ &     $6746$ &        $650$ &           $66$ &      $3.1$ &        $1.6$ &          $0.6$ \\
 HV205840 &           $-2.65$ &                $0.07$ &                $0.22$ &   $-1.23$ &      $0.31$ &        $0.13$ &  $6623$ &       $215$ &     $7025$ &        $488$ &          $164$ &      $3.5$ &        $0.3$ &          $0.6$ \\ 
      J023001 &           $-2.49$ &                $0.07$ &                $0.22$ &   $-1.80$ &      $0.10$ &          -- &  $6963$ &       $195$ &     $7131$ &        $583$ &           $13$ &      $3.5$ &        $1.6$ &          $0.6$ \\
HV210205 &           $-2.10$ &                $0.10$ &                $0.28$ &   $-1.05$ &      $0.45$ &        $0.02$ &  $6200$ &       $157$ &     $6870$ &        $645$ &          $114$ &      $3.3$ &        $1.5$ &          $0.3$ \\
 HV204704 &           $-2.12$ &                $0.09$ &                $0.27$ &   $-1.18$ &      $0.39$ &        $0.04$ &  $7013$ &       $241$ &     $6526$ &        $563$ &           $48$ &      $2.4$ &        $1.3$ &          $0.2$ \\
 HV210918 &           $-2.49$ &                $0.08$ &                $0.20$ &   $-1.64$ &      $0.49$ &        $0.30$ &  $6730$ &       $222$ &     $6296$ &        $743$ &          $397$ &      $3.4$ &        $1.7$ &          $0.4$ \\
  HD 76483 &                -- &                    -- &                    -- &   $-0.32$ &      $0.47$ &        $0.16$ &      -- &          -- &     $8482$ &        $579$ &          $224$ &      $3.5$ &        $0.5$ &          $0.5$ \\

\hline 
 \multicolumn{15}{|c|}{ Secondary sample } \\

\hline
  J054653 &           $-2.00$ &                $0.09$ &                $0.20$ &   $-1.37$ &      $0.85$ &        $0.00$ &  $4515$ &        $45$ &     $6447$ &       $1456$ &            $0$ &      $1.8$ &        $2.5$ &          $0.0$ \\
    J044339 &           $-2.73$ &                $0.08$ &                $0.01$ &   $-2.11$ &      $0.40$ &        $0.90$ &  $6635$ &       $131$ &     $6179$ &        $524$ &          $940$ &      $3.7$ &        $1.7$ &          $0.5$ \\
    J051213 &           $-1.95$ &                $0.10$ &                $0.09$ &   $-1.57$ &      $0.87$ &        $0.00$ &  $6489$ &       $178$ &     $6182$ &       $1295$ &            $0$ &      $1.6$ &        $1.4$ &          $0.0$ \\
 HiTS091050 &           $-2.19$ &                $0.07$ &                $0.27$ &   $-1.60$ &      $0.86$ &        $0.24$ &  $6525$ &       $141$ &     $5931$ &       $1111$ &          $332$ &      $2.9$ &        $2.5$ &          $0.8$ \\

\hline

\end{tabular}
\end{center}
\end{table*}


\section{Sample selection and observations}
\label{sec:sample}

\subsection{Sample selection}

We selected a subsample of halo RRLs from previous studies, namely the Catalina surveys \citep{Drake14,Drake17,Torrealba15}, the HiTS survey \citep{Medina2018}, and the Halo Outskirts with Variable Stars survey \citep[HOWVAST;][]{Medina2021}. 
The Catalina surveys consist of an extensive database of $V-$band photometry for thousands of variable sources observed over $\sim$ 33,000\,sq.\,deg, carried out with three dedicated telescopes (of 0.5, 0.7, and 1.5\,m primary mirror diameter). 
HiTS and HOWVAST, on the other hand, observed a combined total of $\sim$ 350\,sq.\,deg. of the halo with deep $g$ and $r$ images using the Dark Energy Camera \citep[DECam;][]{Flaugher15}, which is mounted on the Blanco 4\,m telescope at the Cerro Tololo Inter-American Observatory. 
HiTS was originally planned to look for supernova explosions in real time, whereas HOWVAST was specifically designed 
to detect RRLs in the outskirts of the MW.
We focused on RRLs with estimated heliocentric distances ranging from 15 to 165\,kpc, 
classified as ab-type and with pulsation periods longer than 0.48\,days. 
Our target sample consists of nine stars from the Catalina surveys, seven stars from HiTS, and four stars from HOWVAST. 
Thus, we observed a total of 20 RRab. 
Their main properties are summarized in Table~\ref{tab:Targets}.

We designed our observations to avoid pulsation phases $\phi$ in which the stars were predicted to be close to their minimum radii, where major distortions are expected to affect their spectral features. 
Thus, we preferred phases close to the quiescent stages of their atmospheres, at around 0.4 and 0.8 \citep{Kolenberg10}. 
For the faintest subsample, in particular, we selected $\phi$ near maximum radii, in the descending branch of the light curves ($\phi$ between 0.15 and 0.60).  
However, in a few cases, pulsation phases involving rapid changes in the targets' atmospheres were difficult to avoid.
In addition, a long-period modulation that leads to variations in the period and amplitude of RRLs, the so-called Blazhko effect \citep{Blazhko1907}, could affect part of our sample, since it is thought to be a common effect among ab-type RRLs. 
These modulations modify the amplitude of pulsation of RRLs by a few tenths of magnitudes and are observed in 20-30\,per\,cent of the RRab stars, with typical periods of weeks to months. 
High-precision studies have revealed, however, that the true incidence rate of the Blazhko effect is closer to 50\,per\,cent \citep[see e.g.,][]{Jurcsik2009,Benko2014}, with modulation time scales as short as a few days and as long as a few decades \citep[e.g.,][]{Jurcsik2006,Jurcsik2016}.  
It is noteworthy that the periods obtained by \citet{Medina2018} were computed using a relatively low number of observations (from 20 to 30 data points), and the ephemerides of the stars from the Catalina survey might have slightly varied since their times of observation. 
Therefore, our phase predictions might in addition suffer deviations from the actual pulsation values. 
The RRL light curves, and the estimated pulsation phases in which we performed our observations are provided in Table~\ref{tab:Targets} and displayed in Figure~\ref{fig:lcs}.

\subsection{Observation and data reduction}

The spectroscopic observations took place on two separate runs, carried out on 2019 January 14 and 15, and on 2020 November 9 and 11 (four nights in total), 
with the Magellan Inamori Kyocera Echelle (MIKE) double echelle spectrograph mounted at the 6.5-m Clay Magellan telescope at the Las Campanas Observatory in Chile. 
The wavelength interval covered by this instrument ranges from $\sim$ 3,500 to  
$\sim$ 9,500\,\AA\ with a few gaps at the reddest wavelengths. 
The wavelength range covered by MIKE allows us to study spectral regions with absorption lines of interest for the characterization of RRLs, such as the Ca II triplet (at 8,498\,\AA, 8,542\,\AA, and 8,662\,\AA), the Mg I triplet (5,167\,\AA, 5,173\,\AA, and 5,183\,\AA), and the Balmer lines (H$_\alpha$ at 6,563\,\AA, H$_\beta$ at 4,861\,\AA, H$_\gamma$ at 4,340\,\AA, and H$_\delta$ at 4,102\,\AA). 
For the targets observed during the first run, we used a 1\,arcsec slit, with which MIKE provides a resolution of $\sim$ 19,000 and 25,000 
in the red and the blue arm of the spectrograph, respectively.
Since the targets observed during the first run are the faintest in our sample, 
we selected a configuration with a slow readout time and a strong binning in the spectral direction ($\times$8) in order to increase the signal-to-noise ratio (S/N). 
Thus, we decreased the resolution with respect to the values assumed for a 1\,arcsec slit with MIKE resulting in $R\sim$ 2,000 and 3,000.

For the second run, we adopted a different observing strategy by increasing the slit width to 1.5\,arcsec 
with which, in principle, a resolution of $\sim$ 15,000 and 18,000 is achieved in the red and the blue, respectively. 
However, for this run we used a 2$\times$2 binning, leading to half of the respective resolutions.

In order to obtain a reasonable S/N per star yet avoiding spectral line smearing owing to the pulsations, we observed the RRLs with 900 and 1,200\,s exposures in the first and second observing campaigns, respectively, and proceeded to stack the individual (consecutive) spectra  later on.
The fraction of the periods of the stars observed in a single exposure during both campaigns range from 1.3 to 2.3\,per cent.  
To avoid the smearing of the spectral lines, we stacked only 2--6 spectra for each star, 
with which a total of 3.9 to 10.8\,per cent of the star's periods are covered by our observations.
The S/N of the continuum in the order in which the H$_\alpha$ absorption line lies, resulting from the coadded spectra, are provided in the last column of Table~\ref{tab:Targets} and displayed in Figure~\ref{fig:snrs} (colour-coded by the estimated $\phi$), showing that higher S/N was achieved for brighter (closer) RRLs.  

For wavelength calibration, thorium-argon comparison lamp exposures were obtained  at each star position during each night.
Two reference stars were also observed in order to be used as radial velocity standards (CS~22874-042 and HD~76483).

The data reduction, and the flux and wavelength calibrations were executed using the Carnegie Python tools pipeline \citep[CarPy; ][]{Kelson2000,Kelson2003}.
This pipeline produces spectra separated in orders covering $<$ 100\,\AA\ in both detectors. 
These orders are not merged afterwards by CarPy.
Finally, the spectra were normalized in an order-by-order basis, and shifted in wavelength, as explained in Section~\ref{sec:spectral}.
Part of the spectra of a subsample of our program stars is displayed in Figure~\ref{fig:Halphas}.

\section{Spectral analysis}
\label{sec:spectral}

\subsection{Radial Velocities}
\label{sec:RVs}

To determine the radial velocities to be used for posterior spectral analysis, we used the tools available in the Image Reduction and Analysis Facility \citep[IRAF; ][]{Tody1993} software.
We used the blue metal-poor star CS 22874 \citep{Preston2000}, obtained during our first run, 
and the star HD 76483 \citep{Layden94}, obtained in the second run, as standards for the radial velocity shifts as these two stars have been widely studied in the past, and their spectra should resemble those of RRLs. 
We ran the IRAF cross-correlation function routine {\sc fxcor} and {\sc rvcorrect} to determine the radial velocity shift of the spectra, 
using both the blue and red arms of the detector, and focusing on three different orders on average in each arm. 
Because our targets are remote and metal-poor, we only rely on orders with fairly strong and well-defined lines.
In this regard, the most valuable regions were those containing Balmer lines (mainly H$_\alpha$ and H$_\beta$), the Ca II triplet, and the Mg I triplet. 
The final radial velocity used to shift the spectra resulted from averaging the shifts measured from those orders individually, when available. 

The outcome of the radial velocity determinations is shown in Table~\ref{tab:RVsTable}, where the propagated uncertainties e $v_{\rm los}$ and the scatter in the measured radial velocities from different orders $\sigma_{RV}$ are displayed as a reference. 
It is worth mentioning that the scatter found when comparing velocities computed from different orders (lines), for the stars with several measurements available is in broad agreement with \citet{Sesar2012} (Figure~3 in his work), who measured the scatter of the radial velocities when using different spectral lines. 
\citet{Sesar2012} showed that such scatter generally ranges between 1 and 18\,km s$^{-1}$, and is smaller for phases $<$ 0.6 and for RRLs with larger light curve amplitudes. 

\textbf{\begin{figure}
\includegraphics[angle=0,scale=.38]{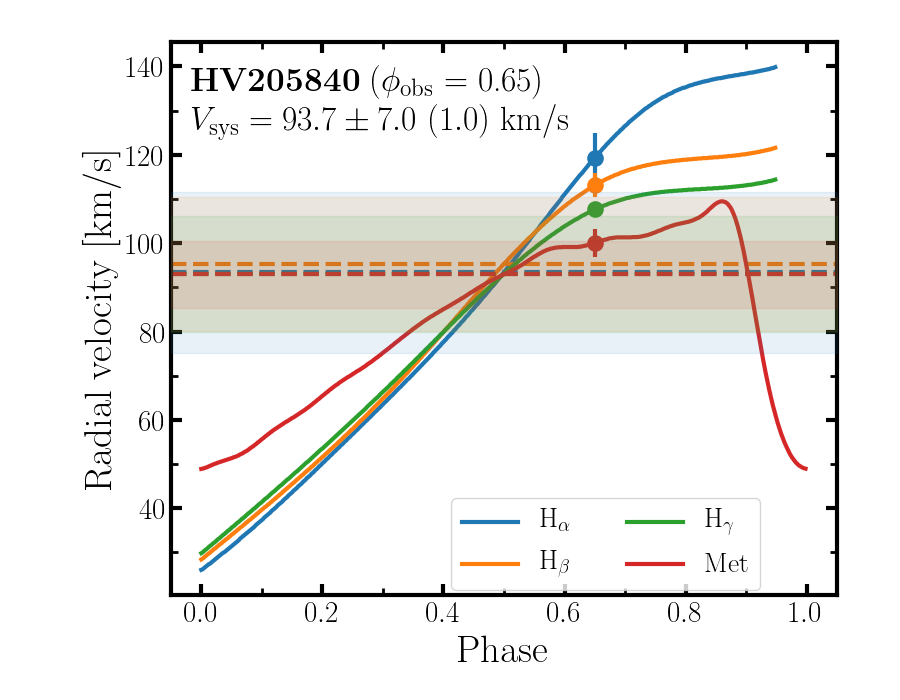} 
\caption{Example of the determination of the systemic velocity $V_{\rm sys}$ for one star, 
based on Balmer and metallic lines. The model radial velocity curves are those provided by \citet{Sesar2012} derived from H$_\alpha$ (blue), H$_\beta$ (orange), H$_\gamma$ (green), and metallic lines (red).
The radial velocities measured at the observed phase ($v_{\rm los}$) are plotted with filled circles. 
The systemic velocities derived from the use of each set of absorption lines are shown with horizontal dashed lines, with shaded regions representing their uncertainties.
This figure illustrates that, for a given $V_{\rm sys}$, the difference between line-of-sight velocities from Balmer and metallic lines depends on the observed phase, and can be up to $\sim$ 40\,km\,s$^{-1}$ even for $\phi<0.8$. 
}
\label{fig:Vsys}
\end{figure}}

We note that for the star J023001 the radial velocity shifts obtained from orders containing Balmer lines significantly differ from those obtained using orders with metallic features (with differences of $\sim$ 40\,km\,s$^{-1}$) as expected for RRLs in phases of rapid atmospheric changes \citep[$<0.2$ and $>0.8$;][]{For2011b}, although this also depends on the amplitude of the stars' light curve \citep{Sesar2012}.
In fact, differences of the order of 40\,km\,s$^{-1}$ can easily be found for halo RRLs observed at phases $>0.80$. 
Thus, the difference in velocities measured for J023001 
might indicate that our initial phase estimation ($\phi \sim 0.70$) is slightly off. 
In this specific case, taking advantage of the relatively high S/N of J023001, additional orders were used to better constrain its metallic-line-based radial velocity.  
We only used these additional orders for the radial velocity correction of J023001.

If one wishes to use the spectra of RRLs to perform kinematic and orbital analyses, it is necessary to subtract the velocity associated with the pulsations to obtain their so-called systemic (centre-of-mass) velocity $v_{\rm sys}$.
In this sense, it is important to consider that the line-of-sight velocities of RRLs obtained from their spectra depend on the lines used to determine them, and their depth in the stellar atmosphere.  
Thus, the amplitude and shape of the line-of-sight velocity curves vary depending on what lines are measured.
Larger line-of-sight velocity variations are expected for lines formed in the upper atmosphere (such as the Balmer lines), in comparison with metallic lines formed deeper in the atmospheres \citep[see e.g.][]{Liu91,Sesar2012}.
To estimate $v_{\rm sys}$ for our sample, we used the line-of-sight velocity templates provided by \citet{Sesar2012} in addition to our knowledge of the observed pulsation phases of our targets. 
Because those velocity templates scale with the $V-$band amplitude of the RRLs pulsation, we transformed the $g$ and $r$ light curve amplitudes to the $V-$band 
using the transformations provided by \citet{Sesar2012} ($A_V = 0.9\ A_g$, $A_V = 1.21\ A_r$).  
The final $v_{\rm sys}$ were obtained by using the measured line-of-sight velocities and the corresponding templates of the Balmer lines, Ca triplet, and Mg triplet (when available) independently, and minimizing the scatter of the resulting $v_{\rm sys}$ after allowing for small shifts in $\phi$ ($\pm \ 0.1$ around the expected phase).
Figure~\ref{fig:Vsys} depicts the differences between $v_{\rm los}$ and $v_{\rm sys}$ for one of the RRLs in our sample. 
It is clear from the figure that observing an RRL at a quiescent phase (e.g., $\phi \sim 0.4$) results in a smaller scatter in the final value of $v_{\rm sys}$.

\textbf{\begin{figure}
\includegraphics[angle=0,scale=0.35]{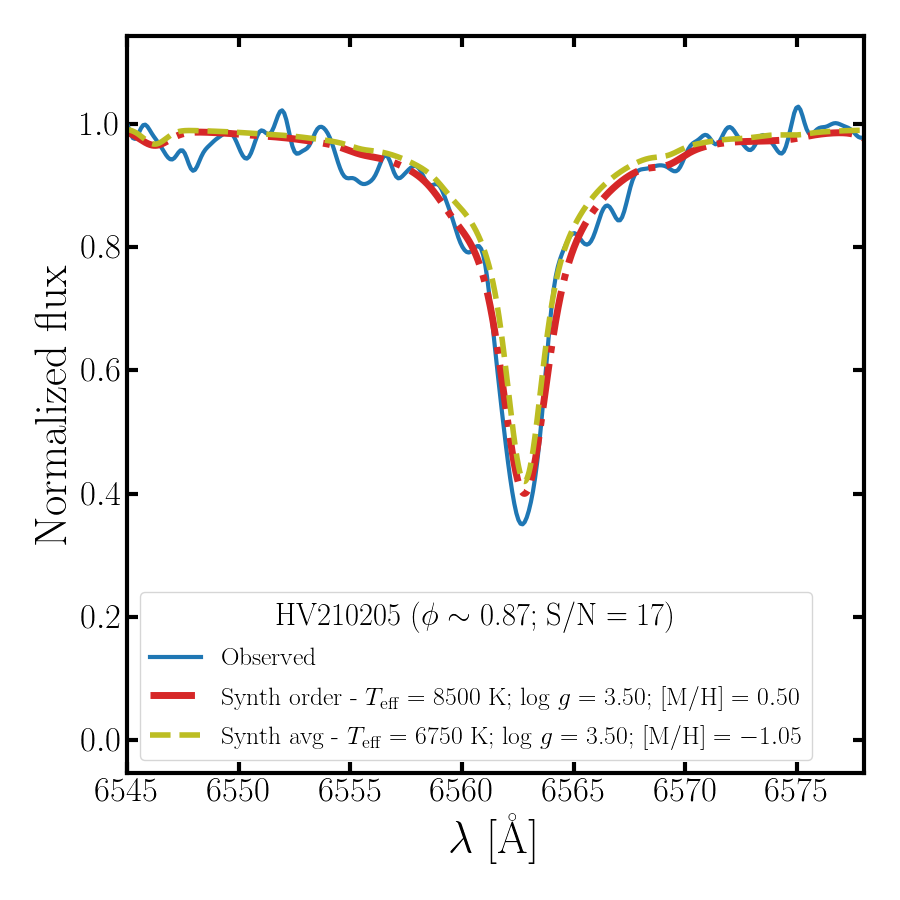} 
\includegraphics[angle=0,scale=0.35]{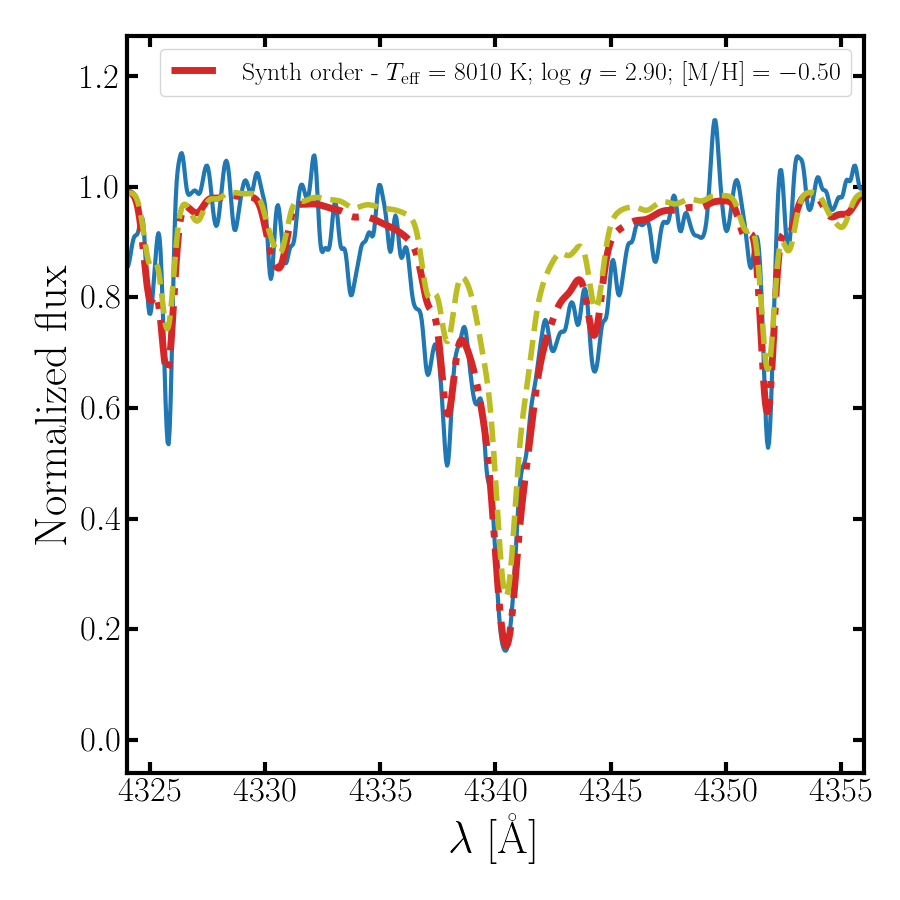} 
\caption{
H$_\alpha$ and H$_\gamma$ profiles of an RRL from our primary sample (HV210205).
In these panels we overplot, for comparison, the best stellar atmosphere fit for the selected orders (red lines) and the average from all the used orders (yellow dashed lines).
This figure shows the discrepancies that can be observed in the 
stellar parameters computed from different (single) orders. 
}
\label{fig:synth}
\end{figure}}

\subsection{Spectroscopic stellar parameters}
\label{sec:synthesis}

In order to determine the atmospheric parameters (e.g., effective temperatures and metallicities) of our program stars, we first derived initial estimates using various techniques. 
Estimating the gravity (log $g$) turned out to be the hardest as is often the case (see e.g. \citealt{Jofre2010} or \citealt{Hanke2020a}).
First we used photometry and the Infra-Red Flux method (IRFM) to determine the temperature, and parallaxes to compute gravities (following \citealt{Nissen97}). 
The metallicity was computed using different spectroscopic tracers and empirical methods, including the $\Delta$S technique. 
However, deriving the stellar parameters in these faint, variable stars turned out to be challenging and the above mentioned methods and their results were instead used as initial guesses in a purely spectroscopic approach.
For a detailed description of this process we refer the reader to Section~\ref{sec:atms} in the Appendix.

We employed the spectral analysis tool iSpec \citep{BlancoCuaresma14a,BlancoCuaresma19}, which outputs the best fitting parameters based on a $\chi^2$ minimization criterion.
For this, we worked with synthetic spectra generated by MOOG \citep[][version 2019]{Sneden1973} and ATLAS9 model atmospheres \citep{Castelli03}, relying on the Vienna Atomic Line Database (VALD) atomic line lists in the range 3,000 -- 11,000\,\AA\  \citep{Piskunov95,Ryabchikova15}. 

The spectroscopic atmospheric parameters were determined following different approaches.
In the first case, considering that the orders of our spectra are not merged, 
we chose five non-contiguous orders at different wavelengths in each side of the detector (ten in total), in which strong parameter-sensitive lines are present.
Due to the narrow wavelength range covered in each order, however, the statistics of the lines fit remains poor, which negatively affects the resulting parameters, especially when the (few) targeted lines are affected by noise or stellar pulsations.
Moreover, using a single order to determine stellar parameters can bias the metallicity due to the small number of metal tracers in the order's wavelength range. 
These biases might also affect the estimation of $T_{\rm eff}$, as it can easily differ by 400-500\,K when determined from, e.g., H$_\alpha$ and H$_{\beta}$. 
Thus, we ran iSpec on the concatenation of the aforementioned orders. 
In the second approach, we used a selection of orders numbered with odd and even identifiers, in order to cover a broader wavelength range and in turn more lines to improve the precision of the parameter determination. Separating the orders into odd and even allows for the use of the overlapping regions between orders without having to merge the orders. 
Because $\sim$ 70\,per cent of the entire wavelength range covered by MIKE is measured in more than one (contiguous) order, the resulting atmospheric parameters from odd and even concatenations are not independent from each other. 
From hereon, we use the results from the concatenation of even orders as derived parameters, given that they resemble those from the ten selected orders but with smaller uncertainties overall. 
The only exceptions for this are HV210205, for which using the odd orders results in significantly smaller uncertainties (due to the presence of distorted lines in the even orders, at $\phi \sim 0.87$), and J051213, for which neither even nor odd orders provide sensible solutions (due to the low number of visible lines at $\phi \sim 0.72$ and low S/N). 
Thus, for these two stars we use the odd and ten selected orders instead, respectively.

The fitting process was carried out with a maximum of six iterations of the code leaving the effective temperatures, metallicity, and log $g$ as free parameters, while fixing the rest of the required parameters  (micro/macro turbulence velocities, rotation, resolution, and limb darkening coefficient). 
We used the values obtained in Section~\ref{sec:fehs} and \ref{sec:teffs} as initial estimations for the [Fe/H] and effective temperatures. 
For all the stars we adopted a fixed value for the limb darkening coefficient (0.6) and $v\, {\rm sin }\, i=2.0$\,km\,s$^{-1}$. 
Figure~\ref{fig:synth} shows an example of spectra with the best fitting parameters from using a single order, and the average of all the orders considered.

An exception in this treatment is made for J023001 and J040422.
For these RRLs, the use of our method results in a metallicity of $-1.23\pm0.30$ and $-0.61\pm0.23$, which is not compatible 
with the visible Fe lines in the range 3,800-5,000\,\AA\ (including Fraunhofer lines), after a visual comparison with synthetic spectra.
This might be attributable to their observed phases (between 0.70 and 0.80 for J023001, and $>0.90$ for J040422). 
Therefore, we re-estimated J023001's and J040422's [Fe/H] by following the EW approach (Section~\ref{sec:Abund}) on clean Fe lines in this wavelength range, from which we obtain [Fe/H] $=-1.80\pm0.10$ and $-1.56\pm0.17$, respectively.
We adopted these values for the rest of their analysis.

Our results are shown in Figure~\ref{fig:atms}, and summarized in Table~\ref{tab:atms}. 
We note in passing that the parameters derived in this section represent the atmosphere of the stars at the moment of the observations, which in most cases corresponds to phases of atmospheric contraction, with a decrease in luminosity, and a plateau in $T_{\rm eff}$
\citep[$\phi$ between 0.40 and 0.85;][]{For2011b, Kolenberg10}.

\subsection{Stellar parameters and uncertainties}
\label{sec:StellarParam}

The observed phase and S/N of an RRL define its suitability for the method described above, which might result in unreliable stellar parameters and/or large uncertainties. 
Thus, we subdivide our target stars into two groups depending on their spectrum quality and phase:
a primary sample, containing stars with relatively high S/N, ideal observing phases, and well constrained atmospheric parameters, and a secondary sample for which the spectra were not observed in optimal conditions and/or low S/N, which resulted in loosely constrained atmospheric parameters.
We include J054653 in the secondary sample since, albeit its relatively high S/N ($\gtrsim$15), it was observed close to maximum light. 
J044339, J051213, and HiTS091050 are included in the secondary sample due to their low S/N ($\lesssim$10) and phase of observation $\sim$ 0.75, at a stage of abrupt atmospheric kinetic energy changes \citep{Kolenberg10}.

For both the primary and secondary sample, the errors in the atmospheric parameters are computed by propagating the uncertainties resulting from the iSpec routine only if the derived values are well defined for a given method. 
Hence if a parameter does not make physical sense or does not return a reasonable uncertainty, we define it as a limit or flag the values.
Table~\ref{tab:atms} shows the atmospheric parameters together with their uncertainties, and the scatter $\sigma$ originating from the empirical relations employed and the spectroscopic measurements.

The uncertainties derived for the stellar [Fe/H] from spectrum synthesis range from  $\sim0.10$ to $\sim0.49$\,dex 
with a mean error of 0.34\,dex and low scatter overall (0.02-0.30\,dex), for the stars in the primary sample. 
For the stars in the secondary sample, the uncertainties are larger in general, with a mean of $\sim0.70$\,dex. 
In contrast, the [Fe/H] values obtained from empirical models (Section~\ref{sec:fehs}) display a mean propagated uncertainty of $\sim$ 0.10\,dex for both the primary and the secondary sample, displaying a scatter between methods that ranges from 0.02 to 0.42\,dex. 
We note that the [Fe/H] values derived from spectrum synthesis are systematically more metal-rich than those from scaling methods, regardless of the metallicity scale used.

The resulting propagated error on the temperature is typically 400-600\,K, whereas the scatter of the derived $T_{\rm eff}$ for this sample ranges from  10\,K to 400\,K, and lies below 70\,K for three stars. 
The stars in the secondary sample display, on the other hand, a mean uncertainty of about twice that of the primary sample.
We note in passing that for J054653 neither 
the estimations of Section~\ref{sec:teffs} nor the spectrum synthesis comparison gave satisfactory results. 
We thus consider it a secondary sample star.

We observe a mean difference of $\sim$ 100\,K when comparing our derived spectroscopic and photometric temperatures (the latter being hotter), with a standard deviation of 460\,K. 
This is consistent with the results of \citet{Mucciarelli2020}, who found the spectroscopic $T_{\rm eff}$ from giant stars to be lower than the photometric ones, with discrepancies increasing with decreasing metallicity, and reaching differences of $\sim 350$\,K at [Fe/H] $\sim-2.5$\,dex (as shown in Figure~9 in their work). 
\citet{Mucciarelli2020} concluded that these differences cannot be interpreted as being due to systematic errors, and are likely due to the physics adopted for the spectroscopic studies. 
Thus, they rely on the choice between temperatures that reproduce the stellar flux and those that reproduce the depths of individual metallic lines. 

Finally, we stress that the temperatures obtained here represent those at the moment of observation, and that the range of $T_{\rm eff}$ that an RRL can exhibit can easily reach 2,000\,K \citep{Pena09, For2011b}. 
For later reference, we list the observed phases in Table~\ref{tab:Targets}.

In the case of the stellar surface gravities, we were only able to derive them using spectrum synthesis. Thus, we use those values hereafter. 
The typical uncertainty on the derived log $g$ for the stars in the primary sample is 0.8\,dex (mean), and this group displays standard deviations (from the use of different order combinations) ranging from 0.2--0.6\,dex. 
The uncertainties are significantly larger for the secondary sample, where the mean error exceeds 1\,dex. 
As the overall uncertainties in the surface gravities are rather large, we highlight that the derived values of log $g$ should be taken with caution. 
It is worth having in mind that, in general, the typical minimum-to-maximum variations of log $g$ of RRLs throughout their pulsation cycle can reach values of $\sim$ 1.5\,dex \citep{Pena09, For2011b}. 
Additionally, \citet{Mucciarelli2020} found that even for non-variable stars the discrepancies between spectroscopic and photometric surface gravities can be as high as $\sim$ 1\,dex.  
As a result, we loosely fix the log $g$ of the secondary sample to $\sim 2$\,dex \citep[which is a reasonable assumption considering the work by][]{For2011b} and only use this to complete our stellar parameters and metallicities. No abundances have been computed for the secondary sample owing to the large uncertainties in stellar parameters.
In Figure~\ref{fig:atms} we display the derived stellar parameters and associated uncertainties for the stars in both samples.

\textbf{\begin{figure*}
\includegraphics[angle=0,scale=.235]{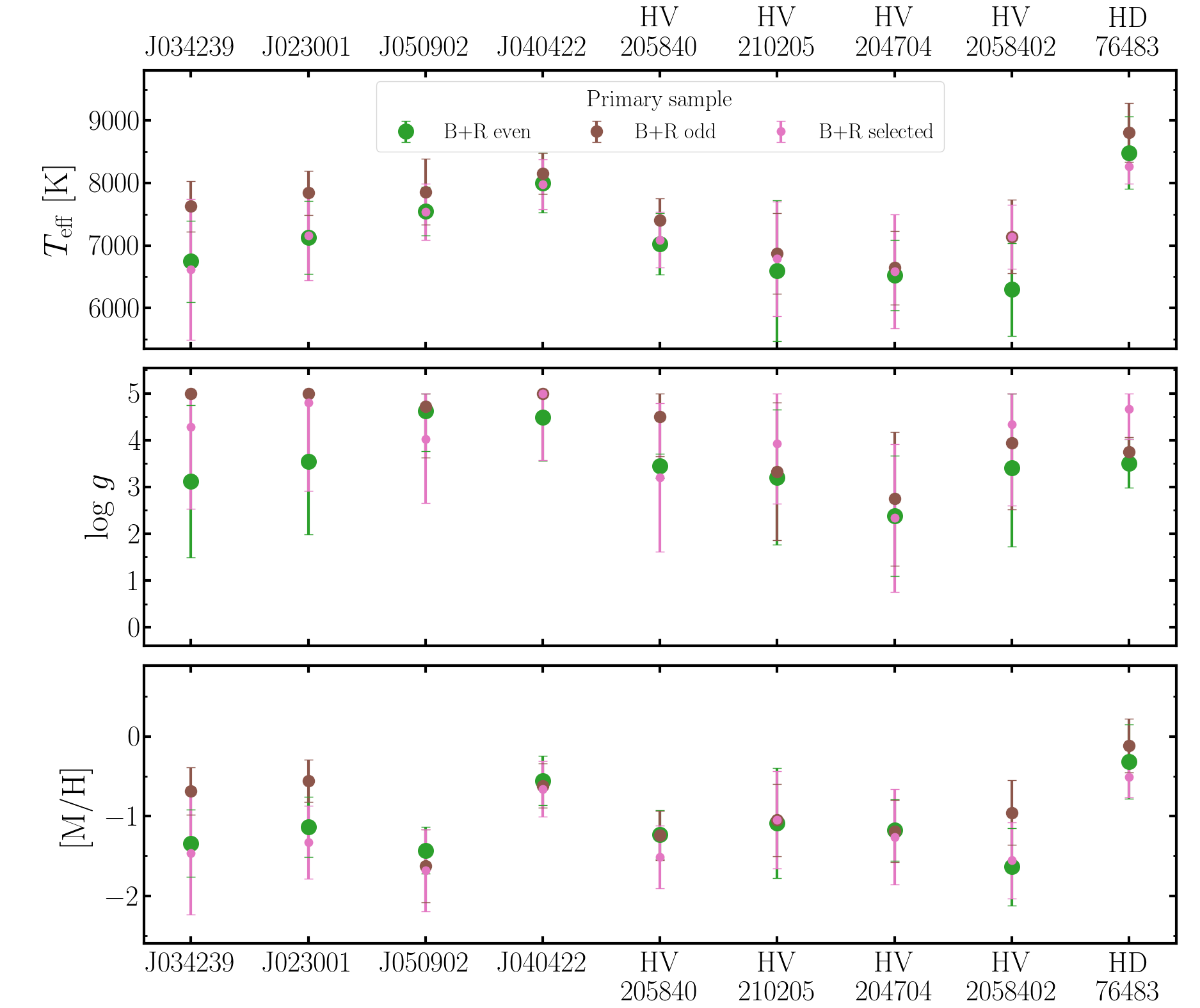} 
\includegraphics[angle=0,scale=.235]{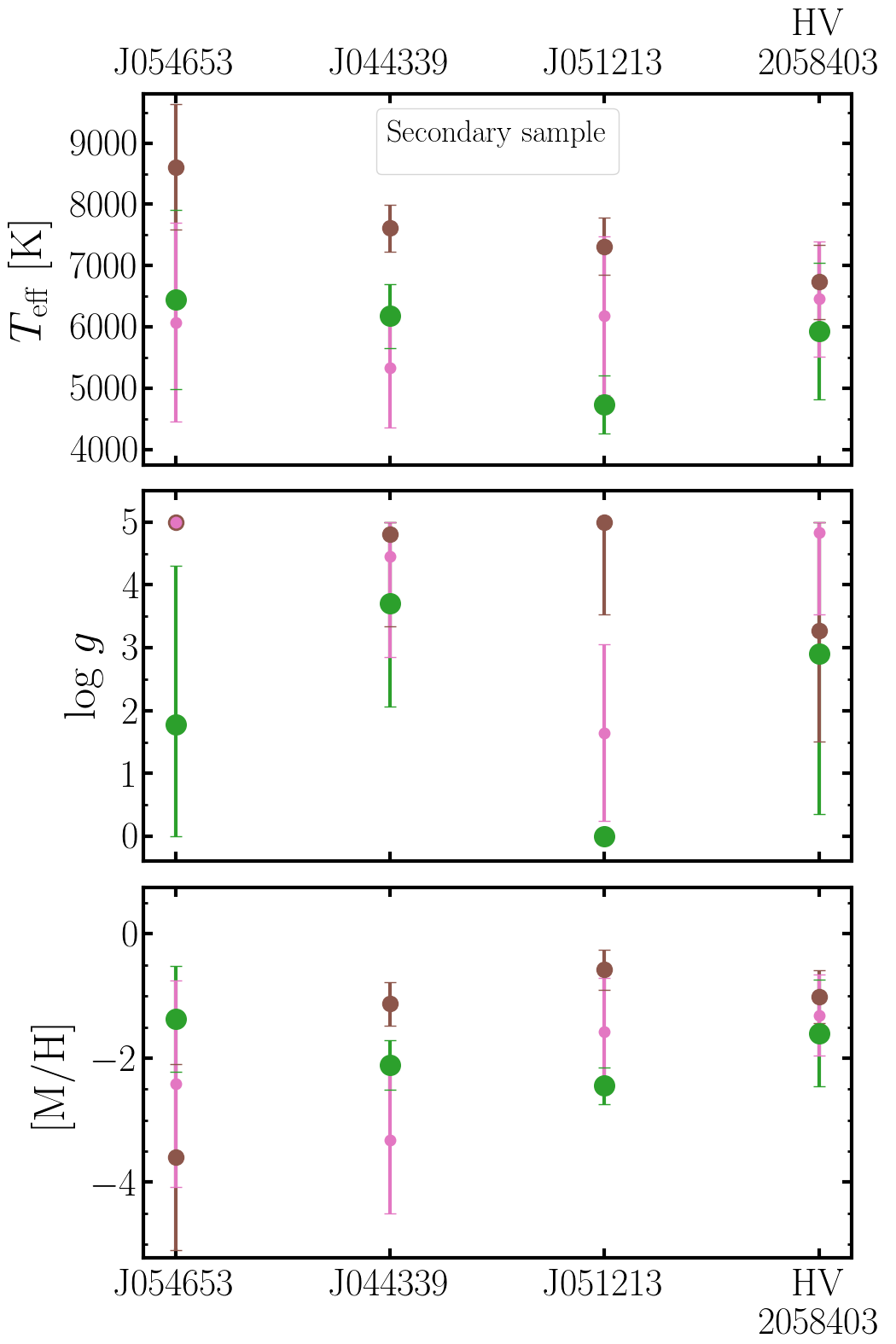} 
\caption{Stellar parameters (as returned by iSpec) and uncertainties of the subsample of our program stars with higher resolution (including the radial velocity standard HD~76483). 
The values derived from the use of even/odd/ten selected orders in both sides of the detector (B+R) are shown with green/brown/pink symbols.
The {\it left} panels depict the RRLs that we consider part of our primary sample, whereas those in our secondary sample are plotted on the {\it right}, respectively. 
These plots illustrate (in-)consistencies in the stellar parameters obtained when using different sets of orders, as well as the the overall higher uncertainties obtained for the secondary sample. 
} 
\label{fig:atms}
\end{figure*}}

\begin{sidewaystable*}\small
\caption{
Summary of the abundance ratios obtained in Section~\ref{sec:Abund}, for seven RRLs in our sample. 
For each element measured, the weighted mean is shown, in addition to its propagated uncertainty, standard deviation, and number of lines used. 
We use the symbol $<$ to flag abundances that we consider limits. }
\label{tab:abs}
\begin{center}

\begin{tabular}{|c|c|c|c|c|c|c|c|c}
\hline
 ID &  [$\alpha $/Fe] &  e [$\alpha $/Fe] &  $\sigma$ [$\alpha $/Fe] &  N $\alpha $ & [Na/Fe] & e [Na/Fe] & $\sigma$ [Na/Fe] & N  Na     \\
\hline
      J023001 &    $0.35$ &           $0.25$ &                  $0.15$ &        $12$ &     0.00 &      0.15 &             0.15 &     2  \\
     J034239 &     $0.40$ &           $0.30$ &                  $0.10$ &         $8$ &     -- &      -- &             -- &    --  \\
     J050902 &     $-0.05$ &           $0.20$ &                  $0.15$ &         $8$ &  $<-0.50$ &     $0.20$ &             -- &     1    \\
  HV205840 &      $-0.20$ &           $0.25$ &                  $0.20$ &         $6$ &    $-0.60$ &      0.25 &             0.15 &     2     \\
    J040422 &      $0.60$ &           $0.40$ &                  $0.10$ &         $8$ &     -- &      -- &             -- &    --     \\
  HV210205 &          $0.40$ &           $0.15$ &                  $0.10$ &        $10$ &   $0.25$ &    $0.15$ &             -- &     1  \\
 HV204704 &     $0.25$ &           $0.10$ &                  $0.15$ &        $11$ &     0.30 &      0.10 &             0.10 &     2  \\

\hline
\end{tabular}

\vspace{25pt}

\begin{tabular}{c|c|c|c|c|c|c|c|c|c|c|c|c|c|c|c}
\hline
   [O/Fe] & e [O/Fe] & $\sigma$ [O/Fe] & N  O &  [Mg/Fe] &  e [Mg/Fe] & $\sigma$ [Mg/Fe] & N  Mg & [Ca/Fe] & e [Ca/Fe] & $\sigma$ [Ca/Fe] & N  Ca & [Ti/Fe] & e [Ti/Fe] & $\sigma$ [Ti/Fe] & N  Ti \\
\hline
        $<0.65$ &    $0.3$ &          $0.15$ &    2 &    0.25 &      0.20 &             0.05 &     5 &  $<0.45$ &    $0.15$ &             -- &     1 &    0.35 &      0.30 &             0.05 &     4  \\
       $0.45$ &    $0.30$ &            -- &    1 &   $0.30$ &    $0.35$ &             -- &     1 &   $<0.10$ &     $0.10$ &             -- &     1 &    0.40 &      0.30 &             0.10 &     5 \\
       $-0.10$ &    $0.20$ &            -- &    1 &   $-0.05$ &      0.20 &             0.15 &     3 &    $-0.10$ &      0.20 &             0.20 &     2 &    0.10 &      0.20 &             0.00 &     2  \\
     -- &     -- &            -- &   -- &   $-0.25$ &      0.25 &             0.15 &     5 &     -- &      -- &             -- &    -- &  $0.05$ &    $0.35$ &             -- &     1 \\
      0.60 &     0.35 &            0.10 &    2 &    0.55 &      0.40 &             0.20 &     2 &     -- &      -- &             -- &    -- &    0.65 &      0.40 &             0.05 &     4  \\
  0.25 &     0.15 &            0.05 &    2 &    0.45 &      0.20 &             0.10 &     5 &     0.45 &      0.10 &             0.15 &     3 &    -- &      -- &             -- &    --  \\
     0.30 &     0.15 &            0.15 &    3 &    0.30 &      0.05 &             0.15 &     3 &     0.05 &      0.05 &             0.15 &     2 &    0.30 &      0.10 &             0.15 &     3  \\

\hline
\end{tabular}

\vspace{25pt}

\begin{tabular}{c|c|c|c|c|c|c|c|}
\hline
      [Sr/Fe] & e [Sr/Fe] & $\sigma$ [Sr/Fe] & N  Sr & [Ba/Fe] & e [Ba/Fe] & $\sigma$ [Ba/Fe] & N  Ba  \\
\hline
     -- &      -- &             -- &    -- &   $0.15$ &    $0.25$ &             -- &     1  \\
   $<0.45$ &    $0.15$ &             -- &     1 &     0.35 &      0.25 &             0.30 &     2  \\
  $<0.10$ &     $0.20$ &             -- &     1 &   $<0.20$ &     $0.20$ &             -- &     1  \\
 $<1.15$ &    $0.25$ &           $0.05$ &     2 &     -- &      -- &             -- &    --  \\
 0.40 &      0.40 &             0.05 &     2 &   $0.55$ &     $0.40$ &             -- &     1  \\
  $<1.40$ &    $0.15$ &           $0.05$ &     2 &  $<1.45$ &    $0.15$ &            $0.2$ &     2  \\
 $<1.45$ &     $0.10$ &             -- &     1 &  $<0.25$ &     $0.10$ &             -- &     1  \\

\hline
\end{tabular}

\end{center}
\end{sidewaystable*}


\subsection{Abundances}
\label{sec:Abund}

In addition to determining the stellar parameters described above, we derive local thermodynamic equilibrium (LTE) abundances for O, Na, Mg, Ca, Ti II,  Sr II, and Ba II employing the EW method and spectrum synthesis.
For this purpose, we used the direct integration of Gaussian profiles fitted to the observed lines using IRAF. 
We only use the lines if they appear clean, free of emission owing to the variability, and have a clearly defined continuum on at least one side of the line. The line list is taken from \citet{Hansen2011} and complemented with information from the National Institute of Standards and Technology (NIST) atomic spectra database\footnote{\url{https://www.nist.gov/pml/atomic-spectra-database}}. 
The abundandes were derived using a Python implementation of MOOG with the input parameters reported in Section~\ref{sec:synthesis}. 
Given that our program stars are faint, and as a consequence of  overall low S/N, we are forced to mainly use strong lines (e.g., Fraunhofer lines).

We then checked the consistency of the results from the EW approach with spectrum syntheses from MOOG. 
For Ba II and Sr II, however, we relied on the use of synthetic spectra only, to handle the hyperfine splitting and isotopic substructure of the atomic lines considered.
Figure~\ref{fig:BaFe} shows examples of the results of the abundance determination 
of Ba II in two of our stars, via synthetic specta. 
In general, we disregarded lines with an EW larger than 300\,m\AA, full-width-at-half-maximum outside the range 0.1--0.5, with shifts in central wavelength greater than 0.3\,\AA, or the ones that were based on unclear features, due to the noise or blends. 

To estimate the uncertainties of individual line measurements, 
we varied the input parameters of the models ([Fe/H], $T_{\rm eff}$, and log $g$) one by one based on their scatter 
while keeping the others fixed, and finally adding the abundance variations in quadrature.
On top of that, from the manual inspection of our spectra, we find that determining abundances at a level better than 0.1--0.15\,dex is, in general, unfeasible.
Therefore, a systematic uncertainty, or a ground level for the errors, should be considered.

To correct for LTE departures, we used the online interface hosted by the Max-Planck-Institut f{\"u}r Astronomie (MPIA)\footnote{\url{https://nlte.mpia.de/gui-siuAC\_secE.php}}. These non-LTE (NLTE) corrections apply for O \citep{Sitnova13}, Mg \citep{Bergemann17}, Ca \citep{Mashonkina07}, and Ti \citep{Bergemann11}. 

We note in passing that the RRLs' abundance ratios are not expected to vary significantly throughout their pulsation cycles \citep[see e.g. Figure 13 from][]{For2011b}, even if the changes in the RRLs' effective temperatures amount to $\sim800$\,K. 
For spectra taken at phase $\phi\sim0.35$, however, atomic lines are expected to suffer from minimal blending, and are therefore best suited for chemical composition analyses.
On the other hand, observing RRLs on the fainter end of the descending branch is beneficial, for instance, 
for metal lines with low excitation potentials (that saturate at cooler parts of RRLs cycles), as they are weaker at hotter phases.

We compute abundance ratios [X/Fe] relative to the solar abundances of \citet{Asplund2009}.    
We note in passing that adopting a different solar mixture, such as that recommended by \citet{Magg2022}, 
does not change significantly our reported results.
In fact, the absolute abundance differences are equal or smaller than 0.05\,dex for all the elements analysed in this work with the exception of O, for which the abundance reported by \citet{Magg2022} is 0.08\,dex higher. 
Therefore, this would just lead to a systematic offset in the Galactic chemical evolution models well within our estimated errors, not affecting our results and conclusions. 
In Table~\ref{tab:abs} we provide the averaged abundance ratios for each element, weighted by our confidence in the line measurement (limits and saturated lines were given half weight and flagged in the list). 
Additionally, we list the dispersion of the abundances when more than one line is available and passed the aforementioned cuts.
The line-by-line atomic data of the line list used is presented in Table~\ref{tab:lines}, including excitation potentials, log $gf$, EWs, individual abundances, and NLTE corrections.

\textbf{\begin{figure}
\includegraphics[angle=0,scale=0.40]{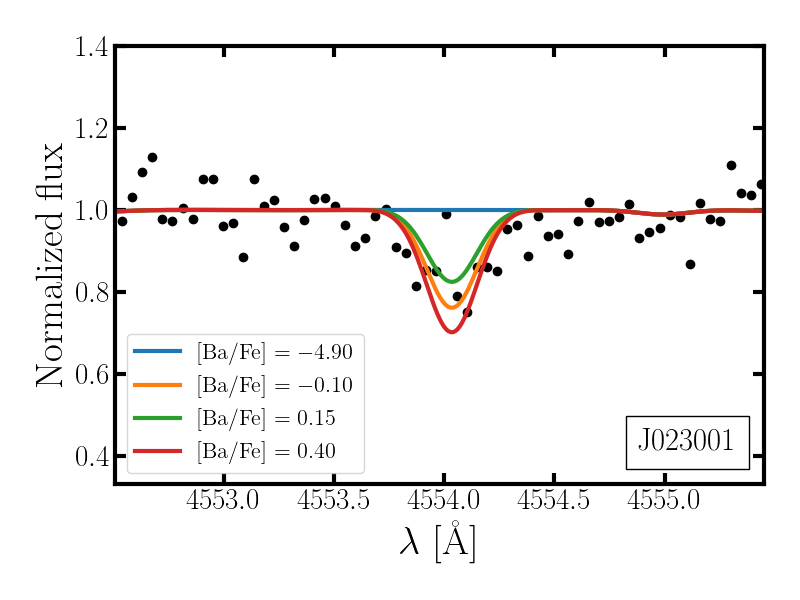} 
\includegraphics[angle=0,scale=0.40]{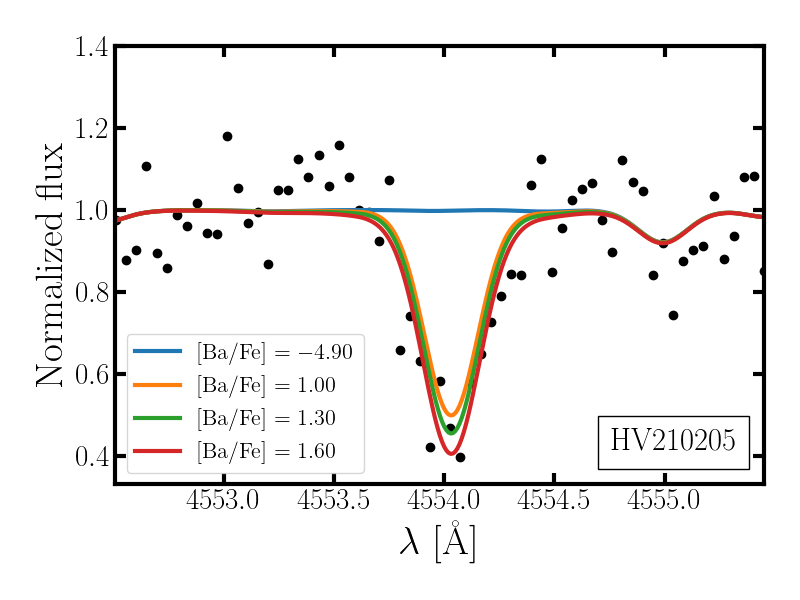} 
\caption{Ba absorption line at 4554.03\,\AA\ for two of the program RRLs (J023001 and HV210205 in the {\it top} and {\it bottom} panels, respectively), displaying the estimated [Ba/Fe], illustrative variations around the estimated values (of 0.25\,dex in the {\it top} panel and 0.30\,dex in the  {\it bottom} panel),
and continuum levels. 
}
\label{fig:BaFe}
\end{figure}}

\textbf{\begin{figure*}
\includegraphics[angle=0,scale=.31]{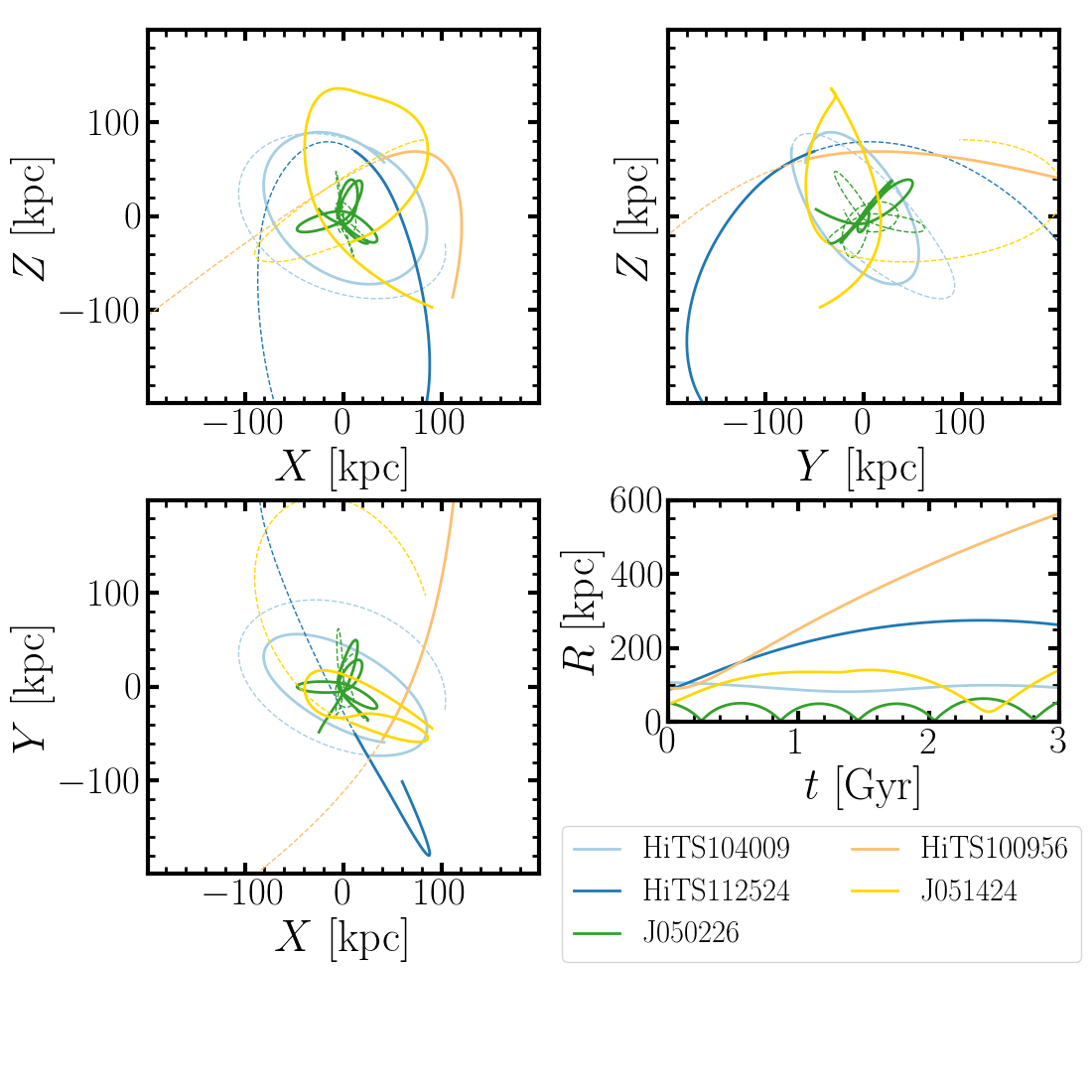} 
\includegraphics[angle=0,scale=.31]{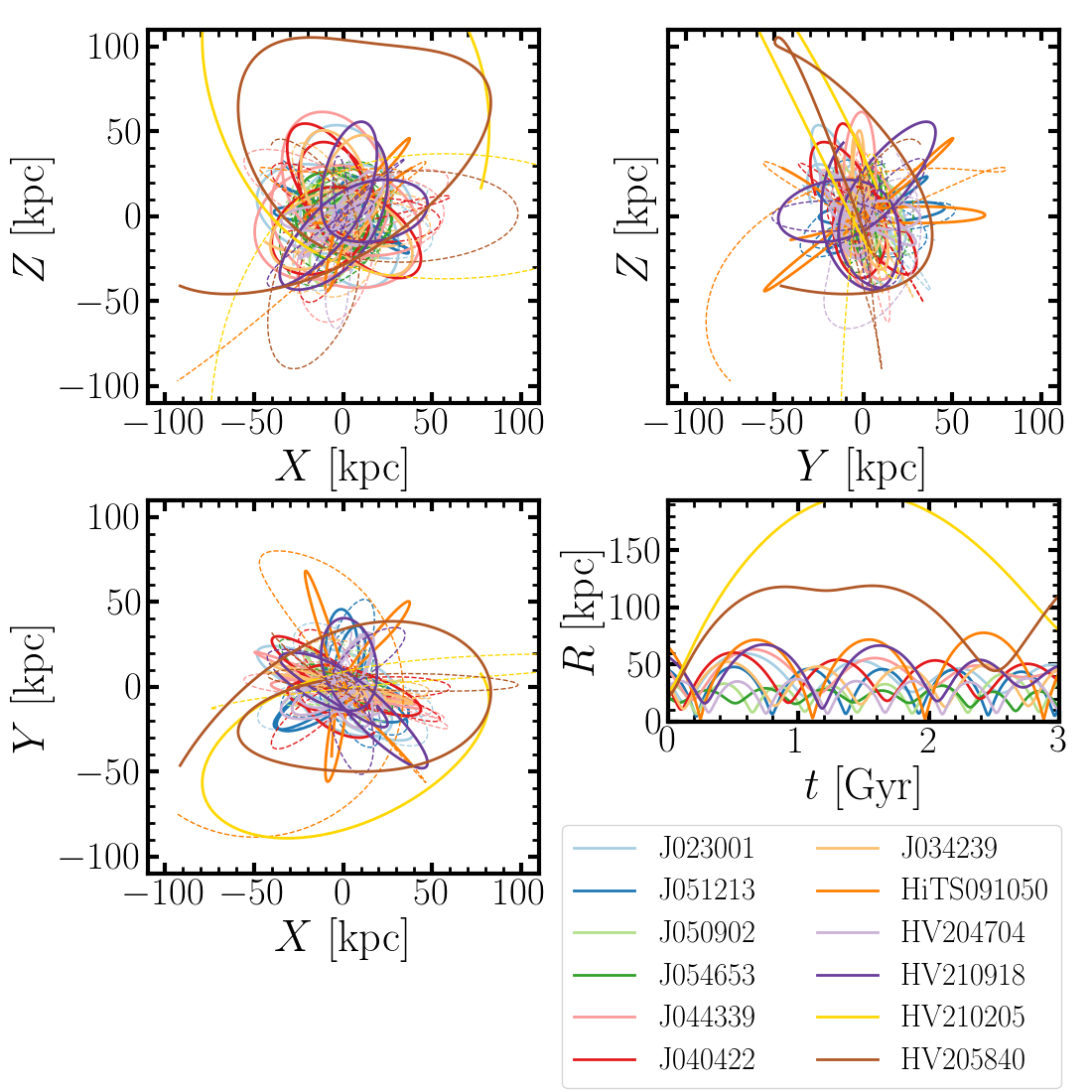} 
\caption{Orbits of the 17 stars for which we possess 3D kinematic information, computed with GALPY, accounting for the LMC infall, and integrated for 3\,Gyr forward and backward (solid and dashed lines, respectively). 
The {\it left} panels show the five RRLs observed in our first run (with lower spectral resolution and S/N), and the {\it right} panels show the RRLs from our second run.
These figures illustrate that the gravitational perturbation of the LMC can affect the orbit of the stars significantly in some cases (e.g., J051424 and HV205840). 
The orbit and position of the LMC relative to three of our RRLs is displayed in Figures~\ref{fig:abundOrbits1} to \ref{fig:abundOrbits3} to illustrate the effect of the LMC on individual orbits. 
}
\label{fig:orbitsAll}
\end{figure*}}

\subsection{Orbital analysis}
\label{sec:Orbital}

We combine the radial velocities derived in Section~\ref{sec:RVs} with proper motions in right ascension and declination ($\mu^*_\alpha$ and $\mu_\delta$, respectively) from the {\it Gaia} third data release (DR3) to estimate the orbital parameters and history of our sample of halo RRLs. 
We exclude three of the stars from this part of the analysis (HiTS101243, HiTS103943, and HiTS102414, at $d_{\rm H} \sim 90$, $110$, and $165$\,kpc, respectively) as they are not listed in the {\it Gaia} catalogue.  
Together with the kinematic information of our targets, we take advantage of the RRLs being standard candles to obtain a precise distance estimation for each star. 
We estimated the heliocentric distance $d_{\rm H}$ of our RRLs by adopting the period-luminosity-metallicity (PLZ) relation from \citet{Sesar2017}, using the periods from Table~\ref{tab:Targets} and the metallicities from Section~\ref{sec:synthesis}. 
For the stars with no metallicity information, 
we adopted [Fe/H] $= -1.5$ as a reasonable representation of the Galactic halo metallicity distribution function 
\citep[see e.g. ][]{Suntzeff1991,Prantzos2008,Liu2018,Conroy2019}.  

In order to integrate the stellar orbits, we use the python package \textsc{GALPY} \citep{Bovy15}\footnote{\url{http://github.com/jobovy/galpy}}, 
adopting an \textit{isolated} model MW potential consisting of a spherical nucleus and bulge (Hernquist potential),  a Miyamoto-Nagai disc model, 
and a spherical Navarro-Frenk-White (NFW) dark matter halo. 
In \textsc{GALPY} this corresponds to the built-in \textit{MWPotential2014} preset. 
We used a second potential (hereafter called \textit{perturbed} potential) that takes into account the growing evidence of the perturbations caused by a massive LMC to the MW gravitational potential \citep{vanderMarel14,Laporte18, Erkal18,Erkal20,Vasiliev21,Cunningham20}. 
For the LMC, we assumed the equatorial coordinates $\alpha = 78.77$\,deg and $\delta = -69.01$\,deg, a distance $d_{\rm LMC} = 49.6$\,kpc \citep{Pietrzynski2019}, proper motions $\mu_{\alpha^*} = 1.85$\,mas\ yr$^{-1}$ and $\mu_\delta = 0.234$\,mas\ yr$^{-1}$ \citep{GaiaHelmi18}, 
and 262.2\,km\ s$^{-1}$ as its systemic line-of sight velocity \citep{vanderMarel02}. 
Additionally, we multiply the MW halo mass by 1.5 for the isolated and the perturbed potentials to correct for the fact that the LMC is unbound in \textit{MWPotential2014}\footnote{Following \url{https://docs.galpy.org/en/v1.5.0/orbit.html}}. 

For the mass of the LMC, M$_{\rm LMC}$, we used $1.88 \times 10^{11}$\,M$_\odot$, based on the recent estimation of \citet{Shipp21} from stellar streams.
In addition, we assumed a scale length $a_{\rm LMC} = 20.22$\,kpc, with which the input parameters used match the observed circular velocity 91.7\,km\ s$^{-1}$ at 8.7\,kpc from the LMC centre \citep{vanderMarel14}.
Given that the LMC is a massive MW satellite, we decided to take the Chandrasekhar dynamical friction into account for its orbit integration. 
In this work, we ignore the impact of other massive perturbers of the MW potential, such as the Sagittarius (Sgr) dwarf spheroidal galaxy. 

For the isolated potential, we integrate the orbits for 10\,Gyr backward and forward, with a step size of 1\,Myr.
In the case of the perturbed potential, the integrations are limited to 3\,Gyr in both directions, assuming that the perturbations in the MW potential beyond these limits are likely not significant.
In order to obtain uncertainties for the derived orbital parameters, we draw 100 input parameters (systemic velocities, heliocentric distances, and proper motions) assuming Gaussian distributions and using the covariance matrices of the stars from the {\it Gaia} DR3 catalogue.
Then, we select the median value of the resulting parameters, and the 16 and 84 percentiles as errors to represent the asymmetry of their distributions. 
We find that estimating the orbital parameters from the distribution of 100 drawings is sufficient to reach convergence for the majority of stars and orbital parameters, as the shape of the resulting orbits and the parameters values remain consistent (within their uncertainties) when repeating this experiment. 
It is worth noting, however, that for the most distant stars in our sample parameters such as vertical angular momentum and apocentric distance result in poorly constrained estimates, as a consequence of their uncertain proper motions. 
Finally, we treat the fraction of bound solutions over the total number of integrated orbits as a proxy of the bound likelihood for each star.

Figure~\ref{fig:orbitsAll} displays the resulting orbits for the \textit{MWPotential2014} perturbed by the infall of the LMC.
The figure illustrates that the majority of the stars' orbits are not significantly affected by the choice of the potential, with a few exceptions (HV205840, HV210918, and J051424).
Figure~\ref{fig:orbitsAll} also shows RRLs with loosely bound orbits, namely HiTS112524 and HiTS100956.
Tables~\ref{tab:orbits1} and ~\ref{tab:orbits2} summarize the main orbital parameters of our target stars.

\textbf{\begin{figure}
\includegraphics[angle=0,scale=.42]{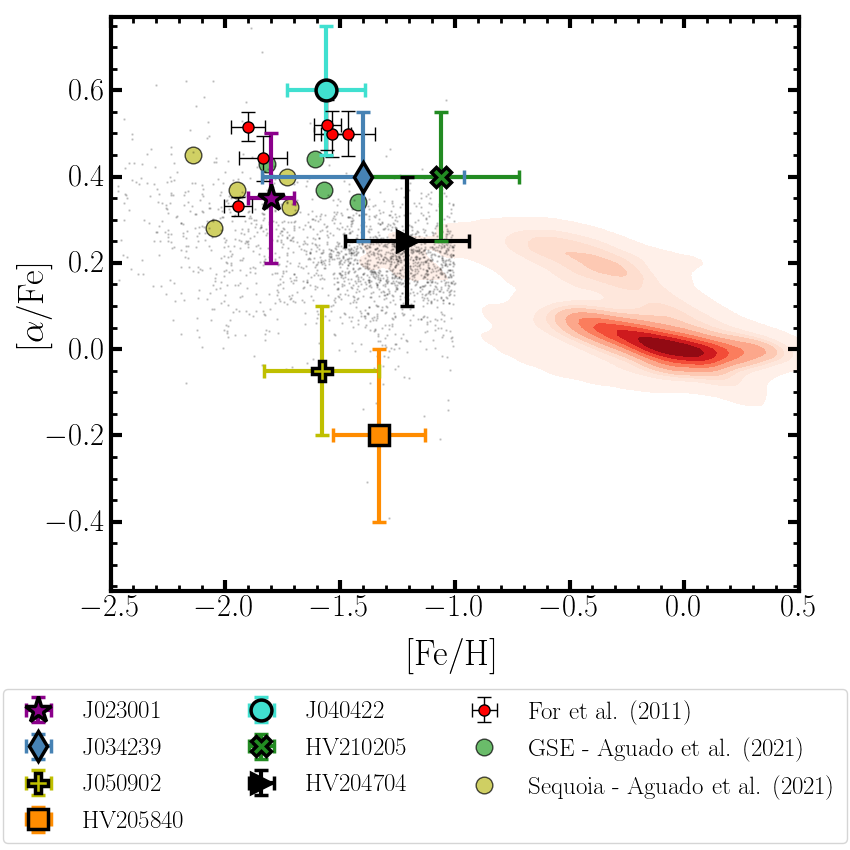} 
\caption{Relative abundance of $\alpha$-elements as a function of [Fe/H] for our target stars, computed assuming LTE. APOGEE-based abundance patterns of the MW non-halo and halo populations are displayed as red contours and grey dots, respectively. The red filled circles represent the chemical abundances of the RRLs studied by \citet{For2011b}, whereas those corresponding to stars in GSE and Sequoia \citep[from ][]{Aguado2021} are plotted in green and yellow, respectively. 
The error bars in the abundances represent their standard deviations $\sigma$ from Table~\ref{tab:abs}. 
It is clear from this figure that two of our RRLs, HV205840 and J050902, display an underabundance of $\alpha$-elements for their metallicity. 
}
\label{fig:alphas}
\end{figure}}

\section{Results}
\label{sec:results}

Given that the dynamical time-scales in the outer halo are long, the partial phase-mixing of accreted systems allows for the detection of tidal disruption signatures as overdensities \citep[e.g.,][]{Haywood2018}. 
Thus, halo stars with a common origin (same progenitor) tend to share similar orbital parameters and kinematics \citep[e.g.,][]{Belokurov2018b}, even if these stars do not lie close to each other when projected on-sky \citep[e.g.,][]{Hanke2020b}.
In addition, ``chemical tagging" allows one to trace back the stars' origins to their parent populations, based on their shared elemental abundances pattern \citep[e.g.,][]{Freeman2002,Buder2022}. 
Assessing especially the stellar [$\alpha$/Fe] vs [Fe/H] is a useful chemical diagnostic, as the level of the $\alpha-$element abundance ratio can help separate in-situ from ex-situ formed stars. 
Thus, with multi-dimensional stellar chemodynamics it is possible to hypothesize about the origin of a star by looking at tentative associations with known substructures.

Here we compare the resulting chemical abundances and orbital parameters of our target RRLs to those of MW halo stars. 
For these comparisons, we use data from the Apache Point Observatory Galactic Evolution Experiment \citep[APOGEE, DR16;][]{Ahumada2020} combined with distances from the Bayesian isochrone-fitting code StarHorse \citep{Queiroz2020}, 
to select stars from the halo program with metallicities [M/H] between $-5.0$ and $-1.0$, and Cartesian coordinates $Z > 2$\,kpc.   
Subsequently, we computed the orbits of these stars following the procedure described in Section~\ref{sec:Orbital}.

\subsection{Chemical comparison}

In this section we compare our stellar abundances to other studies to expand on the chemical origin of our metal-poor RRLs. 
As tracers we have selected elements with strong lines that carry information on the early nucleosynthesis in the MW's outskirts. 
Using $\alpha$-elements (O, Mg, Ca, and Ti), we can comment on the mass of the previous supernova progenitors as well as the stellar origin (in situ vs. accreted). 
Combining these abundances with heavy element abundances of neutron-capture elements (Sr and Ba) 
from intrinsically strong lines, we can assess the early production of such heavy elements in the most remote parts of the Galaxy.

\begin{figure*}

\includegraphics[angle=0,scale=.327]{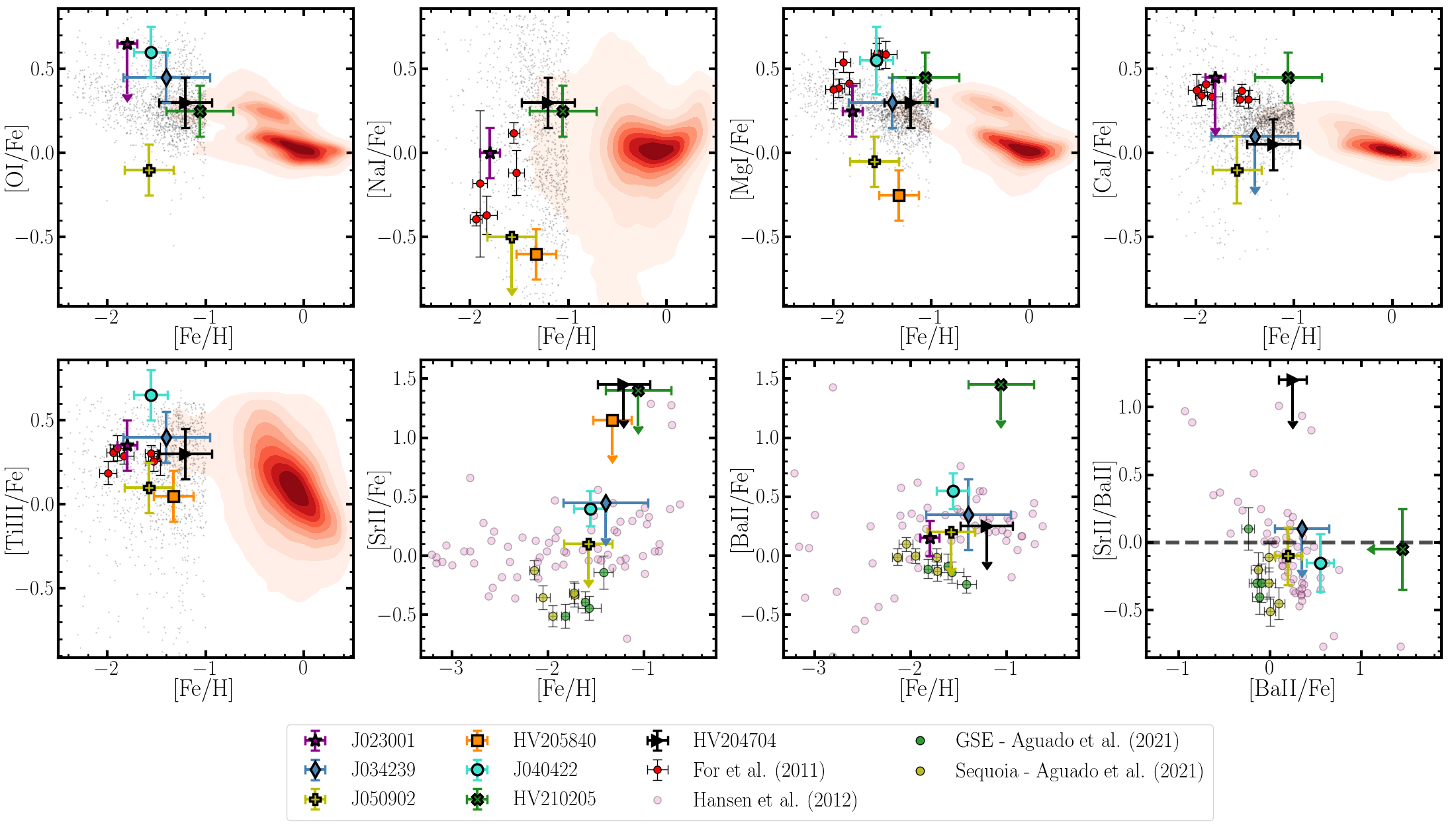} 
\caption{
Same as Figure~\ref{fig:alphas}, but showing the abundance ratios of individual elements as a function of iron abundance.
The {\it bottom right} panel includes a dashed horizontal line representing Solar values as a reference.
An arrow of arbitrary length (0.3) is used instead of error bars for the abundances considered upper limits.
These panels show the similarities in abundance patterns between different groups of RRLs (e.g., J050902 and HV205840, and HV210205 and HV204704), which we use in Section~\ref{sec:accreted} to speculate about their origins.
}
\label{fig:abunds}
\end{figure*}

Figure~\ref{fig:alphas} depicts the abundance of $\alpha$-elements with respect to Fe as a function of [Fe/H].
These quantities are computed as the average of O, Mg, Ca, and Ti, weighted by their uncertainties. 
In Figure~\ref{fig:abunds}, we display individual element abundance ratios ($\alpha$, Na, Ba, and Sr), and compare them with those from field RRLs \citep{For2011b}, halo stars (from APOGEE and \citealt{Hansen2012}), and stars from massive mergers \citep[GSE and Sequoia;][]{Aguado2021}.

\citet{Nissen10}, \citet{Ivezic2012}, and several Galactic chemical evolution studies have shown that the Galactic halo follows a bimodal $\alpha-$element abundance trend ([$\alpha$/Fe]). 
The outer halo shows, owing to it being predominantly accreted, lower $\alpha$-element abundances while the inner halo tends to exhibit higher abundances of the $\alpha$-elements also at higher metallicities. 
This stems from a less efficient star formation history typically causing fewer (and/or less) massive supernova (SN) events owing to the poorer gas reservoir and/or stronger gas loss due to winds in the accreted dwarf galaxies. 
Thus, the mass of the SN directly correlates with the amount of ejected $\alpha$-elements \citep{Tinsley1978}, and tidal remnants tend to show lower [$\alpha$/Fe] ratios at a given (low) [Fe/H] than stars formed in-situ \citep[see e.g.][]{Lanfranchi2008,Sakari2019}.

In our sample, J050902 and HV205840 show lower $\alpha$ abundance ratios as compared with normal halo stars at their metallicity, thus suggesting an accreted origin. 
We note that taking into account NLTE corrections for O, Mg, and Ti does not significantly change the apparent $\alpha$-poor nature of these stars. J034239, J023001, J040422, HV210205, and HV204704, on the other hand, show enhanced Na and $\alpha$-abundances (Mg, Ca, O, and Ti) that are comparable with halo stars, which suggests an in-situ formation.

As shown in numerous other studies, the heavy neutron-capture elements show a large star-to-star scatter in the halo \citep[see e.g.][]{Hansen2012,Aoki2013}. 
We measure the abundance of two such elements, Sr and Ba, owing to the strong transitions in the blue/visual part of the spectra (as seen in Figure~\ref{fig:BalmerProfiles}). 
Both Sr and Ba can be produced by either of the neutron-capture processes - namely the slow and the rapid n-capture process (where slow/rapid refers to the capture rate with respect to the following $\beta-$decay time). 
However, keeping in mind that our RRL sample focuses on old, low-mass stars, it is unlikely that we are tracing s-process yields from AGB stars \citep{Kaepeller2011,Karakas2014}, but rather see the early production made by the r-process. 
The r-process is associated with neutron star mergers or rare magneto-hydrodynamic supernova explosions (MHD SN), 
as shown in the recent review by \citet{Cowan2021}. 
When comparing the n-capture process element abundances of our sample, Sr and Ba (those that are not merely limits), to the halo sample of \citet{Hansen2012} we find that J034239 and J050902 follow the halo-like trend for both elements.
This is depicted in Figure~\ref{fig:SrBaH}. 
For HV204704, even though we are only able to derive upper limits for Sr, the high [Sr/Ba] ratio could be indicative of yet another origin of these heavy elements, namely fast rotating massive stars that produce (relatively) more Sr than Ba in an early s-process \citep{Frischknecht2016,Choplin2018}. 
The [Sr/Ba] ratio also carries important nucleosynthetic information on metal-poor stars, as Sr and Ba can be produced in a number of different formation sites, as detailed above. 
This ratio has also been used to trace carbon-enhanced metal-poor (CEMP) stars, as these are bona fide second generation stars, and as such provide pure insight into the pristine gas composition and ejecta from the first stars. 
With a [Sr/Ba] $<1.2$, HV204704 might belong to the CEMP-no group \citep[CEMP stars with low abundances of n-capture elements;][]{Beers2005,Yong2013,Hansen2019}, however, the spectrum quality and high temperature of HV204704 prevented a detection of carbon. 
Another CEMP RRL candidate is HV210205. 
In this case, however, both the Sr and Ba detections are considered upper limits. 
Thus, a detailed study of HV204704 and HV210205 is required to test their Sr and Ba formation scenario, along the same lines as the study of \citet{Kennedy2014} (who analysed CEMP RRLs).

\subsection{Kinematics}

The Toomre diagram illustrates the orbital velocity $V_Y$ of stars against their velocity perpendicular to the Galactic rotation $\sqrt{V_X^2+V_Z^2}$, where $V_X$, $V_Y$, and $V_Z$ are the Cartesian velocity components relative to the local standard of rest (LSR).  
This diagram is usually used to distinguish between thin disc/thick disc, and halo populations based on their kinematics \citep[e.g.,][]{Bensby2003,Venn2004,Nissen10}, with a total velocity between 180 and 220\,km\,s$^{-1}$ often used as a discriminant \citep[e.g.,][]{Bonaca2017,Amarsi2019,Buder2019}. 
Figure~\ref{fig:toomre} ({\it right} panel) depicts the Toomre diagram for the studied RRLs and halo stars from APOGEE, and shows that the studied RRLs are roughly consistent with being halo stars.

It is also worth noticing that MW halo stars' orbits consistent with satellite accretion or with retrograde motion could be an indication of an extragalactic origin \citep[][]{Roederer2018,Sakari2018,Sakari2019}. 
Figure~\ref{fig:toomre} ({\it right} panel) shows that approximately half of our sample consists of stars with retrograde orbits.
We note that one of the stars with $\alpha$-abundances below Solar (J050902) 
also displays a retrograde orbit, which supports the hypothesis of its accreted origin.

Lastly, the total velocity as a function of Galactocentric distance of the stars in our sample place them within the escape velocity limit for the MW when using both of the adopted potentials (Figure~\ref{fig:toomre}).
Thus, the bound likelihood of the stars determined as the fraction of computed orbits with valid solutions is most likely only associated with their large proper motion uncertainties, and does not necessarily represent their actual bound/unbound status.

\textbf{\begin{figure}
\includegraphics[angle=0,scale=.42]{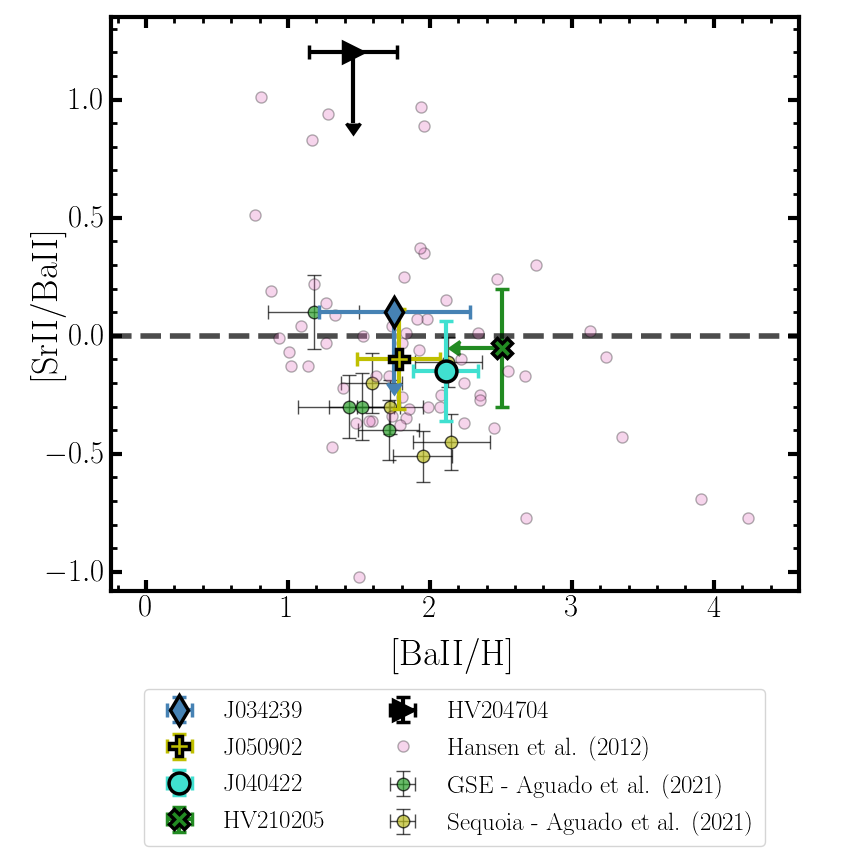} 
\caption{Abundance ratios of the n-capture process elements Sr and Ba of our stars and those from \citet{Aguado2021} and  \citet{Hansen2012}.
The solar [Sr II/Ba II] value is depicted as a dashed horizontal line for reference.
An arrow of arbitrary length (0.3) is used instead of error bars for HV204704 and J034239, to represent that their Sr measurement are upper limits.
For HV204704, the high [Sr/Ba] ratio could be indicative of a fast rotating massive star origin \citep{Frischknecht2016}. 
}
\label{fig:SrBaH}
\end{figure}}

\begin{figure*}
\includegraphics[angle=0,scale=.38]{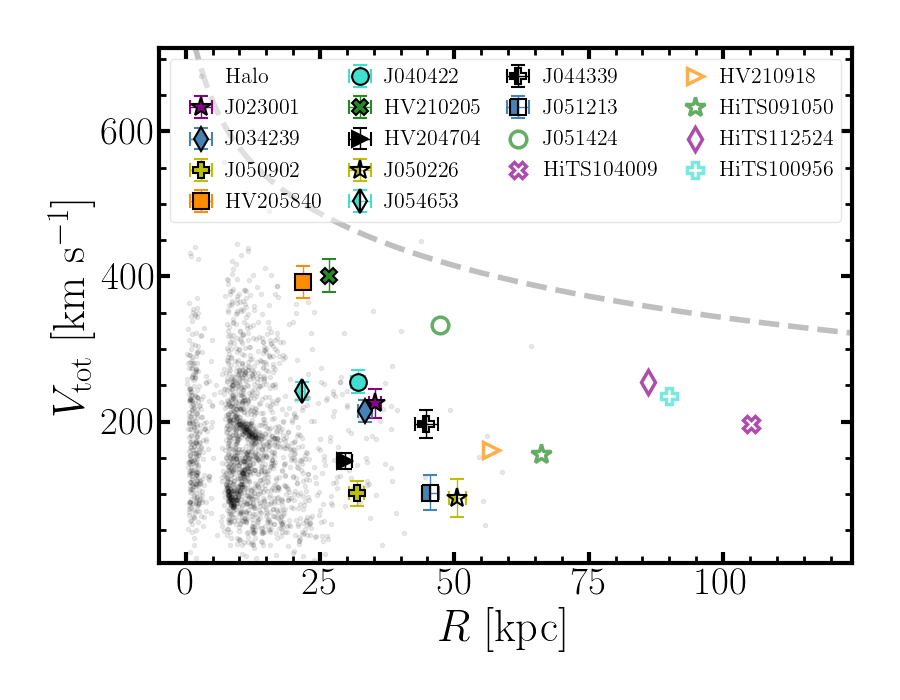} 
\includegraphics[angle=0,scale=.38]{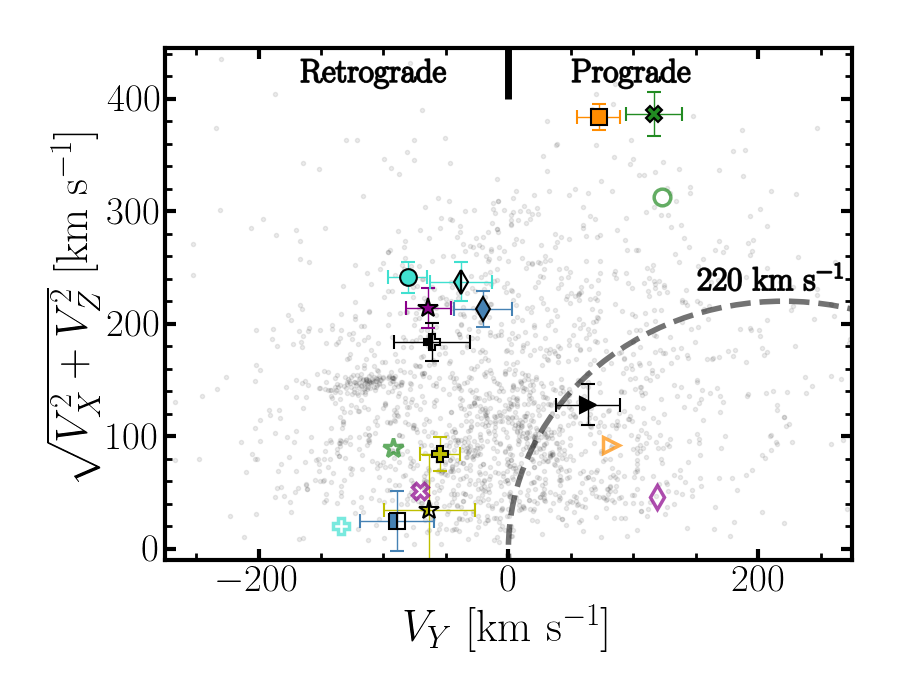} 
\caption{ {\it Left}: Total velocity as a function of Galactocentric distance for the stars in our sample (large symbols) 
and halo stars from APOGEE (black dots), using the symbols defined in Figure~\ref{fig:alphas}. The dashed line represents the escape velocity from the MW when assuming a \textit{MWPotential2014} as a reference. 
We see in this figure that all the stars in our sample have total velocities consistent with them being gravitationally bound to the MW.
{\it Right}: Toomre diagram depicting $\sqrt{V_X^2+V_Z^2}$ as a function of the orbital velocity $V_Y$. 
The dashed line marks the regions with $\sqrt{V_X^2 + \left( V_Y - 220 \right)^2 + V_Z^2} > 220$\,km\ s$^{-1}$, as a kinematic discriminant between MW halo and disc components. 
Stars from our primary sample are plotted with large (fully) filled symbols, whereas open symbols represent the stars with loosely constrained parameters.  
For clarity, the (large) uncertainties of the latter are not shown in these panels.  
From this figure we conclude that approximately half of our sample display retrograde orbits, suggesting accreted origins. 
}
\label{fig:toomre}
\end{figure*}

\begin{figure*}
\includegraphics[angle=0,scale=.38]{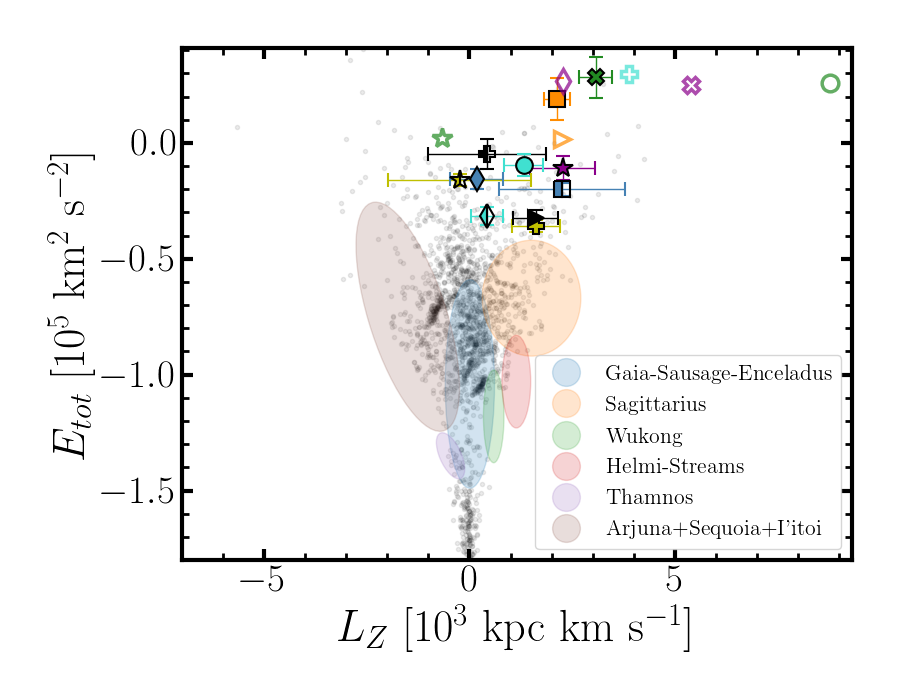} 
\includegraphics[angle=0,scale=.38]{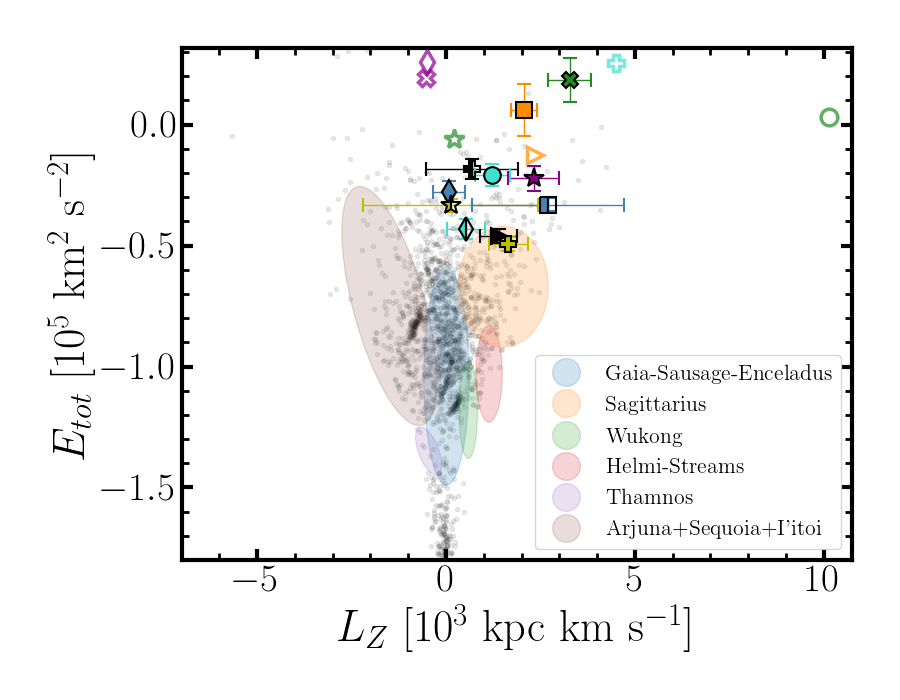} 
\caption{Energy versus vertical angular momentum for the stars in our sample assuming an isolated potential ({\it left}) and a perturbed potential ({\it right}), using the symbol scheme of Figure~\ref{fig:toomre}.
Shaded regions represent an approximation of the area occupied by the substructures studied by \citet{Naidu2020}.
These plots show that, regardless of the adopted potential, our RRLs display total energies higher than those of the considered substructures, and that their main uncertainty is in their vertical angular momenta.
}
\label{fig:LzE}
\end{figure*}

\begin{figure*}
\includegraphics[angle=0,scale=.475]{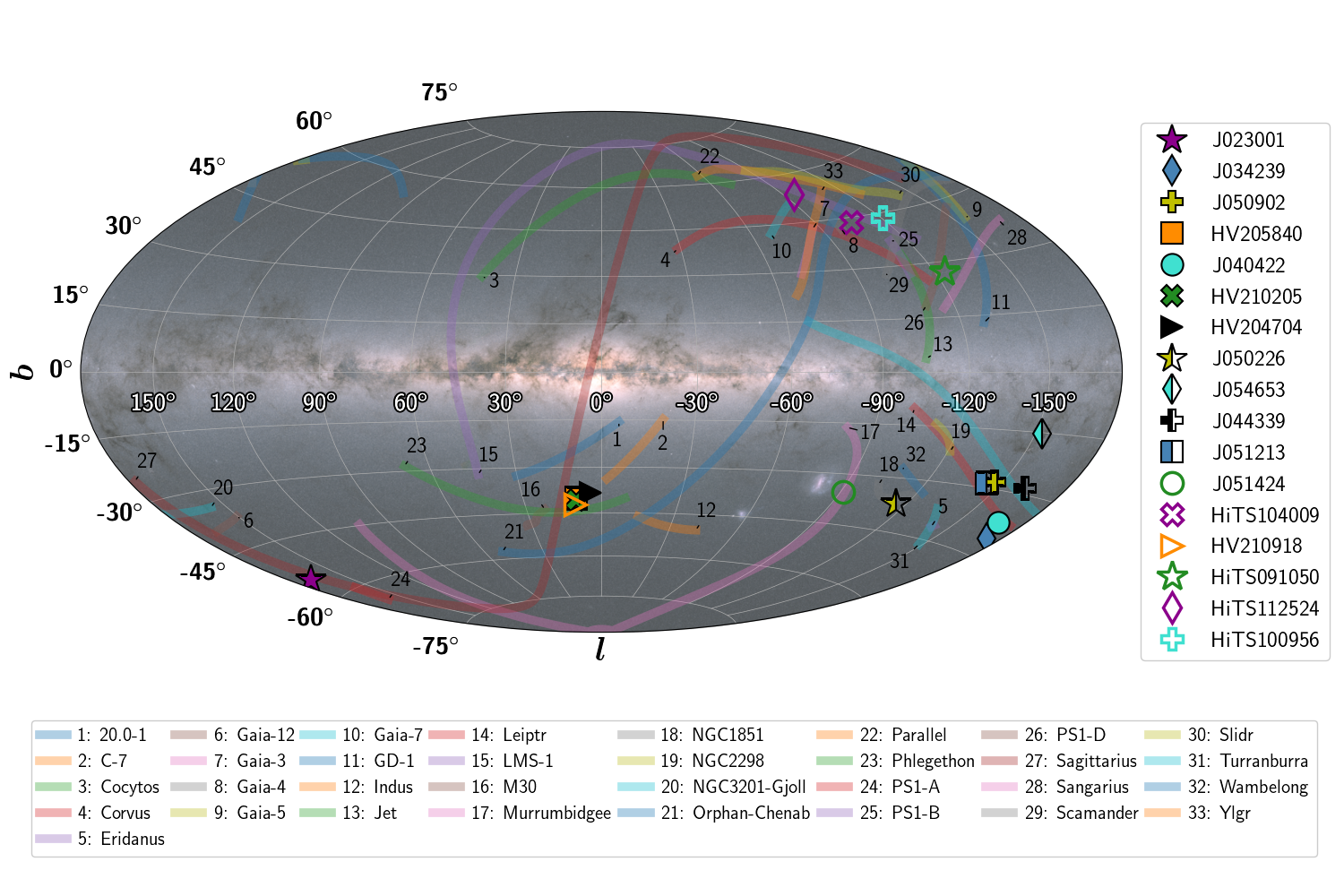} 
\caption{Footprint of 33 of the stream tracks in {\it galstreams} \citep{Mateu2022} discussed in Section~\ref{sec:streams}. 
The positions of our target RRLs are represented with filled, semi-filled, and open symbols. The {\it Gaia} all-sky map is shown in the background as a reference. \textit{Image Credit: Gaia Data Processing and Analysis Consortium (DPAC); A. Moitinho / A. F. Silva / M. Barros / C. Barata, University of Lisbon, Portugal; H. Savietto, Fork Research, Portugal.}
}
\label{fig:streams}
\end{figure*}

\begin{figure*}
\includegraphics[angle=0,scale=.45]{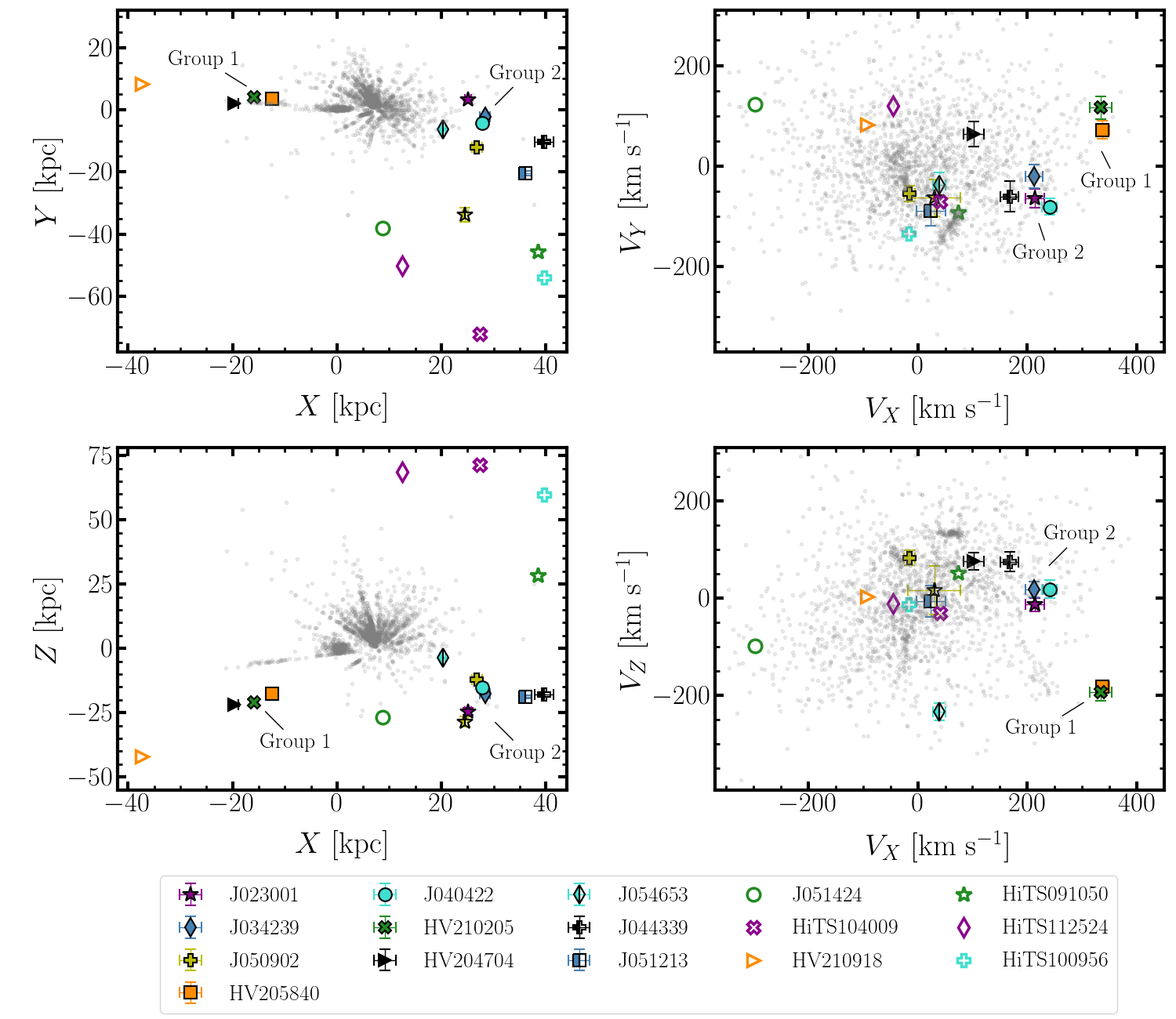} 
\caption{Positions and velocities of our sample in Galactic Cartesian coordinates.  
The location of the two groups of RRLs with similar positions and velocities (discussed in Section~\ref{sec:streams}) is marked in each panel.  
}
\label{fig:PosVels}
\end{figure*}

\section{Discussion}

\label{sec:accreted}

One of the criteria used for the selection of our target RRLs was the length of their pulsation periods (see Section~\ref{sec:sample}). 
Selecting RRLs based on their periods does not only allow for optimizing the observing strategy (e.g., allowing longer integration times), but can also be valuable to assess their connection with the accretion of MW satellites through the Oosterhoff classification \citep[Oo;][]{Oosterhoff1939}. 
The Oosterhoff dichotomy was originally observed among globular cluster RRLs, and separates them according to their mean periods and the ratio between the number of RRab and RRc stars in the clusters.
In terms of periods, Oo~I and Oo~II systems contain RRLs with mean periods of $\sim$ 0.55 and 0.65\,d, respectively \citep{Smith1995}, whereas Oo~intermediate (Oo-int) systems possess properties between Oo~I and Oo~II.
This dichotomy, however, is not commonly observed in satellite galaxies \citep[e.g.,][]{Catelan2009,Clementini2014,Martinez2017,Vivas2022}, which are often classified as Oo-int.  
Most RRLs in the halo general population (within $80$\,kpc) lie near the locus of the Oo~I group \citep[see e.g. ][]{Zinn14}. 
Thus, although this is not a unique criterion and works best on large statistical samples, the agreement between the period of our RRLs and those of the Oo groups can provide additional hints about their origin.

Our sample consists of RRab stars with periods between 0.48 and 0.76\,d, with a majority displaying pulsations longer than 0.60\,d (85\,per\,cent), and five stars (25\,per\,cent) with periods $\geq$ 0.65\,d. 
Thus, most of our stars could be classified as Oo-int or Oo~II. 
This is confirmed by considering their amplitudes of pulsation (transformed into the $V-$band when necessary) and their positions with respect to the Oo~I and Oo~II fiducial lines provided by \citet{Fabrizio2021} in the period-amplitude diagram. 
Here we inspect their derived chemical abundances and positions in phase space and compare them with those of known satellites and streams, looking for evidence of an accreted origin.

\subsection{J051424 and the LMC}

One of our stars, J051424, is located at the outskirts of the LMC, and has an heliocentric distance compatible with that of the LMC ($\sim$ 48\,kpc). 
Here we analyze the possibility that J051424 is a member of this satellite galaxy. 

J051424 lies close to the limits of the LMC's asymmetrical (with respect to its distribution and dynamical centre) extended halo at $\sim8$\,kpc (9\,deg) from its centre, in the region opposing the Magellanic bridge \citep{Jacyszyn2017}. 
This star has been observed in the LMC fields of the Optical Gravitational Lensing Experiment survey \citep[OGLE;][]{Soszynski2016}\footnote{\url{www.astrouw.edu.pl/ogle/ogle4/OCVS/lmc/rrlyr}}, and was catalogued under the name of LMC581.17.130 (or alternatively OGLE-LMC-RRLYR-30511).  
The proper motions of J051424, as listed in the {\it Gaia} DR3 catalogue ($\mu^*_\alpha=1.42\pm0.27$\,mas\ yr$^{-1}$, $\mu_\delta=1.20\pm0.28$\,mas\ yr$^{-1}$), are consistent with the proper motion distribution of the LMC ($1.76\pm0.45$\,mas\ yr$^{-1}$, $0.30\pm0.64$\,mas\ yr$^{-1}$; \citealt{GaiaLuri2021}).
There is also a rough agreement between J051424's systemic line-of-sight velocity ($228.2\pm5.5$\,km\,s$^{-1}$) with the velocity distribution of the LMC ($262.2\pm3.4$\,km\,s$^{-1}$; \citealt{vanderMarel02}).
The observed radial velocity difference might be attributable to the small number of observations (in phase) used to determine J051424's systemic velocity.

We find a [Fe/H] from $-2.34$ to $-2.57$\,dex
when using the $\Delta S$ method on J051424 \citep[based on][]{Crestani21b} from different line combinations, which is consistent with the spectroscopic metallicity range of LMC RRLs found by \citet{Haschke2012b} (from $-1.97$ to $-2.67$\,dex). 
Additionally, from \citet{Haschke2012a}, the metallicity distribution of the old component of the LMC from the Fourier decomposition of RRLs light curves is Gaussian, with a mean [Fe/H] of $-1.22$ and $-1.49$\,dex (on the \citealt{Jurcsik1995} and \citealt{Zinn1984} scales, respectively), and a dispersion of 0.26\,dex. 
Using a recent photometric metallicity calibration based on a large sample of RRLs with high-resolution metallicities (on data from \citealt{Crestani21a}, \citealt{For2011b}, \citealt{Chadid2017}, and \citealt{Sneden2017}), \citet{Dekany2021} found these [Fe/H] values to be systematically overestimated by up to 0.4\,dex.
These authors find that the mode of the RRLs' metallicity distribution function in the LMC is $-1.83$\,dex, with a mean absolute error for individual stars of 0.16\,dex from their calibration. 
We are not able to reliably estimate J051424's [Fe/H] with other methods, nor other element abundances in this work to confirm/reject membership.
However, we consider that J051424 is likely an LMC member, lying in the more metal-poor end of its metallicity distribution function.

\subsection{J023001, Whiting 1, and Sagittarius}
\label{sec:cat7Sgr}

We discuss here the possibility that the RRL J023001 is a member of either the Galactic halo globular cluster Whiting~1 or the Sgr stream. 

J023001 is located at $\sim30.5\pm1.5$\,kpc from the Sun, similar to the cluster Whiting~1, 
which is thought to be a former member of the disrupting Sgr dwarf spheroidal galaxy, and which is located at a heliocentric distance of $\sim30.6\pm1.2$\,kpc \citep{Baumgardt2021}. 
\citet{Carraro2005} estimated Whiting~1's metallicity to be $\sim-1.2$\,dex, whereas \citet{Carraro2007} concluded that its [Fe/H] probably lies within the range of $-0.4$ to $-1.1$.
Both of these works determined the cluster's metallicity by fitting isochrones to its colour-magnitude diagrams. 
This shows that the consistency of J023001's metallicity, i.e.,
$-1.2\pm0.3$\,dex from synthesis and $-1.8\pm0.1$\,dex from EW measurements with that of Whiting~1 strongly depends on the literature value and the method used for the comparison. 
In any case, J023001 is located at 7\,deg from Whiting~1's centre, a comparatively large angular distance given the cluster's angular radius \citep[$\sim$ 0.5\,arcmin;][]{Dias2002,Carraro2005}.
Additionally, the cluster age estimated by \citet{Carraro2007} is $6.5^{+1.0}_{-0.5}$\,Gyr, which makes it one of the youngest globular clusters in the halo, and is incompatible with it hosting RRLs.
The mean $V$ magnitude of J023001 is 18.2, also incompatible with the horizontal branch position in the cluster's colour-magnitude diagrams displayed in the aforementioned studies.

The RRL J023001 also lies relatively close to the footprint of the Sgr stream, with a latitude-like  coordinate ($B_{\rm Sgr}$) of $\sim$ 3\,deg in the Sgr stream coordinate system \citep{Majewski2003}. 
Several authors have studied the chemical abundance patterns and orbital parameters of Sgr \citep[e.g.,][]{Chou2007,Carlin2018,Hansen2018,Hayes2020,Johnson2020,delPino2021,Hasselquist2021}, including the identification of Sgr stream stars using APOGEE \citep{Hasselquist2019}. 
These studies have shown that the bulk of the [Fe/H] distribution spans between $-0.8$ and $0$\,dex for Sgr's core, that the Sgr stream's [Fe/H] is $\sim0.5$\,dex more metal-poor than its main body, and that the more metal-rich stars in Sgr display [X/Fe] abundances below the MW abundance trends. 
Recently, \citet{Hasselquist2021} reported Sgr's [$\alpha$/Fe] to smoothly decline from the MW halo-like high-$\alpha$ plateau at the metal-poor end of its [Fe/H] distribution ([$\alpha$/Fe] $>0.2$ for [Fe/H] $<-1.5$) to below the MW low-$\alpha$ disc trend at [Fe/H] $>-1.0$\footnote{It is worth noticing that metal-poor members of Sgr with high $\alpha-$abundances have also been identified in the literature \citep[see e.g.][]{Hansen2018}. }. 
For J023001 we obtain [$\alpha$/Fe] $=0.35\pm0.25$ (with a scatter of 0.15\,dex). 
Additionally, \citet{Hasselquist2019} found the locus of the eccentricity $e$ distribution of Sgr stream members to be between 0.4 and 0.7, and their apocentric distances $r_{\rm apo}$ to vary between 30 and 80\,kpc (see their Figure~6).
For J023001, we find $e\sim0.55$ regardless of the adopted potential ($0.52^{+0.05}_{-0.05}$ for the perturbed model), and $r_{\rm apo}\sim70$\,kpc ($68^{+15}_{-8}$\,kpc).
Thus, our results indicate that J023001 is more likely to be associated with Sgr than Whiting~1.

\subsection{Association with known substructures}
\label{sec:substructures}

Various studies have suggested that the bulk of the halo is built from accreted satellite systems and the heating of the disc \citep[see e.g.][]{Robertson2005,Font2006,Naidu2020,Ibata2021}, and evidence of these substructures can be observed in the energy-vertical angular momentum space ($E$-$L_Z$).

Figure~\ref{fig:LzE} shows the position of our RRLs and halo stars in the $E$-$L_Z$ diagram determined for orbits under an isolated and a perturbed MW potential.
The figure highlights the regions where the majority of stellar distributions of known substructures are located (for GSE, Sgr, Wukong, the Helmi streams, Thamnos, and Arjuna+Sequoia+I'itoi), as shown by \citet{Naidu2020}. 
We point out that the energies and $L_Z$ computed from both models are only significant for the stars without loosely constrained orbits. 
It is clear from the figure that the RRLs display total energies higher than those of the considered substructures (at least at the high end of the distributions), and that their main uncertainty is in $L_Z$.

In terms of $\alpha$-abundances, even though J034239, HV210205, and HV204704 show an enhancement compared with normal halo stars from APOGEE (Figure~\ref{fig:alphas} and Figure~\ref{fig:abunds}), their [$\alpha$/Fe] ratios are also compatible with those of the GSE and Sequoia stars analysed by \citet{Aguado2021} (see their Figure 4), with $0.2 < $ [$\alpha$/Fe$] < 0.4$.

Our stars show, however, [Sr/Fe] and [Ba/Fe] that are overall higher than those of the GSE and Sequoia stars analysed by \citet{Aguado2021}, who reported an underabundance of Sr as compared to halo stars, and Ba around the solar value for both GSE and Sequoia stars.
This is clearly visible in Figure~\ref{fig:abunds}.
Furthermore, their distances are not compatible with those expected from these merger events.

\subsection{Streams and other associations}
\label{sec:streams}

Overdensities in the phase-space (positions and velocities) can, in principle, be used to trace recently accreted substructures, and even undiscovered satellites. 
To investigate the connection between our RRLs and the past and ongoing tidal dissolution of satellites, we used the recently updated Python library {\it galstreams} \citep{Mateu2022}, which contains celestial, distance, proper motion, and radial velocity information for 125 stream tracks corresponding to 97 distinct stellar streams (proper motions and velocities when available).
Figure~\ref{fig:streams} shows 33 tracks from {\it galstreams} located in the proximity of our targets in equatorial coordinates.

For a given stream, we examine case by case all the RRLs in our sample that lie within 15\,deg (on-sky projection) from the its footprint. 
The following streams pass this filter:  
Phlegethon, C-7, Ylgr, NGC~3201-Gjoll, Leiptr, NGC~1851 \citep{Ibata2019,Ibata2021},
M30 \citep{Sollima2020,Harris1996},
LMS-1 \citep{Yuan2020}, 
Orphan-Chenab \citep{Grillmair2006,Shipp2018,Koposov2019},
Gaia-4 \citep{Malhan2018a},
Corvus \citep{Mateu2018},
Scamander, Sangarius \citep{Grillmair2017a},
PS1-D \citep{Bernard2016},
Murrumbidgee \citep{Grillmair2017south}.
Most of these streams, however, are located in the inner Galaxy, i.e., with distances $<10$\,kpc and with proper motions clearly dissimilar to those of our RRLs.

Among the cases of interest, we confirm J023001 as a likely Sgr stream member (as discussed in Section~\ref{sec:cat7Sgr}), and find four other RRLs possibly associated with the stream (including proper motions and radial velocities within the expected ranges; see \citealt{Mateu2022} and references therein), 
namely HV205840, HV210205, HV204704, and HV210918. 
Additionally, the position of these stars in the $E$-$L_Z$ space is consistent with the region that contains the bulk of stars from the Sgr stream. 
Furthermore, given their longitudes in the Sgr stream coordinate system ($\Lambda_{\rm Sgr}$) defined by \citet{Majewski2003} ($\Lambda_{\rm Sgr} \sim $ 109.79, 26.17, 26.89, 23.49, and 28.41\,deg for J023001, HV205840, HV210205, HV204704, and HV210918), 
their velocities are compatible with them being part of the Sgr trailing arm \citep[see e.g.][]{Johnson2020,Hasselquist2019}. 
In addition, the latitude of these stars with respect to the Sgr stream is $\lessapprox 4.3$\,deg, with the exception of HV204704 for which $B_{\rm Sgr}$ is $\sim $ 8\,deg.  
Of these stars, however, only J023001 and HV205840 are observed with $\alpha$-element ratios compatible with those of the Sgr $\alpha$-abundance  trends at their metallicity. 
We do not possess abundance information for HV210918 to check its concordance with the stars in the stream and the dwarf galaxy.
Nonetheless, this does not rule out their potential connection with the stream, given the wide range of Sagittarius' [$\alpha$/Fe]  \citep[see Figure~5 from][]{Hasselquist2021}.

Additionally, we find two RRLs (HiTS112524 and HiTS104009) relatively close to the Orphan-Chenab stream projected in the sky.
However, given the large physical distance of both of these stars to the stream, and their inconsistency with the proper motions and predicted systemic velocities trends for Orphan-Chenab's members at right ascension between 160 and 170 ($\mu^*_\alpha\sim-1.6$\,mas\ yr$^{-1}$, $\mu_\delta\sim0.8$\,mas\ yr$^{-1}$, $v_{\rm sys}\sim200$\,km\ s$^{-1}$; \citealt{Koposov2019,Prudil2021}), we consider their association unlikely. \\

As part of our target selection process, we selected stars located close to each other in equatorial coordinates, which might contribute to the detection of unknown overdensities.  
From the phase-space location of our stars and their abundances (when available), displayed in Figure~\ref{fig:PosVels}, we identify two groups containing stars not only coherent in Cartesian coordinates, but also displaying similar velocities. 

The first group consists of HV205840, HV210205, and HV204704, which display similar Cartesian coordinates $X$, $Y$, and $Z$ (with a dispersion in these coordinates $<2$\,kpc). 
HV205840 and HV210205 have also similar total velocities ($390\pm16$ and $405\pm24$\,km\,s$^{-1}$, respectively), which differ from that of HV204704 ($147\pm12$\,km\,s$^{-1}$).
Additionally, the element ratios of HV210205 and HV204704 are similar (see Figure~\ref{fig:abunds}). 
Moreover, the orbits of HV205840 and HV210205 have pericentres within 5\,kpc, similarly large apocentres (both uncertain), and eccentricities that are alike ($\sim$ 0.75, with a difference of 0.03), and more certain.
We note that these stars are part of the group potentially associated with the Sgr stream based on their on-sky proximity to the stream. 
However, only HV205840 displays clear indications of a low $\alpha$-to-Fe abundance ratio ($-0.2$ at [Fe/H] $=-1.23$) compatible with accreted stars from Sgr.

The other group includes J050902, J040422, and J034239.
These three stars are not only coincident in the phase-space, but also share similar n-capture process element abundances. 
However, only J034239 and J040422 are alike in their $\alpha$-elements abundances, whereas J050902 displays a significantly lower [$\alpha$/Fe] ($\sim$ 0.5\,dex lower).
Additionally, the apocentric and pericentric distance, and the eccentricities of J034239 and J040422 are substantially congruent (see Figure~\ref{fig:periapoe}), with differences $\Delta r_{\rm peri}\lesssim2$\,kpc,  $\Delta r_{\rm apo}\lesssim10$\,kpc, and $\Delta e\lesssim0.05$
regardless of the model adopted to determine them.
This is an indication of a common origin for these two RRLs. 
We note in passing that these RRLs are preferentially located towards the locus of the Oo-int line defined by \citet{Fabrizio2021} in the period-amplitude space, and have [Fe/H] ratios expected for RRLs in the long-period edge of the Oosterhoff gap \citep[see e.g.][]{Catelan2015,Monelli2022}.

\par 

For the rest of the stars in our sample we do not find clear indications of associations. 
As previously mentioned, the stars J040422 and J034239 (at $d_{\rm H} \sim $ 25 and 27\,kpc, respectively) are coincident in phase space and have $\alpha$-abundances comparable to those of the sample of halo stars from APOGEE. 
This might also be interpreted as an indication of them having been formed in-situ. 
Along the same lines, the pulsation periods of two of the RRLs with lower resolution spectra in our sample (J050226 and HiTS101243, at $d_{\rm H} \sim $ 48 and 91\,kpc, and periods $\sim$ 0.53\,d) are similar to the periods typical of RRLs in the Oo~I group.
Because the majority of the RRLs in the general halo population follow the locus of the Oo~I, this might be a hint of their origins being in agreement with the main trend of nearby field RRLs. 
Without chemical abundances and more precise orbital parameters, however, it becomes challenging to set solid constraints on the formation conditions of the most distant stars in our sample (including J050226 and HiTS101243).

\section{Summary and conclusions}
\label{sec:conclusions}

We have conducted a pilot study to characterize spectroscopically remote halo RR Lyrae stars, to better understand the liming factors in determining their stellar parameters, abundances, and kinematics, and to explore their role in understanding the Milky Way's accretion history.
We have obtained MIKE@Magellan medium and low resolution optical spectroscopy for a sample of 20 halo RRLs with precise heliocentric distance information, between 15 and 165\,kpc. 
These stars were selected from the HiTS, HOWVAST, and Catalina surveys, based on their pulsating properties (ab-type RRLs with periods $\gtrsim0.5$\,d).
Given the combination of distance and variable nature of our targets, the signal-to-noise of our coadded spectra ranges from $\sim$ 5 to 20.

We derived (systemic) radial velocities for our whole sample with typical uncertainties of $\sim$ 5--10\,km\,s$^{-1}$. 
By combining proper motions from {\it Gaia} DR3 with these velocities and period-luminosity-based distances we computed orbital parameters for more than half of our sample (with great precision out to 50\,kpc from the Galactic centre), and estimated their iron abundances by following various approaches. 

For computing their orbital parameters, we considered two models: one assuming an isolated evolution of the MW potential, and one taking the gravitational effects of the infall of the LMC into account, in line with recent studies \citep[e.g.,][]{Vasiliev21}.
Based on the number of valid solutions resulting from determining the orbits of our RRLs we compute their likelihood to be gravitationally bound to the MW, and find two stars with loosely constrained orbits. 
Interestingly, the velocity of the potentially unbound stars lie within the bounds of the MW escape velocity curve at their respective distances. 
From the data at hand, we conclude that the biggest limitations in exploring the full six-dimensional phase-space, and the bound likelihood of the orbits, come from the large uncertainties in the proper motions of stars beyond 30\,kpc.

We derive atmospheric parameters and chemical abundances (including $\alpha$-element abundance ratios and n-capture elements, and considering NLTE corrections) for seven stars in our sample that have distances between 20\,kpc and 40\,kpc.
We find the estimated atmospheric parameters consistent with their observed phases of pulsation, and the spread of their spectroscopic [Fe/H] values (from $-1.80$ to $-1.05$\,dex) in general agreement with the peak of the halo metallicity distribution.

By combining the stars' orbital parameters and their derived chemical abundances, we speculate about their origin and associate them with potential parent populations, including the LMC.
We find two RRLs with an underabundance of $\alpha$-elements for their metallicity (HV205840 and J050902),
which is not compatible with in-situ formed MW stars and suggests an accreted origin. 
Applying NLTE corrections does not change the abundances significantly. 
Furthermore, we deduce the early production of two n-capture process elements (via the r-process) for three of our stars, two of which follow the expected halo-like trend (J034239 and J050902).
For the third star (HV204704), we find a [Sr/Ba] ratio that suggests a CEMP classification, which could be explained by pollution from a fast rotating massive star. 
Further studies are required to confirm this classification. 
Additionally, about half of our sample is found in counter-rotating orbits, which might indicate an extragalactic origin.
We are able to confirm one of our stars (J051424) as an LMC outskirts member, and find a likely association of another RRL with Sagittarius (J023001).
We also find other RRLs for which additional data are required to confirm an association with Sagittarius (HV205840, HV210205, HV204704, HV210918). 
We analyse other substructures, including major merger events (e.g., GSE and Sequoia) and streams, but do not find 
convincing evidence of their connections with our RRLs. 
Observing larger samples of RRLs with dedicated high/medium resolution spectroscopy at large aperture telescopes (throughout their pulsation cycles or at specific phases) might render it possible to associate single halo RRLs with known or yet undiscovered substructures, and is required to recover a more complete scenario of their origins, together with dedicated spectroscopic studies of satellites and streams in the halo \citep[e.g.,][]{Ji2020,Ji2021,Li2021,Martin2022a,Martin2022b}. 
Our results indicate that a S/N $>15$ is sufficient for determining the RRLs' systemic velocities and abundances to assess these associations.

Studying the systemic velocities of distant halo RRLs will also contribute to placing our Galaxy in a proper cosmological context through precise estimations of its total mass, as current models (and their comparison with observations) are highly sensitive to the halo mass \citep[e.g.,][]{Geha2017}. 
Currently, MW mass estimates beyond the disc rely on different groups of dynamical tracers (e.g., globular clusters, dwarf galaxies, stellar streams).  
Because the widest dispersion in the mass estimations of the MW correspond to the most distant tracers \citep{Eadie2016,Deason2019,Wang2020,Deason2021,Rodriguez2021}, even single remote stars ($>100$\,kpc) with precise distance determinations and velocities could provide valuable insights into the full MW gravitational potential \citep{Watkins2010}. Our pilot study lays the groundwork for surveys aiming to build a larger sample of outer-halo RRLs in the near future, with which robust constraints of the MW mass profile will be possible.

Spectroscopically characterizing faint halo RRLs is an inherently challenging task, and will continue to be so in the near future, given the conflict between the long exposure times required and their short pulsation periods.
Only the chemodynamical analysis of large numbers of halo RRLs as part of the next generation surveys will help us unveil the real nature of the distant halo, and its connection with old populations in dwarf galaxies and streams, which will result from the synergies between large-sky photometric/spectroscopic surveys and dedicated efforts. 
The complementarity of these surveys and upcoming instruments \citep[e.g., the Subaru Prime Focus Spectrograph survey, the 4-metre Multi-Object Spectroscopic Telescope, the Dark Energy Spectroscopic Instrument, and the ten-year Rubin Observatory Legacy Survey of Space and Time;][]{Takada2014,deJong2014,Tamura2016,LSST2009,Levi2019} 
with future {\it Gaia} data releases\footnote{\url{https://www.cosmos.esa.int/web/gaia/release}}, 
with their corresponding improvements in astrometric precision and accuracy, 
will thus be pivotal to unveil the MW history at large radii. \\

\newpage

\section*{Acknowledgements}

We thank the anonymous referee for her/his constructive and insightful report, which helped improve this paper. 
GEM and EKG gratefully acknowledge the support of the Hector Fellow Academy.
CJH would like to acknowledge the ELEMENTS Research Cluster (Project ID 500/10.006) and ChETEC-INFRA (European Union's Horizon 2020 research and innovation programme under grant agreement No. 101008324). 
RRM gratefully acknowledges support by the ANID BASAL project FB210003 and ANID Fondecyt  project 1221695. 
GEM and EKG acknowledge the Deutsche Forschungsgemeinschaft (DFG, German Research Foundation) -- Project-ID 138713538 -- SFB 881 (“The Milky Way System”, subproject A03). 
JLC acknowledges support from National Science Foundation (NSF) grant AST-1816196. 
CEMV is supported by the international Gemini Observatory, a program of NSF's NOIRLab, which is managed by the Association of Universities for Research in Astronomy (AURA) under a cooperative agreement with the National Science Foundation, on behalf of the Gemini partnership of Argentina, Brazil, Canada, Chile, the Republic of Korea, and the United States of America. 
The results of this work were obtained using data from the European Space Agency (ESA) mission {\it Gaia}, processed by the {\it Gaia} Data Processing and Analysis Consortium  (DPAC). 
Funding for the DPAC has been provided by  national institutions, in particular the institutions participating in the {\it Gaia} Multilateral Agreement (MLA).
The {\it Gaia} mission website is \url{https://www.cosmos.esa.int/gaia}. 
The {\it Gaia} archive website is \url{https://archives.esac.esa.int/gaia}.

This research has made use of {\sc pandas} \citep{McKinney10}, {\sc numpy}  \citep{vanderWalt11}, the {\sc Astropy} library \citep{Astropy13,Astropy18}, and the software {\sc TOPCAT} \citep{Taylor05}. 
This research has made use of the VizieR catalogue access tool, CDS, Strasbourg, France. 
The original description of the VizieR service was published in A\&AS 143, 23.
The figures in this paper were produced with {\sc Matplotlib} \citep{Hunter07}.

\section*{Data Availability}

The data underlying this article will be shared upon reasonable request to the corresponding author.

\clearpage


\appendix

\renewcommand{\thefigure}{A\arabic{figure}}
\setcounter{figure}{0}

\renewcommand{\thetable}{A\arabic{table}}
\setcounter{table}{0}

\section{Complementary figures and line list}

\begin{figure}
\includegraphics[angle=0,scale=.28]{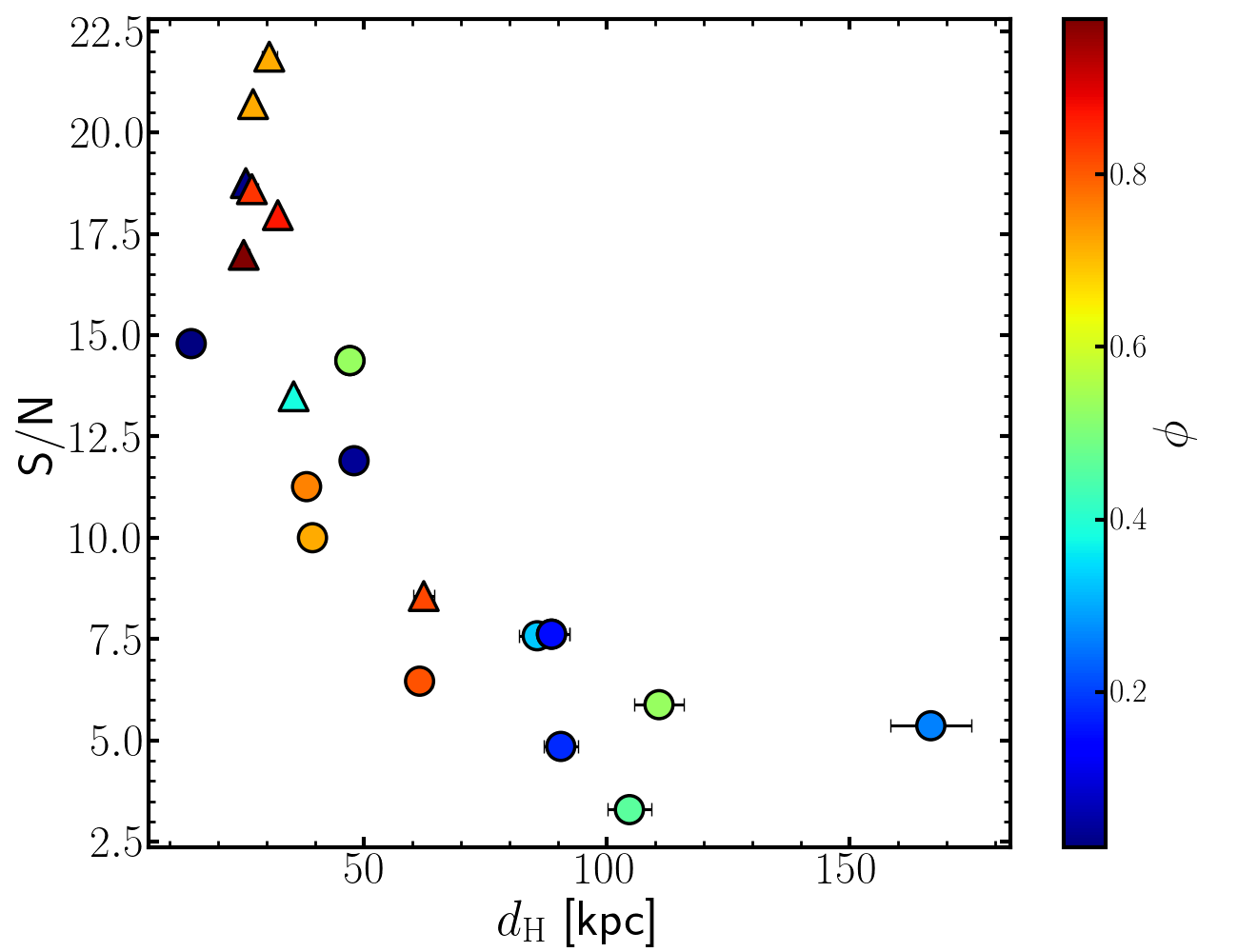} 
\caption{Signal-to-noise ratio of our target stars, as a function of heliocentric distance and colour-coded by their phase of observation. 
Stars for which we obtain spectroscopic atmospheric parameters (the primary sample, as defined in Section~\ref{sec:StellarParam}) 
are plotted with triangle markers. 
}
\label{fig:snrs}
\end{figure}

In this part of the Appendix we provide additional material to complement the content of Sections~\ref{sec:sample} and \ref{sec:spectral}. 
In Figure~\ref{fig:snrs} we show the signal-to-noise ratio of our co-added spectra as a function of the heliocentric distance of our targets. Figures~\ref{fig:BalmerProfiles} and \ref{fig:MetallicProfiles} display the spectral regions surrounding Balmer and metallic lines used in this work, for the RRLs observed in our second run. 
These figures include markers indicating the presence of metallic lines when clearly visible. 
Finally, Table~\ref{tab:lines} lists the absorption lines used in this work, together with their excitation potentials, log $g f$, equivalent widths, and element abundance ratios. 

\textbf{\begin{figure*}
\includegraphics[angle=0,scale=.27]{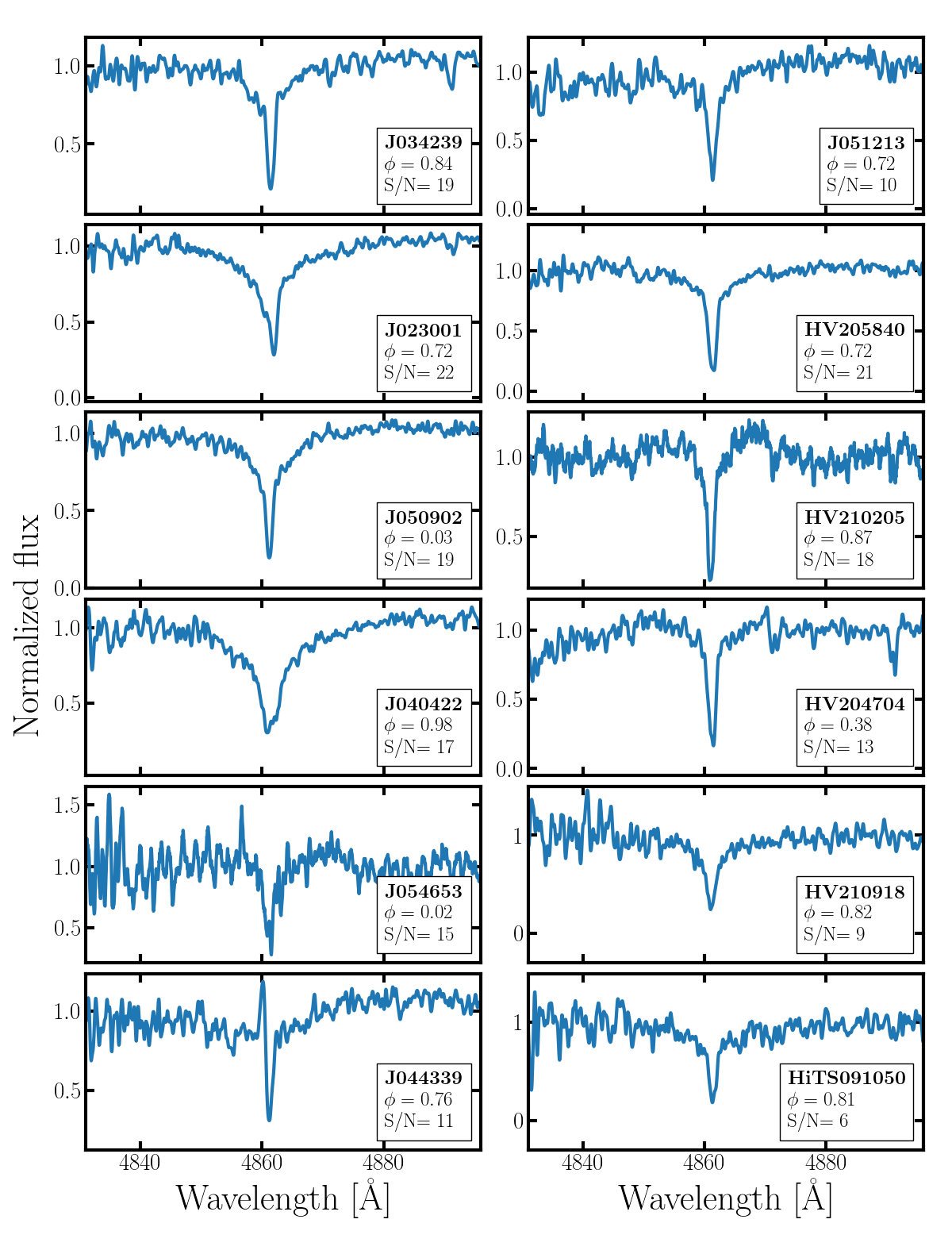} 
\includegraphics[angle=0,scale=.27]{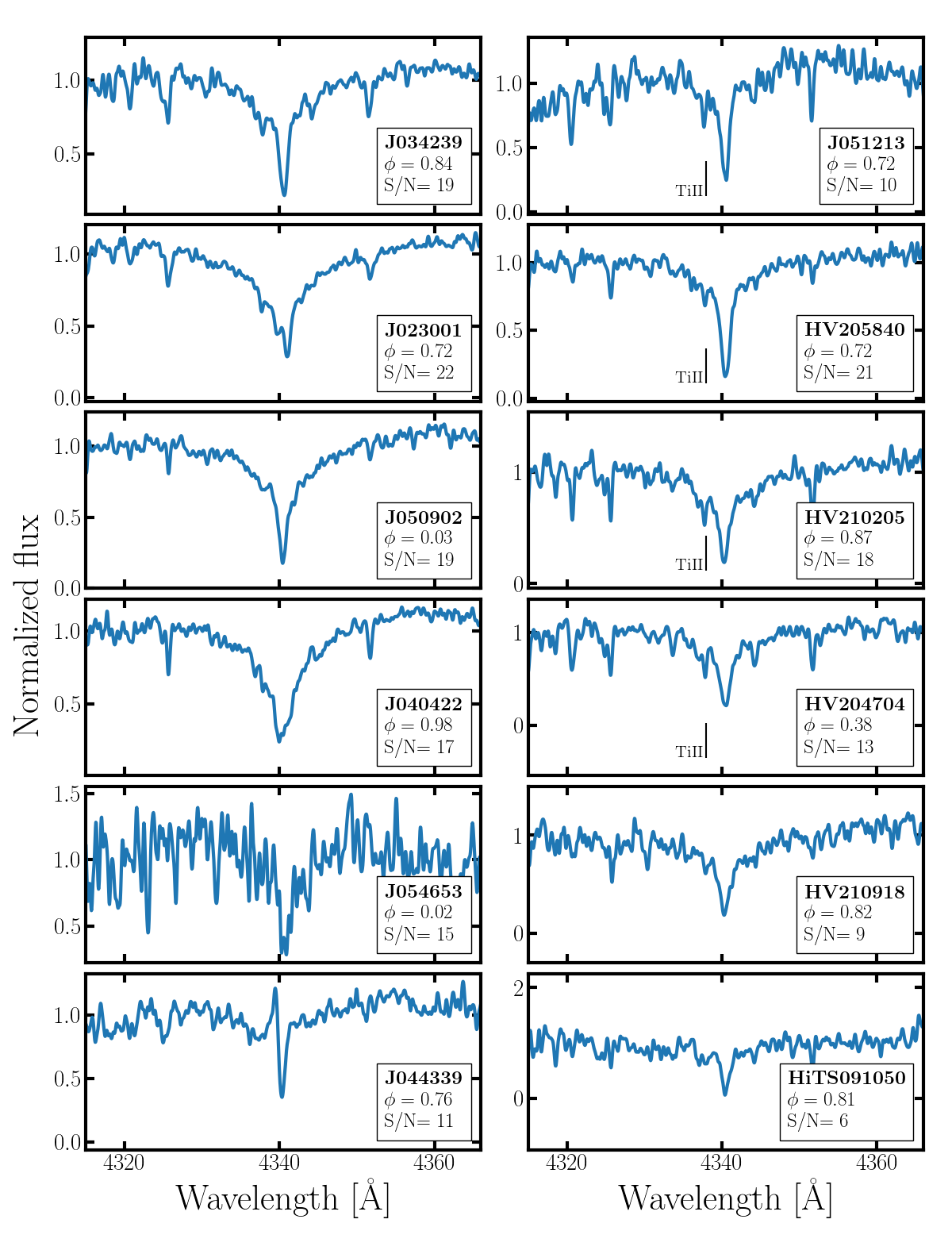} 
\includegraphics[angle=0,scale=.27]{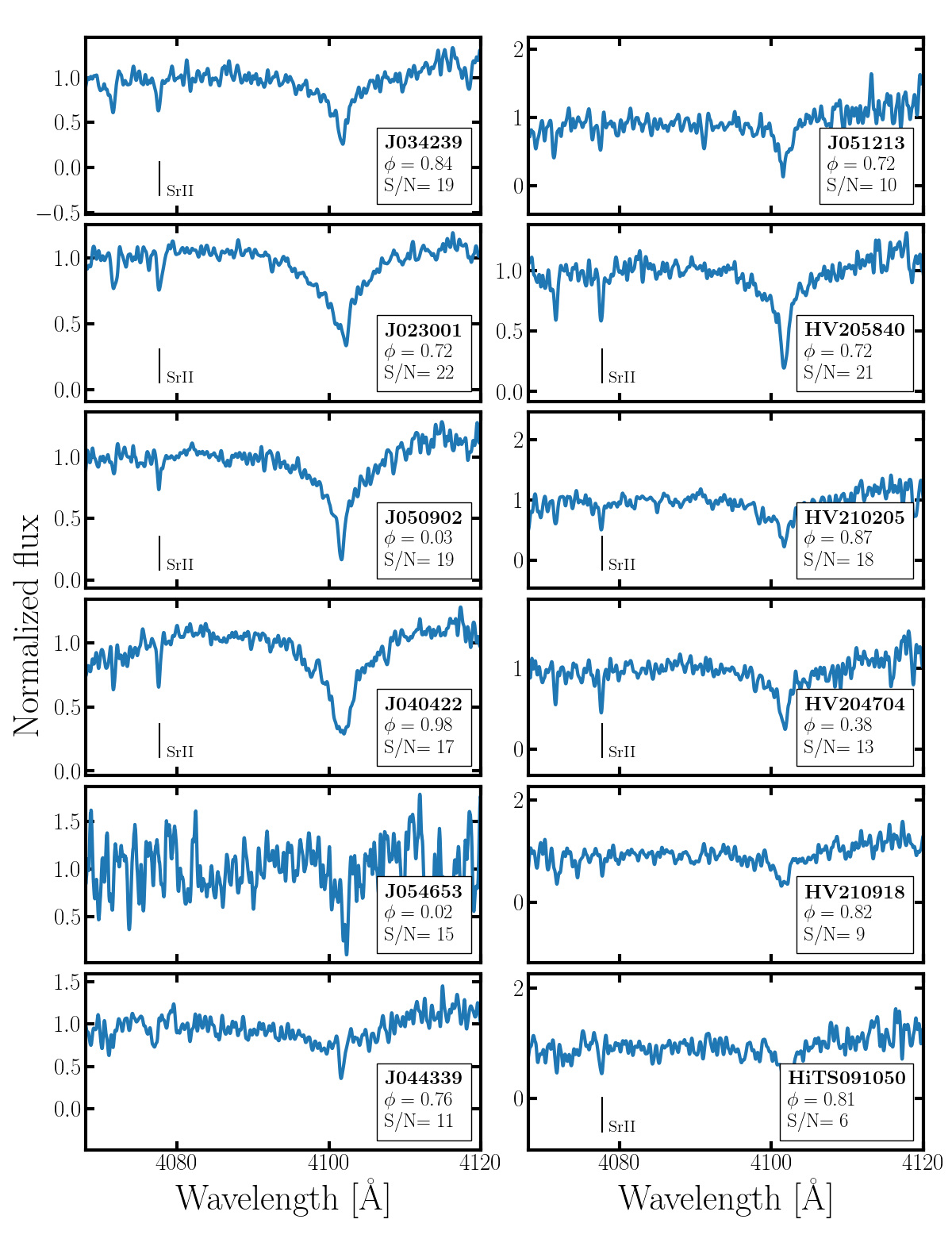} 
\caption{Spectral regions surrounding the Balmer lines for the RRLs observed in our second run (with higher resolution and S/N). 
A Gaussian convolution with $\sigma=3$ was applied to smooth the spectra. Ti II and Sr II lines are marked when clearly visible in the smoothed spectra.}
\label{fig:BalmerProfiles}
\end{figure*}}

\textbf{\begin{figure*}
\includegraphics[angle=0,scale=.28]{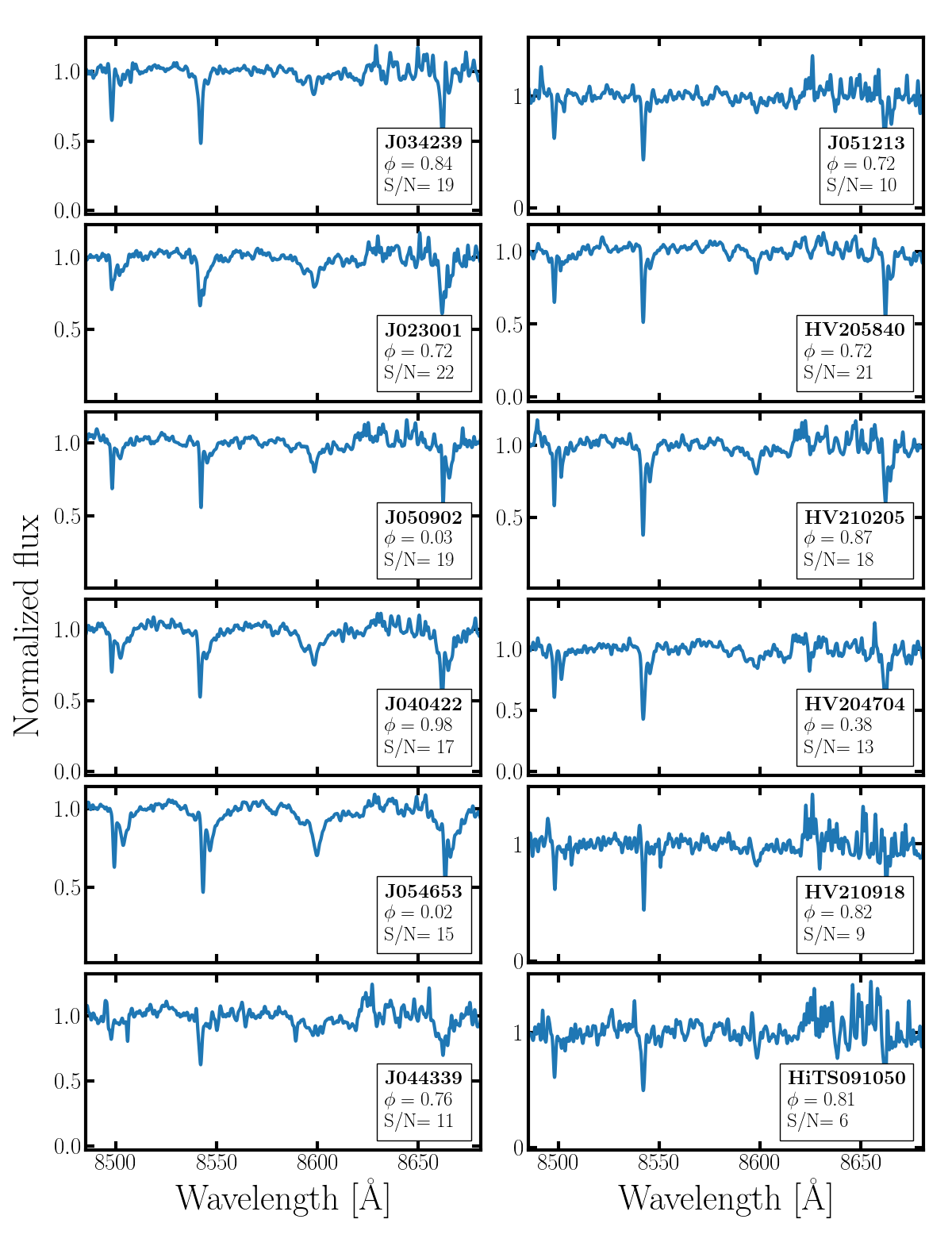} 
\includegraphics[angle=0,scale=.28]{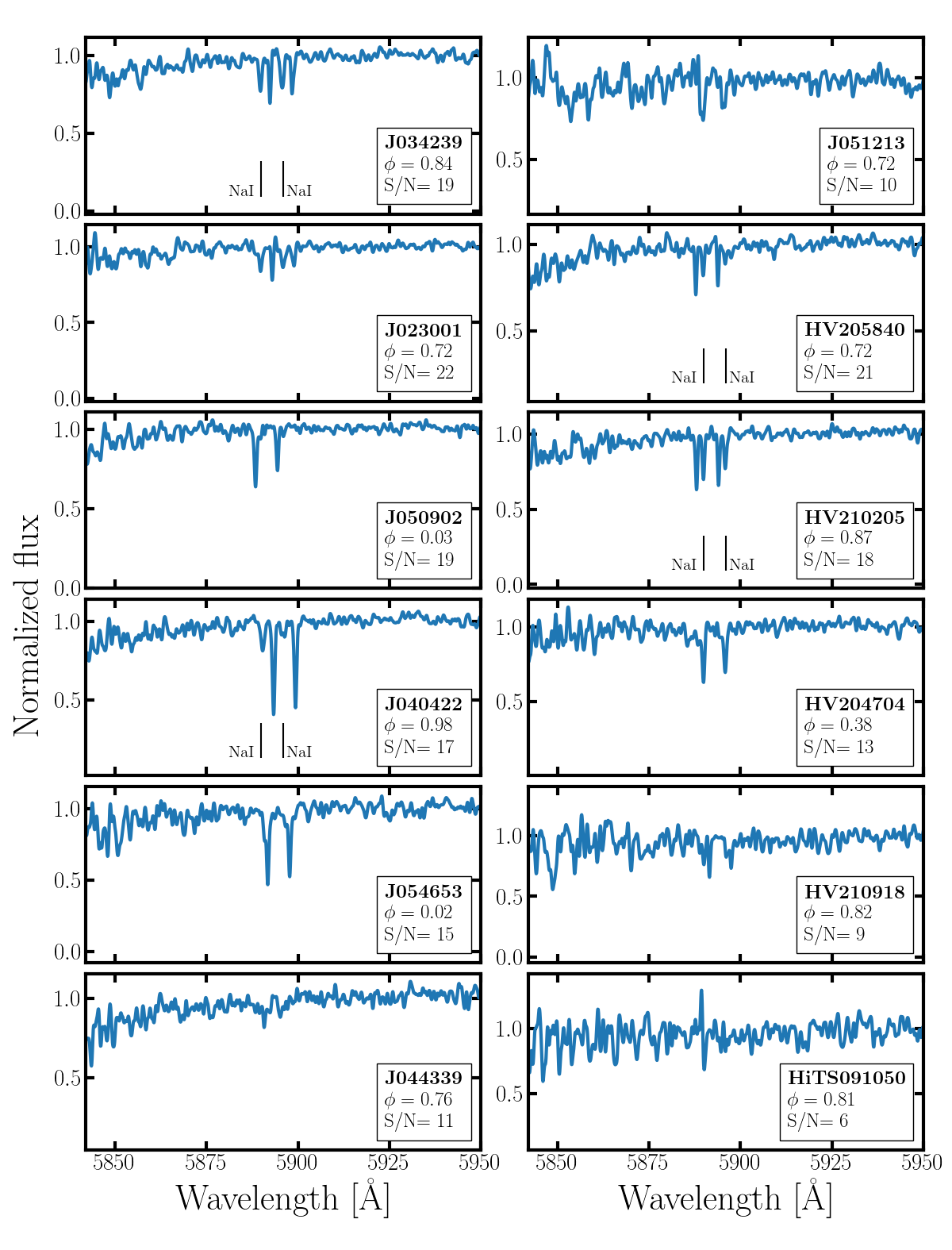}\\ 
\includegraphics[angle=0,scale=.28]{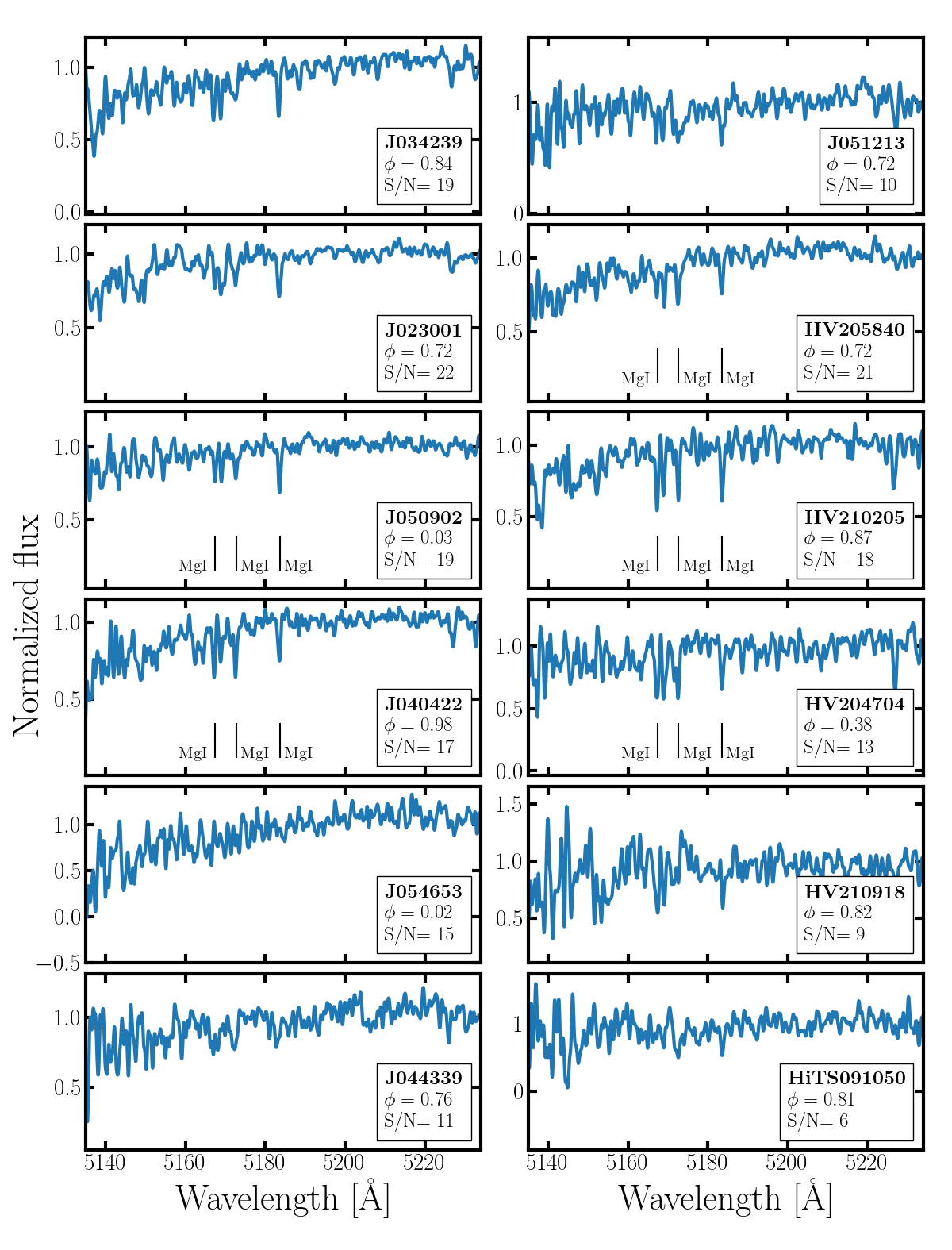} 
\includegraphics[angle=0,scale=.28]{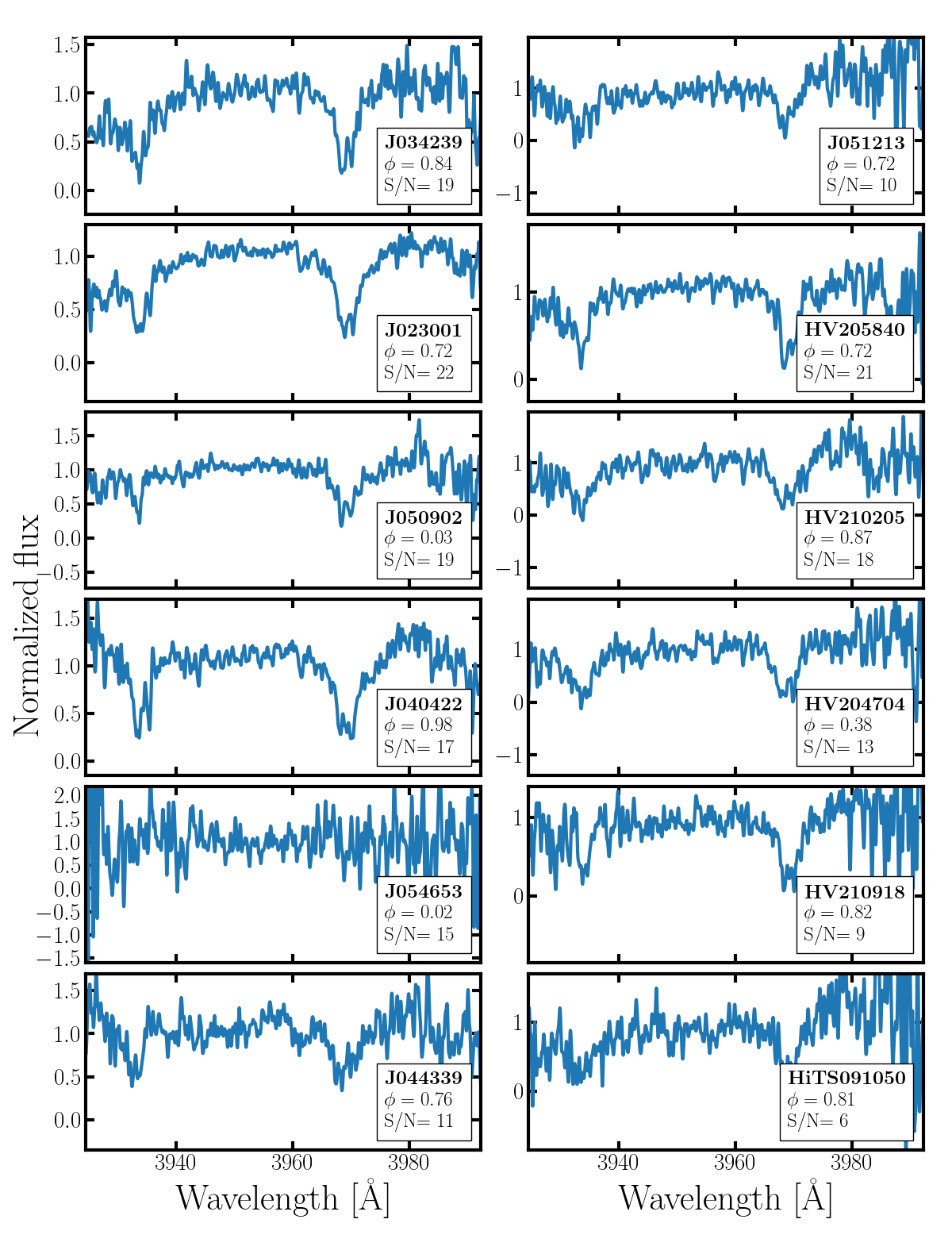} 
\caption{Same as Figure~\ref{fig:BalmerProfiles}, but for the regions containing some of the metallic lines considered in this work. Metallic lines are marked when clearly visible in the smoothed spectra.}
\label{fig:MetallicProfiles}
\end{figure*}}

\begin{sidewaystable*}\tiny
\caption{
List of the absorption lines detected for our RRLs, including their 
wavelength and identification numbers (ID), 
excitation potential (EP), log $gf$, equivalent widths (EW), abundance ratios (ab) including the ratios relative to the Sun (rel ab) and their uncertainties (rel ab e), and NLTE corrections (NLTE corr). 
}  
\label{tab:lines}
\begin{center}

\begin{tabular}{|c|c|c|c|c|c|c|c|c|c|c|c|c|c|c|c|c|c|c}
\hline
 Wavelength [\AA] &    ID &     EP &  log $gf$ & \shortstack{EW \\ J023001} & \shortstack{ab \\ J023001} & \shortstack{rel ab \\ J023001} & \shortstack{rel ab e \\ J023001} & \shortstack{NLTE corr \\ J023001} & \shortstack{EW \\ J034239} & \shortstack{ab \\ J034239} & \shortstack{rel ab \\ J034239} & \shortstack{rel ab e \\ J034239} & \shortstack{NLTE corr \\ J034239} & \shortstack{EW \\ J050902} & \shortstack{ab \\ J050902} & \shortstack{rel ab \\ J050902} & \shortstack{rel ab e \\ J050902} & \shortstack{NLTE corr \\ J050902}     \\

\hline
    7771.94 &   8.0 &  9.146 &  0.002 &   36.7 &   7.65 &     $<$0.75 &           0.30 &          -- &     -- &     -- &          -- &             -- &             -- &     -- &     -- &          -- &             -- &             --  \\
    7774.17 &   8.0 &  9.146 &  0.223 &   39.0 &   7.45 &     $<$0.55 &           0.30 &          -- &   52.4 &   7.75 &        0.45 &           0.30 &          $-$0.20 &   26.8 &   7.00 &       $-$0.10 &           0.20 &          $-$0.35  \\
    7775.39 &   8.0 &  9.146 &  0.369 &     -- &     -- &          -- &             -- &             -- &     -- &     -- &          -- &             -- &             -- &     -- &     -- &          -- &             -- &             --       \\
    5889.95 &  11.0 &  0.000 &  0.108 &  109.0 &   4.35 &       $-$0.10 &           0.15 &           -- &     -- &     -- &          -- &             -- &             -- &     -- &     -- &          -- &             -- &             --  \\
    5895.92 &  11.0 &  0.000 & $-$0.194 &  101.4 &   4.55 &        0.10 &           0.15 &           -- &     -- &     -- &          -- &             -- &             -- &   59.8 &   4.15 &    $<$$-$0.50 &           0.20 &           --  \\
    4351.90 &  12.0 &  4.346 & $-$0.583 &   46.2 &   5.95 &        0.15 &           0.15 &           0.05 &     -- &     -- &          -- &             -- &             -- &     -- &     -- &          -- &             -- &             --  \\
    4702.99 &  12.0 &  4.346 & $-$0.440 &   67.9 &   6.10 &        0.30 &           0.15 &           0.05 &     -- &     -- &          -- &             -- &             -- &     -- &     -- &          -- &             -- &             --  \\
    5167.32 &  12.0 &  2.709 & $-$0.870 &  123.0 &   6.00 &        0.20 &           0.15 &           0.03 &     -- &     -- &          -- &             -- &             -- &  152.9 &   6.10 &     $<$0.05 &           0.20 &           0.00  \\
    5172.68 &  12.0 &  2.712 & $-$0.393 &     -- &     -- &          -- &             -- &             -- &     -- &     -- &          -- &             -- &             -- &  178.8 &   5.85 &       $-$0.20 &           0.20 &          $-$0.00  \\
    5183.60 &  12.0 &  2.717 & $-$0.167 &  177.7 &   6.05 &        0.25 &           0.30 &          $-$0.02 &  256.4 &   6.50 &        0.30 &           0.35 &           0.05 &  222.1 &   6.00 &       $-$0.00 &           0.20 &          $-$0.05  \\
    5528.40 &  12.0 &  4.346 & $-$0.498 &   62.0 &   6.10 &        0.30 &           0.15 &           0.05 &     -- &     -- &          -- &             -- &             -- &     -- &     -- &          -- &             -- &             --  \\
    6102.72 &  20.0 &  1.879 & $-$0.790 &   17.2 &   5.00 &     $<$0.45 &           0.15 &           -- &     -- &     -- &          -- &             -- &             -- &     -- &     -- &          -- &             -- &             --  \\
    6122.22 &  20.0 &  1.886 & $-$0.315 &     -- &     -- &          -- &             -- &             -- &   60.7 &   5.05 &     $<$0.10 &           0.10 &           0.20 &   19.1 &   4.80 &     $<$0.05 &           0.20 &          --  \\
    6162.17 &  20.0 &  1.899 & $-$0.089 &     -- &     -- &          -- &             -- &             -- &     -- &     -- &          -- &             -- &             -- &   17.5 &   4.55 &       $-$0.20 &           0.20 &          --  \\
    4337.91 &  22.1 &  1.080 & $-$1.130 &   56.8 &   3.55 &        0.40 &           0.30 &           0.00 &     -- &     -- &          -- &             -- &             -- &     -- &     -- &          -- &             -- &             --  \\
    4443.80 &  22.1 &  1.080 & $-$0.717 &   80.6 &   3.45 &        0.30 &           0.30 &          $-$0.00 &  128.9 &   3.90 &     $<$0.40 &           0.30 &          $-$0.00 &     -- &     -- &          -- &             -- &             --  \\
    4450.48 &  22.1 &  1.084 & $-$1.518 &     -- &     -- &          -- &             -- &             -- &     -- &     -- &          -- &             -- &             -- &     -- &     -- &          -- &             -- &             --  \\
    4468.49 &  22.1 &  1.131 & $-$0.620 &     -- &     -- &          -- &             -- &             -- &  137.5 &   4.00 &        0.50 &           0.30 &          $-$0.00 &     -- &     -- &          -- &             -- &             --  \\
    4501.27 &  22.1 &  1.116 & $-$0.767 &   75.5 &   3.45 &        0.30 &           0.30 &          $-$0.02 &  118.6 &   3.80 &        0.25 &           0.30 &          $-$0.10 &     -- &     -- &          -- &             -- &             -- \\
    4533.96 &  22.1 &  1.237 & $-$0.770 &   67.3 &   3.45 &        0.30 &           0.30 &          $-$0.02 &     -- &     -- &          -- &             -- &             -- &   59.8 &   3.50 &        0.10 &           0.20 &           0.00  \\
    4549.62 &  22.1 &  1.584 & $-$0.105 &     -- &     -- &          -- &             -- &             -- &  148.2 &   4.10 &     $<$0.55 &           0.30 &           0.00 &     -- &     -- &          -- &             -- &             --  \\
    4563.76 &  22.1 &  1.221 & $-$0.960 &     -- &     -- &          -- &             -- &             -- &  112.3 &   4.00 &        0.45 &           0.30 &          $-$0.10 &   43.3 &   3.50 &        0.10 &           0.20 &           0.00  \\
    4077.71 &  38.1 &  0.000 &  0.148 &     -- &     -- &          -- &             -- &             -- &    0.0 &   1.90 &     $<$0.45 &           0.15 &           -- &    0.0 &   1.40 &     $<$0.10 &           0.20 &           --  \\
    4215.52 &  38.1 &  0.000 & $-$0.166 &     -- &     -- &          -- &             -- &             -- &     -- &     -- &          -- &             -- &             -- &     -- &     -- &          -- &             -- &             -- \\
    4554.03 &  56.1 &  0.000 &  0.140 &    0.0 &   0.55 &        0.15 &           0.25 &           -- &    0.0 &   1.00 &        0.20 &           0.25 &           -- &    0.0 &   0.80 &     $<$0.20 &           0.20 &           --  \\
    4934.08 &  56.1 &  0.000 & $-$0.160 &     -- &     -- &          -- &             -- &             -- &    0.0 &   1.40 &     $<$0.60 &           0.25 &           -- &     -- &     -- &          -- &             -- &             -- \\
\hline
\end{tabular}

\vspace{15pt}

\begin{tabular}{c|c|c|c|c|c|c|c|c|c|c|c|c|c|c}
\hline
 \shortstack{EW \\ HV205840} & \shortstack{ab \\ HV205840} & \shortstack{rel ab \\ HV205840} & \shortstack{rel ab e \\ HV205840} & \shortstack{NLTE corr \\ HV205840} & \shortstack{EW \\ J040422} & \shortstack{ab \\ J040422} & \shortstack{rel ab \\ J040422} & \shortstack{rel ab e \\ J040422} & \shortstack{NLTE corr \\ J040422} & \shortstack{EW \\ HV210205} & \shortstack{ab \\ HV210205} & \shortstack{rel ab \\ HV210205} & \shortstack{rel ab e \\ HV210205} & \shortstack{NLTE corr \\ HV210205} \\

\hline
        -- &     -- &          -- &             -- &             -- &      -- &      -- &           -- &              -- &              -- &     -- &     -- &          -- &             -- &             --  \\
      -- &     -- &          -- &             -- &             -- &    92.0 &    7.85 &      $<$0.70 &            0.35 &           -- &   70.1 &   7.95 &        0.30 &           0.15 &          $-$0.35  \\
     -- &     -- &          -- &             -- &             -- &    90.1 &    7.65 &         0.55 &            0.35 &           -- &   76.0 &   7.85 &        0.25 &           0.15 &          $-$0.25 \\
    102.0 &   4.40 &       $-$0.55 &           0.25 &           0.00 &      -- &      -- &           -- &              -- &              -- &     -- &     -- &          -- &             -- &             --  \\
    69.2 &   4.20 &       $-$0.70 &           0.25 &           0.00 &      -- &      -- &           -- &              -- &              -- &  153.8 &   5.45 &        0.25 &           0.15 &           0.00  \\
    -- &     -- &          -- &             -- &             -- &      -- &      -- &           -- &              -- &              -- &     -- &     -- &          -- &             -- &             --  \\
    44.3 &   5.85 &       $-$0.40 &           0.20 &           0.05 &      -- &      -- &           -- &              -- &              -- &  135.5 &   7.00 &        0.45 &           0.15 &           0.05  \\
    111.6 &   6.00 &    $<$$-$0.30 &           0.25 &          $-$0.00 &   160.0 &    6.45 &         0.40 &            0.35 &           0.00 &  192.0 &   7.05 &        0.50 &           0.20 &           0.05  \\
    152.6 &   6.20 &       $-$0.10 &           0.25 &          $-$0.00 &   231.6 &    6.70 &         0.65 &            0.40 &           $-$0.02 &  257.0 &   7.05 &        0.50 &           0.25 &           0.05  \\
     149.5 &   5.90 &       $-$0.35 &           0.25 &          $-$0.00 &      -- &      -- &           -- &              -- &              -- &  277.0 &   6.90 &        0.40 &           0.25 &           0.05  \\
     61.5 &   6.15 &       $-$0.10 &           0.20 &           0.05 &      -- &      -- &           -- &              -- &              -- &  116.6 &   6.85 &        0.35 &           0.10 &           0.00  \\
       -- &     -- &          -- &             -- &             -- &      -- &      -- &           -- &              -- &              -- &   72.6 &   5.90 &        0.65 &           0.10 &           0.40  \\
       -- &     -- &          -- &             -- &             -- &      -- &      -- &           -- &              -- &              -- &   88.0 &   5.70 &        0.45 &           0.10 &           0.40  \\
      -- &     -- &          -- &             -- &             -- &      -- &      -- &           -- &              -- &              -- &   92.1 &   5.55 &        0.30 &           0.10 &           0.40  \\
      -- &     -- &          -- &             -- &             -- &      -- &      -- &           -- &              -- &              -- &     -- &     -- &          -- &             -- &             --  \\
      -- &     -- &          -- &             -- &             -- &   108.5 &    4.00 &         0.60 &            0.40 &           0.00 &     -- &     -- &          -- &             -- &             -- \\
       -- &     -- &          -- &             -- &             -- &      -- &      -- &           -- &              -- &              -- &     -- &     -- &          -- &             -- &             --  \\
       -- &     -- &          -- &             -- &             -- &   123.8 &    4.05 &         0.65 &            0.40 &           0.00 &     -- &     -- &          -- &             -- &             -- \\
       -- &     -- &          -- &             -- &             -- &   103.6 &    4.00 &         0.60 &            0.40 &           0.00 &     -- &     -- &          -- &             -- &             --  \\
        -- &     -- &          -- &             -- &             -- &      -- &      -- &           -- &              -- &              -- &     -- &     -- &          -- &             -- &             -- - \\
       -- &     -- &          -- &             -- &             -- &      -- &      -- &           -- &              -- &              -- &     -- &     -- &          -- &             -- &             --  \\
     74.8 &   3.65 &        0.05 &           0.35 &          $-$0.10 &    80.6 &    4.05 &         0.70 &            0.40 &            0.01 &     -- &     -- &          -- &             -- &             --  \\
       0.0 &   2.75 &     $<$1.20 &           0.25 &           -- &     0.0 &    1.70 &         0.40 &            0.40 &            -- &    0.0 &   3.20 &     $<$1.40 &           0.15 &           --  \\
     0.0 &   2.70 &     $<$1.15 &           0.30 &           -- &     0.0 &    1.65 &         0.35 &            0.40 &            -- &    0.0 &   3.25 &     $<$1.45 &           0.15 &           --  \\
       -- &     -- &          -- &             -- &             -- &     0.0 &    1.15 &         0.55 &            0.40 &            -- &    0.0 &   2.40 &     $<$1.30 &           0.15 &           --  \\
       -- &     -- &          -- &             -- &             -- &      -- &      -- &           -- &              -- &              -- &    0.0 &   2.70 &     $<$1.60 &           0.15 &           --  \\
\hline
\end{tabular}

\vspace{15pt}

\begin{tabular}{c|c|c|c|c|}
\hline
 \shortstack{EW \\ HV204704} & \shortstack{ab \\ HV204704} & \shortstack{rel ab \\ HV204704} & \shortstack{rel ab e \\ HV204704} & \shortstack{NLTE corr \\ HV204704} \\

\hline
    48.0 &    7.55 &      $<$0.05 &            0.10 &           $-$0.45 \\
    89.3 &    7.90 &         0.40 &            0.15 &           $-$0.40 \\
    93.6 &    7.80 &         0.30 &            0.15 &           $-$0.30 \\
   205.2 &    5.40 &         0.40 &            0.10 &            0.00 \\
   175.1 &    5.30 &         0.25 &            0.10 &            0.00 \\
      -- &      -- &           -- &              -- &              -- \\
   134.1 &    6.80 &         0.40 &            0.05 &            0.05 \\
      -- &      -- &           -- &              -- &              -- \\
  219.4 &    6.65 &      $<$0.25 &            0.10 &            0.05 \\
      -- &      -- &           -- &              -- &              -- \\
   107.9 &    6.50 &      $<$0.10 &            0.05 &            0.00 \\
       -- &      -- &           -- &              -- &              -- \\
    84.7 &    5.35 &      $<$0.20 &            0.05 &            0.80 \\
    82.3 &    5.10 &        $-$0.00 &            0.05 &            0.75 \\
    146.7 &    4.15 &         0.40 &            0.10 &           $-$0.10 \\
   163.2 &    4.05 &         0.30 &            0.10 &           $-$0.00 \\
   105.7 &    3.80 &      $<$0.05 &            0.10 &           $-$0.05 \\
     -- &      -- &           -- &              -- &              -- \\
    -- &      -- &           -- &              -- &              -- \\
     -- &      -- &           -- &              -- &              -- \\
      -- &      -- &           -- &              -- &              -- \\
     -- &      -- &           -- &              -- &              -- \\
      -- &      -- &           -- &              -- &              -- \\
     0.0 &    3.10 &      $<$1.45 &            0.10 &            -- \\
     0.0 &    1.20 &      $<$0.25 &            0.10 &            -- \\
     -- &      -- &           -- &              -- &              -- \\
\hline
\end{tabular}

\end{center}
\end{sidewaystable*}

\newpage

\renewcommand{\thefigure}{B\arabic{figure}}
\setcounter{figure}{0}

\renewcommand{\thetable}{B\arabic{table}}
\setcounter{table}{0}

\section{Atmospheric parameters: initial estimates}
\label{sec:atms}

Here we describe the methodology followed to obtain estimates of the stellar parameters (metallicities, effective temperatures, surface gravities, and microturbulence velocities) of our sample. 
We used these values as initial guesses for the purely spectroscopic approach described in Section~\ref{sec:synthesis}. 

\subsection{Metallicities}
\label{sec:fehs}

An initial estimation of the metallicity of our sample followed different approaches, taking advantage of the broad spectral coverage of our spectra.

For the stars with clearly defined Ca~II triplet lines and higher SNR ($\gtrsim 10$), we used the correlation between [Fe/H] and the equivalent width (EW) of the Ca line at 8,498\,\AA\ described by \citet{Wallerstein2012}. 
In some stars, the 8,498\,\AA\ Ca line was not reliably detected, so here we estimated the EW based on a empirical scaling using the 8,498\,\AA\ and the 8,542\,\AA\ obtained from the spectra of CS 22874, HD 76483, and of the metal-poor r-process-rich star HD~20 \citep[see e.g.][]{Hanke2020a}. The latter is a well-known giant for which accurate stellar parameters and high-resolution spectra are available and we use it as a benchmark star.
The scaling coefficient used was EW(8,498\,\AA)/EW(8,542\,\AA) = 0.5119. 
The EW's measured for this approach were obtained using Gaussian profiles, as they better fit the shape of the lines.

\citet{Singh2020} presented empirical relations that predict [Fe/H] and are valid for carbon-enhanced metal-poor (CEMP) stars as well as carbon-normal stars. 
These relations were tested for seven elements with strong features at low metallicities, and showed Cr and Ni as the best [Fe/H] tracers. 
In that work, linear scaling relations were also obtained for the Mg~I absorption lines at 5,173\,\AA\, and 5,184\,\AA, valid for lines with EW $<$ 1,000\,m\AA\ (although with an accuracy within $\sim$ 0.4\,dex).
The empirical Mg I-[Fe/H] relation was obtained via private communication with the authors. 
Thus, we also used these correlations to obtain an estimate of the star's metallicity (iron abundance), by measuring the Mg lines using Gaussians.

\begin{figure*}
\includegraphics[angle=0,scale=.38]{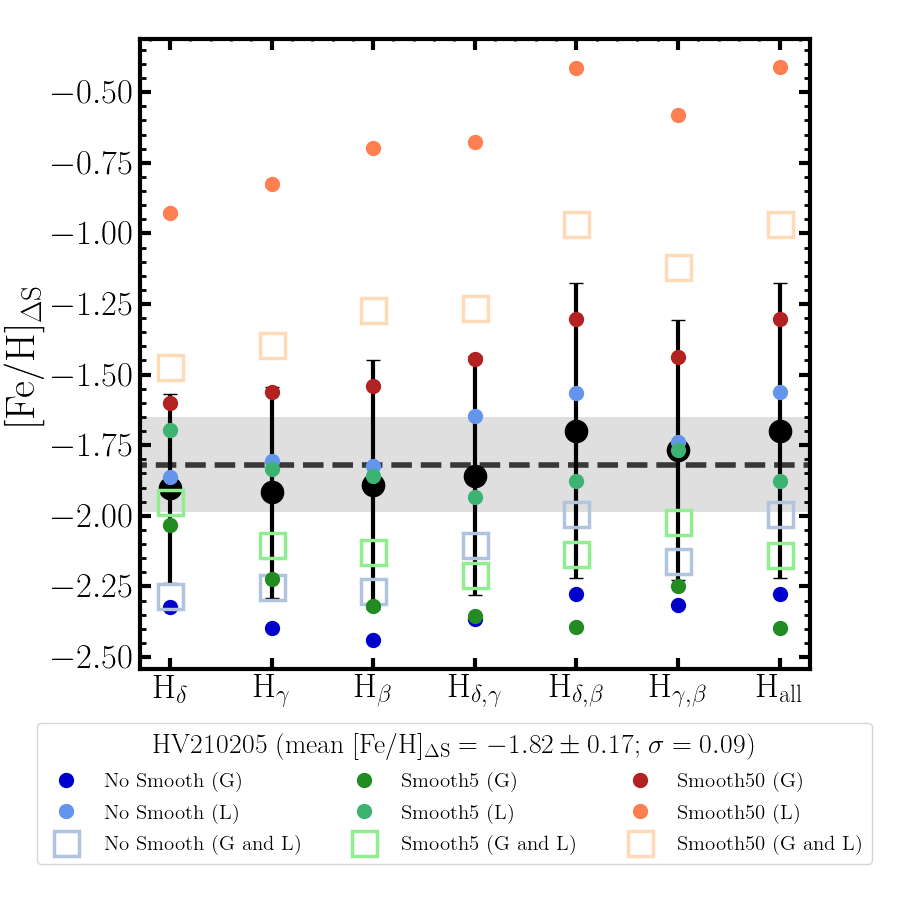} 
\includegraphics[angle=0,scale=.38]{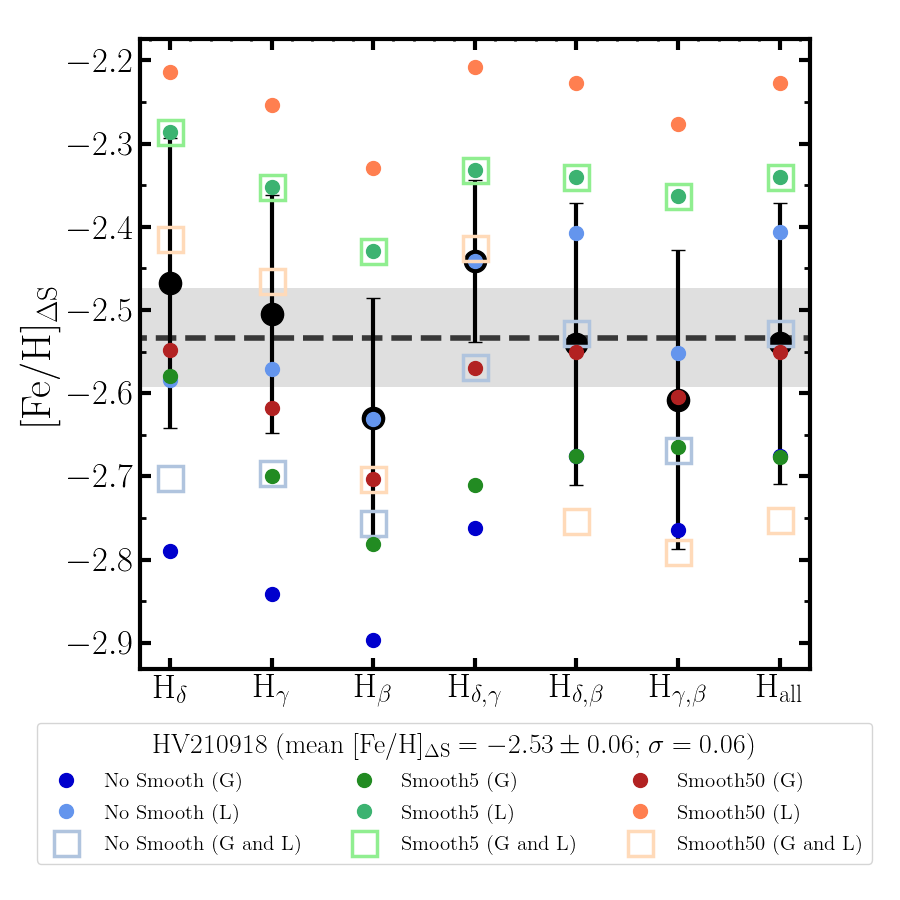} 
\caption{Iron abundance estimations from the $\Delta$S method and different combinations of lines, for two stars in our sample. 
These figures show the strong dependency of the resulting [Fe/H] on the resolution of the input spectra (colour-coded), and the line profiles used, i.e., Gaussian only (G), Lorentzian only (L), or the best-fitting profiles (G and L). The average of the latter is displayed with a black filled circle for each line combination, and the average of the different combinations is represented with a horizontal dashed line for illustrative purposes. The dispersion of this average is shown as a grey region. We caution the use of Lorentzians at low resolution as this leads to inconsistent (overestimated) metallicities. 
}
\label{fig:deltas}
\end{figure*}

Additionally, we estimated the metallicity using an updated version of the $\Delta$S method \citep[][]{Crestani21a}. 
This method relies on the correlations between the EW of the Ca K line (3,933\,\AA\,) and those from the Balmer lines. 
As \citet{Crestani21a} provide correlation coefficients for different combinations of Balmer lines with the Ca K line, we determined the metallicity for each combination, using Gaussian and Lorentzian profiles, and with different levels of spectral convolution to mimic the low resolution of their work, i.e., none, medium (convolution box size of 5\,pixels), and drastic convolution (box size of 50\,pixels). 
Applying medium convolutions results in spectral resolutions close to those of \citet{Crestani21a}. 
Moreover, given the overall higher resolution and lower S/N of our spectra, as compared with those from \citet{Crestani21a}, we reduced the EW integration region around each line from 20\,\AA , as used by the authors, to 10\,\AA . 
In a few cases, the region was even further reduced (down to 5\,\AA\ around the lines) to provide a sensible fit.
For each combination of lines, we selected the line profile that best fit the lines centres, and their wings, for each convolved level. 
This typically corresponds to Lorentzian and Gaussian profiles for the Balmer and Ca lines, respectively. 
An example of the outcome of these calculations is shown in Figure~\ref{fig:deltas}. 
In this figure, the impact of applying the $\Delta$S method to spectra with resolutions significantly different than the one used by \citet{Crestani21a} is clearly visible.
The figure shows that this, in addition to adopting inadequate line profiles in RRLs observed in phases of abrupt atmospheric changes (as is the case of HV210205, observed at $\phi \sim$0.85) can lead to inconsistent metallicity estimations. 
The combination of lines that generally showed lower dispersion between the methods used, of the order of 0.15\,dex, are single comparisons (CaK with either H$_\beta$, H$_\gamma$, or H$_\delta$), whereas using multiple Balmer lines results in [Fe/H] scatter closer to 0.2\,dex.
We also find that, for our RRLs, the metallicity estimates from the $\Delta$S method are on average 0.17\,dex higher than those from \citet{Wallerstein2012}'s correlation.

We note that the metallicity estimates obtained from the relation derived by \citet{Singh2020} are consistently more metal-poor than those obtained from the Ca II triplet and from the $\Delta$S method. 
In fact, they lead to [Fe/H] values that are 0.30 and 0.60\,dex lower on average, respectively.
Given these large differences, and that the validity of the Mg I-Fe relations for the stars in our sample is not certain, we computed the average of the resulting metallicities from the other two methods, as it should represent a sensible range of metallicities to be used as initial guesses for the rest of the analysis. 
We consider this justified since, in theory, it reduces possible biases from the choice of a given metallicity scale, in order to ensure the convergence of the atmospheric parameter determination described in Section~\ref{sec:synthesis}. 
The resulting mean metallicities are displayed in Table~\ref{tab:atms}.

Another method commonly used for deriving the metallicity of RRLs relies on the correlation between their periods, light curve shapes (mainly through the phase parameter $\phi_{31}$), and [Fe/H] 
\citep[see e.g.][]{Jurcsik1996,Smolec2005,Nemec2011,Nemec2013,Dekany2021,Mullen2021,Mullen2022}. 
This approach, however, is inherently sensitive to the uncertainties in the light curve measurements and its phase coverage. 
Because almost half of our targets are taken from the Catalina survey (good phase coverage but large photometric uncertainties), and the other half from our independent surveys (modest phase coverage and small uncertainties), this method is, in principle, not the best suited for our study.
In addition, the [Fe/H] from photometric formulae depends on the metallicity scale used, the [Fe/H] range in which is valid, and can reach a scatter of 0.5\,dex when compared with high-resolution spectra (see e.g. Figure~6 from \citealt{Dekany2021} or Figure~11 from \citealt{Mullen2021}). 
Recently, \citet{Dekany2021} obtained new empirical relations between the iron abundance of RRLs and their light-curve parameters based on near-infrared photometry. 
The training set used by these authors consisted of high-resolution spectra of 80 RRab with [Fe/H] from solar to $\sim -2.5$, collected from the data sets of \citet{Crestani21a}, \citet{For2011b}, \citet{Chadid2017}, and \citet{Sneden2017}.
Also recently, \citet{Mullen2021} reported new period-$\phi_{31}$-[Fe/H] relations in the optical (including $V$), from stars in a similar [Fe/H] range and calibrated with the same metallicity scale. 
For a quick comparison with our work, we note that J051424 is included in the list of Large Magellanic Cloud (LMC) RRLs with $I-$band based [Fe/H] from \citet{Dekany2021}. 
For this star, their model predicts [Fe/H] $=-1.68$, whereas the relation of \citet{Mullen2021} yield $-2.21$ (from Catalina's $V-$band photometry), and the use of the $\Delta S$ method from our spectra results in $-2.34$.
As a second comparison, using the formula from \citet{Mullen2021} on J040422 (for which we possess relatively high signal-to-noise spectra with clearly defined metallic lines) yield [Fe/H] $=-1.51$, while following the $\Delta S$ method and the EW technique (see Section~\ref{sec:Abund}) we obtain [Fe/H]$ =-1.82$ and $-1.56$, respectively.

\subsection{Effective temperatures}
\label{sec:teffs}
As a rough effective temperature estimation for these RRLs we used the photometric colour transformations from \citet{Casagrande2010}, which rely on Johnson-Cousins photometry. 
The stars' estimated [Fe/H] values derived above were used for these transformations. 
Prior to performing any computation, we dereddened the magnitudes of our halo RRLs targets by using the dust maps from \citet{Schlafly11}, adopting R$_V$ $=3.1$ \citep{Schultz75,Cardelli89}.  
For the brighter RRLs in our sample, the usage of the relations from \citet{Casagrande2010} is straightforward given that photometry for their mean magnitudes in the Johnson-Cousins system is available with relatively small uncertainties (from the Catalina survey, for instance).
In the case of the fainter RRLs, we used Pan-STARRS photometry and converted those magnitudes to Johnson-Cousins using the photometric relations provided by \citet{Tonry2012}. 
When neither Johnson-Cousins nor Pan-STARRS magnitudes were directly available, but the star is listed in the {\it Gaia} DR3 catalogue, we adopted the photometric transformations given in the {\it Gaia} documentation to derive Johnson-Cousins magnitudes\footnote{\url{https://gea.esac.esa.int/archive/documentation/GDR2/Data\_processing/chap\_cu5pho/sec\_cu5pho\_calibr/ssec\_cu5pho\_PhotTransf.html}}.

It is worth mentioning that in our study, the effective temperatures and gravities obtained from photometric indices are used as initial guesses for the method described in Section~\ref{sec:synthesis} and mostly as a test, due to their high dependence on the phase in which such photometry was obtained (as stated by, e.g., \citealt{Kolenberg10}), which is in general rather uncertain. 

\subsection{Surface gravity - log $g$}
\label{sec:loggs}
In order to have a rough estimate of the observed surface gravity of RRLs, we first followed the methodology from \citet{Nissen97}, using the bolometric corrections described by \citet{Flower96} and \citet{Torres10}, and the parallaxes from {\it Gaia} DR3, assuming a fixed RRL mass of 0.8\,M$_\odot$ \citep[][]{Simon1989,Clement1997,Catelan2015}.
However, this approach resulted in imprecise gravities overall, mostly due to the stellar parallaxes of our rather distant sources (small parallaxes and large relative uncertainties). 
Thus, we assumed log $g$ = 2.0 as sensible initial estimates \citep[][]{For2011b} for all the RRLs. 

\subsection{Microturbulence - V$_t$}\label{sec:vmic}
The micro/macroturbulence velocities were estimated applying the same empirical relation used by \citet{BlancoCuaresma14b}, based on the {\it Gaia}-ESO Survey (GES) Ultraviolet and Visual Echelle Spectrograph (UVES) data release 1 and the {\it Gaia} FGK benchmark stars \citep{Jofre14}. 
We note that, for a star of T$_{\rm eff}$ = 6,500\,K, [Fe/H] = $-1.5$, and log $g$ = 2.0, using the empirical relation derived by \citet{Mashonkina17} (based on very- and extremely-metal poor stars) results in a microturbulence velocity 0.3\,km\,s$^{-1}$ higher than that based on the {\it Gaia} FGK benchmark stars' scaling.
In the case of the star HD 76483, we used the literature values [Fe/H] = $-0.5$, T$_{\rm eff}$ = 8,600\,K,  and log $g$ = 3.77 as initial estimates \citep{David15}.

\renewcommand{\thefigure}{C\arabic{figure}}
\setcounter{figure}{0}

\renewcommand{\thetable}{C\arabic{table}}
\setcounter{table}{0}

\section{Orbits and kinematics of our RR Lyrae stars}

In this section, we provide the orbital parameters of our sample, as well figures displaying their eccentricity, and apocentric and pericentric distances ($r_{\rm apo}$ and $r_{\rm peri}$, respectively). 
The orbits determined for three of our stars are provided as examples of RRLs at different distances and with different signal-to-noise ratios (J023001, HV205840, and HV204704). 
The effects of the LMC in each of these stars is observed in the asymmetry of the computed orbits.

\textbf{\begin{figure*}
\includegraphics[angle=0,scale=.42]{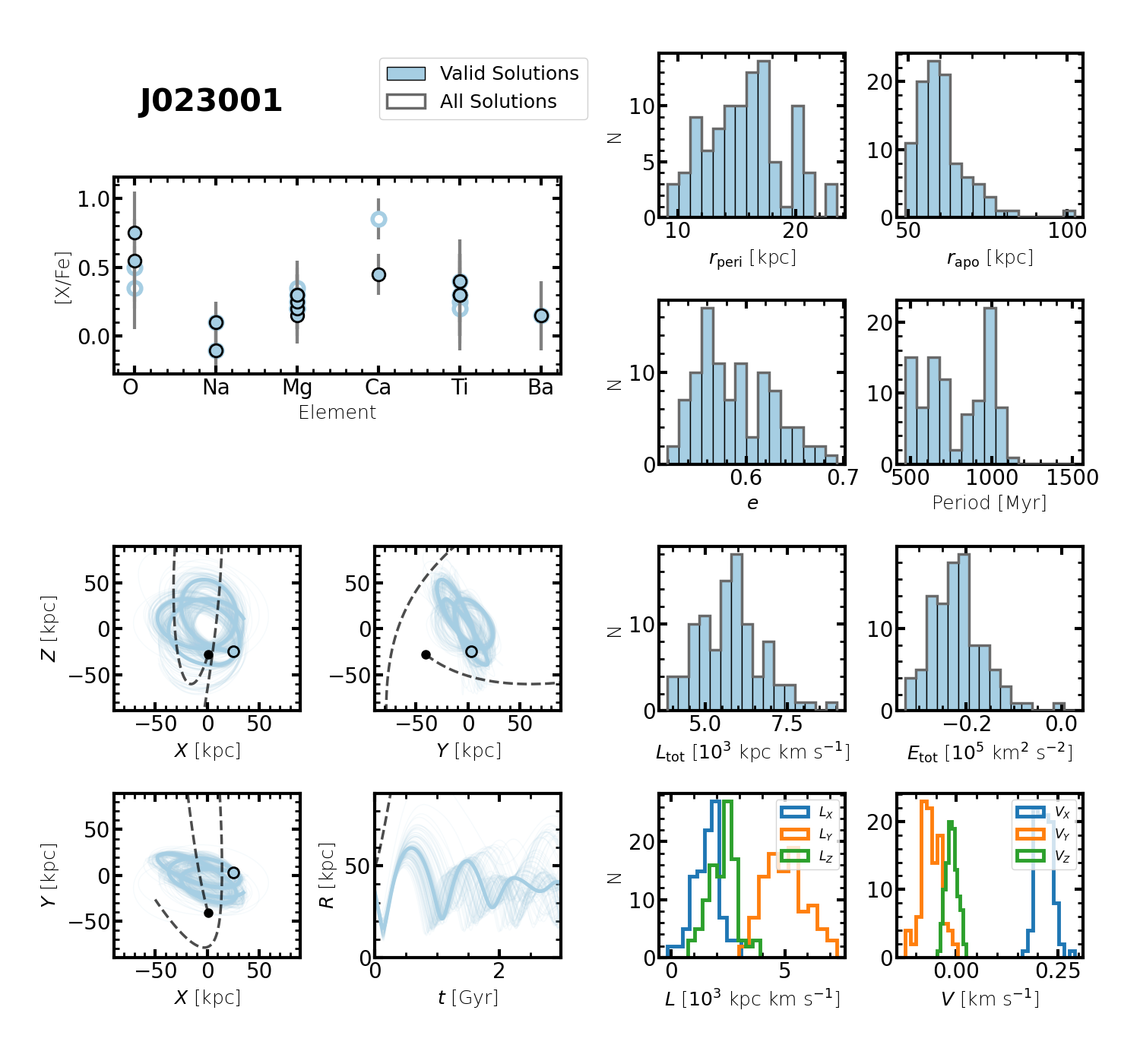} 
\caption{Relative abundances and integrated orbit of the star J034239.
In the {\it top left} panel, open circles represent the abundances after applying NLTE corrections.
The orbits shown are integrated for 3\,Gyr backward using GALPY adopting a perturbed MW potential.
In the {\it bottom left} panels, black dashed lines and a black filled circle represent the orbit and current position of the LMC, respectively.
The histograms display the distribution of the computed orbital parameters from 100 orbit realisations.
These realisations are shown with transparent lines in the {\it bottom left} panels. 
}
\label{fig:abundOrbits1}
\end{figure*}}

\newpage

\textbf{\begin{figure*}
\includegraphics[angle=0,scale=.42]{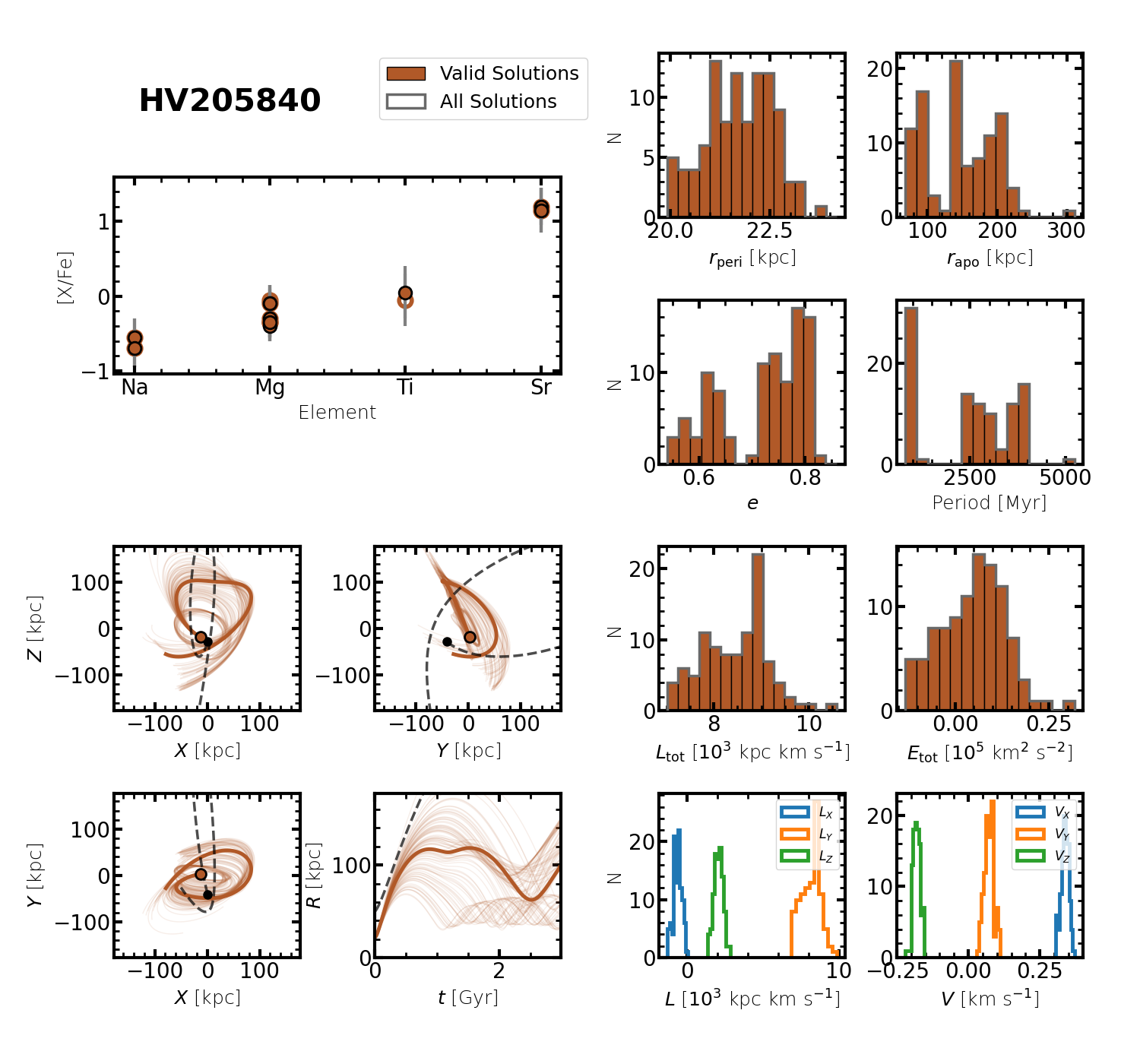} 
\caption{Same as Figure~\ref{fig:abundOrbits1}, but for the star HV205840. The orbits are integrated for 3\,Gyr forward and backward using GALPY.}
\label{fig:abundOrbits2}
\end{figure*}}

\newpage

\textbf{\begin{figure*}
\includegraphics[angle=0,scale=.42]{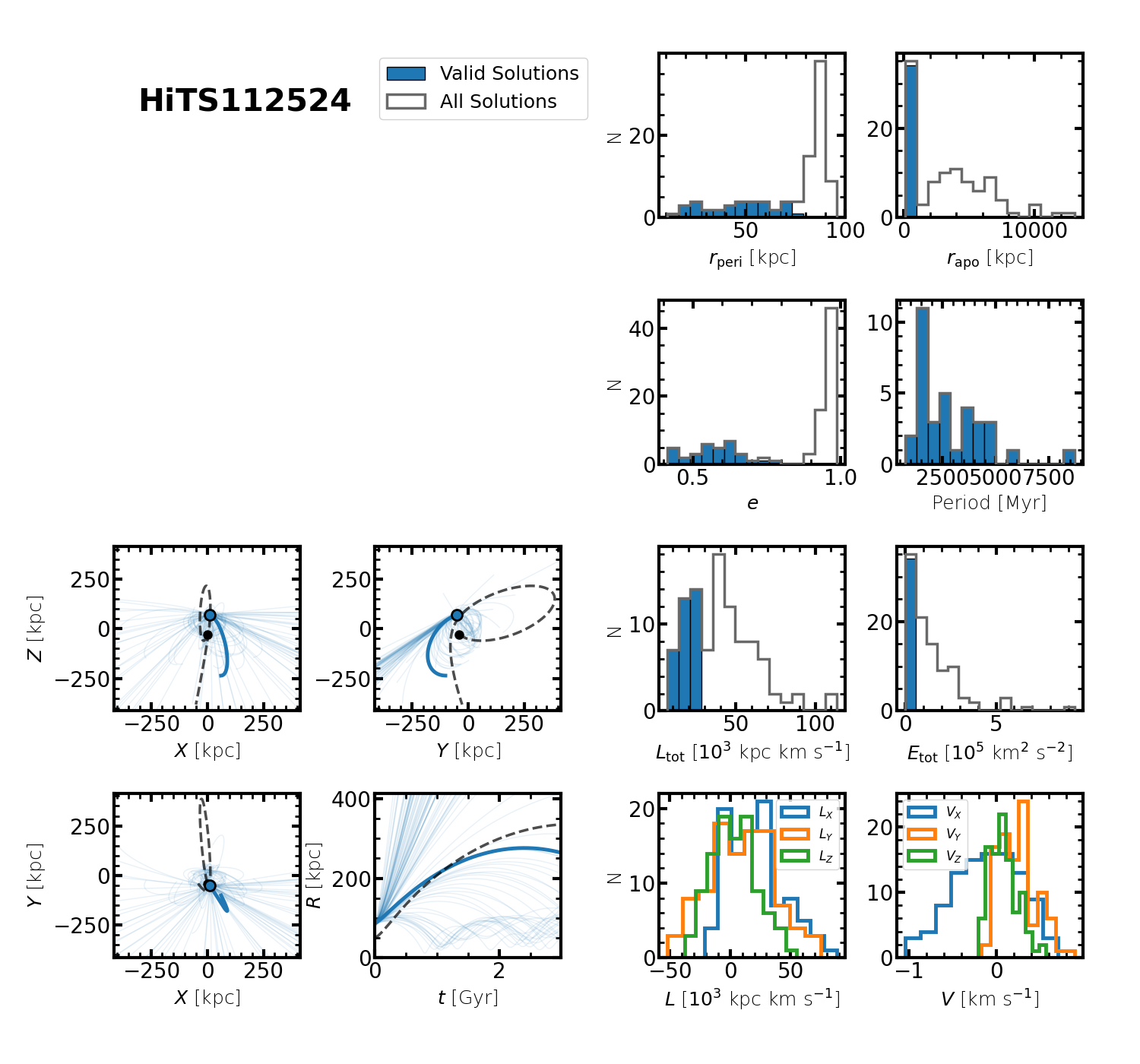} 
\caption{Same as Figure~\ref{fig:abundOrbits1}, but for the star HiTS112524. The orbits are integrated for 3\,Gyr forward and backward using GALPY.}
\label{fig:abundOrbits3}
\end{figure*}}

\begin{figure*}

\includegraphics[angle=0,scale=.34]{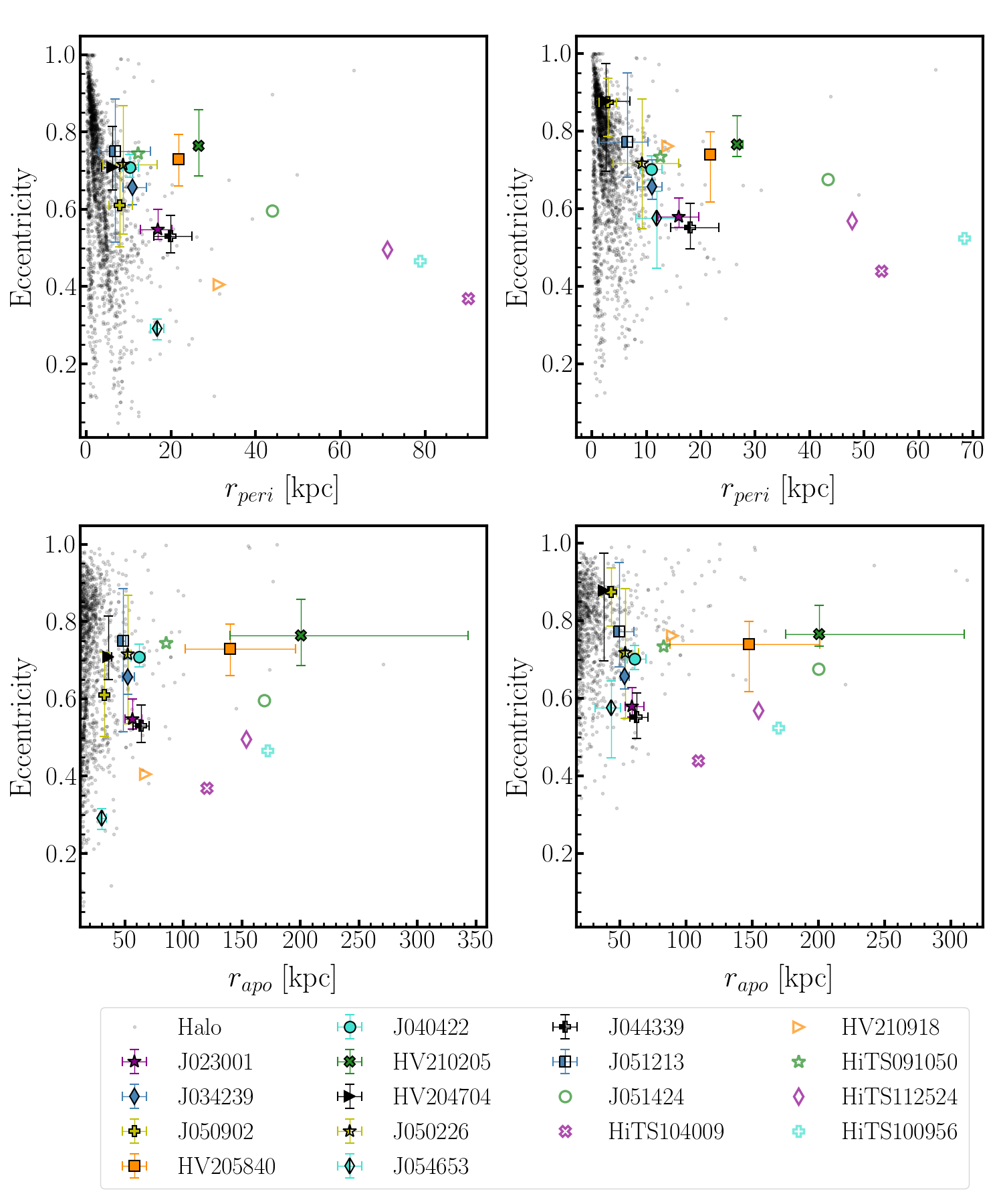} 
\caption{Eccentricity of the integrated orbits of our RRLs (large symbols)
and halo stars (black dots), as a function of pericentric and apocentric distance ({\it upper} and {\it lower} panels, respectively). The {\it left} panels are computed using the isolated MW potential model, and the {\it right} panels consider the perturbed potential.
}
\label{fig:periapoe}
\end{figure*}

\begin{sidewaystable}\small
\caption{
Orbital parameters of the stars in our sample, integrated for 3\,Gyr using GALPY's \textit{MWPotential2014}. 
}
\label{tab:orbits1}
\begin{center}

\begin{tabular}{|c|c|c|c|c|c|c|c|c|c|c|c|c|}
\hline
\multicolumn{1}{|c|}{ID} &
\multicolumn{1}{c|}{$d_{\rm H}$} &
\multicolumn{1}{c|}{$R$} &
\multicolumn{1}{c|}{$\mu_{\alpha^*}$} &
\multicolumn{1}{c|}{$\mu_\delta$} &
\multicolumn{1}{c|}{$V$} &
\multicolumn{1}{c|}{$E$} &
\multicolumn{1}{c|}{$L$} &
\multicolumn{1}{c|}{$r_{\rm peri}$} &
\multicolumn{1}{c|}{$r_{\rm apo}$} &
\multicolumn{1}{c|}{$e$} &
\multicolumn{1}{c|}{Orb. Period} &
\multicolumn{1}{c|}{Bound Likelihood}\\

     &   (kpc) &   (kpc) &   (mas\,yr$^{-1}$) &  (mas\,yr$^{-1}$) &  (km\,s$^{-1}$) &   ($10^{5}$\,km$^{2}$\,s$^{-2}$) &  ($10^{3}$\,kpc\,km\,s$^{-1}$)  &    (kpc) & (kpc) &   & (Gyr) &     \\     

\hline
 HiTS104009 &  $105 \pm 5$ &  $107^{+2}_{-6}$ &  $-0.53 \pm 1.09$ &  $-0.26 \pm 1.25$ &   $168.6^{+85.4}_{-36.8}$ &   $0.25^{+0.15}_{-0.07}$ &   $19.04^{+7.46}_{-4.78}$ &  $90^{+15}_{-36}$ &  $120^{+180}_{-12}$ &  $0.37^{+0.22}_{-0.14}$ &  $2.77^{+3.76}_{-0.73}$ &          $0.13$ \\
 HiTS112524 &   $86 \pm 4$ &   $86^{+4}_{-4}$ &  $-0.84 \pm 0.88$ &  $-0.81 \pm 0.63$ &  $231.7^{+35.6}_{-113.5}$ &   $0.27^{+0.11}_{-0.18}$ &  $18.14^{+5.01}_{-10.40}$ &  $71^{+10}_{-50}$ &  $153^{+131}_{-54}$ &  $0.50^{+0.15}_{-0.13}$ &  $3.02^{+2.80}_{-1.58}$ &          $0.24$ \\
    J050226 &   $48 \pm 2$ &   $51^{+2}_{-2}$ &   $0.39 \pm 0.19$ &  $-0.42 \pm 0.22$ &    $88.4^{+37.5}_{-27.6}$ &  $-0.16^{+0.04}_{-0.03}$ &    $3.66^{+2.22}_{-1.62}$ &     $9^{+8}_{-5}$ &      $53^{+2}_{-2}$ &  $0.72^{+0.15}_{-0.18}$ &  $0.67^{+0.08}_{-0.05}$ &          $1.00$ \\
 HiTS100956 &   $89 \pm 4$ &   $92^{+4}_{-4}$ &  $-0.24 \pm 0.73$ &   $0.38 \pm 0.67$ &   $228.9^{+45.6}_{-93.5}$ &   $0.30^{+0.11}_{-0.18}$ &   $19.87^{+4.90}_{-9.37}$ &   $79^{+8}_{-46}$ &  $172^{+158}_{-70}$ &  $0.47^{+0.22}_{-0.15}$ &  $3.49^{+3.40}_{-1.87}$ &          $0.29$ \\
    J051424 &   $47 \pm 2$ &   $48^{+2}_{-2}$ &   $1.42 \pm 0.27$ &   $1.20 \pm 0.28$ &   $316.7^{+31.6}_{-49.9}$ &   $0.26^{+0.14}_{-0.13}$ &   $14.36^{+2.18}_{-2.58}$ &    $44^{+3}_{-3}$ &  $169^{+178}_{-74}$ &  $0.60^{+0.16}_{-0.21}$ &  $2.87^{+3.65}_{-1.22}$ &          $0.64$ \\
    J023001 &   $30 \pm 2$ &   $36^{+2}_{-1}$ &  $-0.17 \pm 0.18$ &  $-1.57 \pm 0.14$ &   $225.1^{+22.2}_{-23.4}$ &  $-0.11^{+0.06}_{-0.05}$ &    $5.90^{+0.99}_{-1.08}$ &    $17^{+3}_{-4}$ &      $56^{+8}_{-6}$ &  $0.55^{+0.05}_{-0.03}$ &  $0.79^{+0.15}_{-0.10}$ &          $1.00$ \\
    J051213 &   $39 \pm 2$ &   $46^{+2}_{-2}$ &   $0.45 \pm 0.24$ &  $-0.53 \pm 0.22$ &    $93.4^{+40.0}_{-22.0}$ &  $-0.20^{+0.04}_{-0.03}$ &    $3.02^{+2.23}_{-1.40}$ &     $7^{+8}_{-4}$ &      $48^{+2}_{-2}$ &  $0.75^{+0.13}_{-0.24}$ &  $0.61^{+0.07}_{-0.05}$ &          $1.00$ \\
    J050902 &   $26 \pm 1$ &   $32^{+1}_{-1}$ &   $0.29 \pm 0.16$ &  $-1.08 \pm 0.13$ &    $99.3^{+21.1}_{-21.2}$ &  $-0.36^{+0.03}_{-0.03}$ &    $3.09^{+0.73}_{-0.70}$ &     $8^{+3}_{-3}$ &      $32^{+1}_{-1}$ &  $0.61^{+0.11}_{-0.11}$ &  $0.42^{+0.03}_{-0.03}$ &          $1.00$ \\
    J054653 &   $14 \pm 1$ &   $22^{+1}_{-1}$ &   $4.35 \pm 0.31$ &  $-0.81 \pm 0.27$ &   $239.8^{+19.2}_{-11.8}$ &  $-0.31^{+0.06}_{-0.04}$ &    $4.82^{+0.56}_{-0.34}$ &    $17^{+2}_{-2}$ &      $30^{+4}_{-2}$ &  $0.29^{+0.02}_{-0.03}$ &  $0.47^{+0.07}_{-0.04}$ &          $1.00$ \\
    J044339 &   $38 \pm 2$ &   $45^{+2}_{-2}$ &   $0.02 \pm 0.18$ &  $-1.38 \pm 0.17$ &   $203.3^{+19.7}_{-22.5}$ &  $-0.05^{+0.05}_{-0.06}$ &    $6.83^{+1.33}_{-1.08}$ &    $20^{+5}_{-4}$ &      $64^{+7}_{-7}$ &  $0.53^{+0.05}_{-0.04}$ &  $0.93^{+0.14}_{-0.16}$ &          $1.00$ \\
    J040422 &   $25 \pm 1$ &   $31^{+1}_{-1}$ &  $-0.01 \pm 0.17$ &  $-1.98 \pm 0.13$ &   $253.3^{+12.7}_{-14.3}$ &  $-0.10^{+0.04}_{-0.05}$ &    $4.37^{+0.66}_{-0.61}$ &    $10^{+2}_{-2}$ &      $62^{+5}_{-7}$ &  $0.71^{+0.03}_{-0.03}$ &  $0.80^{+0.09}_{-0.10}$ &          $1.00$ \\
    J034239 &   $27 \pm 1$ &   $33^{+1}_{-1}$ &   $0.24 \pm 0.17$ &  $-2.08 \pm 0.13$ &   $219.0^{+16.9}_{-16.0}$ &  $-0.16^{+0.05}_{-0.04}$ &    $4.32^{+1.02}_{-0.62}$ &    $11^{+3}_{-2}$ &      $52^{+6}_{-4}$ &  $0.66^{+0.05}_{-0.05}$ &  $0.69^{+0.11}_{-0.07}$ &          $1.00$ \\
 HiTS091050 &   $61 \pm 3$ &   $66^{+3}_{-3}$ &   $0.02 \pm 0.28$ &  $-0.54 \pm 0.25$ &   $154.3^{+45.7}_{-17.6}$ &   $0.02^{+0.08}_{-0.03}$ &    $5.19^{+4.81}_{-2.07}$ &   $12^{+18}_{-6}$ &     $85^{+14}_{-5}$ &  $0.75^{+0.12}_{-0.20}$ &  $1.13^{+0.38}_{-0.09}$ &          $1.00$ \\
   HV204704 &   $35 \pm 1$ &   $30^{+1}_{-1}$ &  $-0.07 \pm 0.14$ &  $-1.80 \pm 0.11$ &    $149.3^{+13.2}_{-9.5}$ &  $-0.32^{+0.03}_{-0.03}$ &    $2.59^{+0.50}_{-0.89}$ &     $6^{+2}_{-2}$ &      $36^{+2}_{-1}$ &  $0.71^{+0.10}_{-0.06}$ &  $0.45^{+0.03}_{-0.04}$ &          $1.00$ \\
   HV210918 &   $62 \pm 2$ &   $56^{+2}_{-2}$ &   $0.19 \pm 0.39$ &  $-1.04 \pm 0.34$ &   $179.4^{+54.6}_{-50.1}$ &   $0.02^{+0.11}_{-0.10}$ &    $9.07^{+3.79}_{-2.92}$ &  $31^{+18}_{-14}$ &     $67^{+27}_{-8}$ &  $0.41^{+0.26}_{-0.13}$ &  $1.14^{+0.52}_{-0.30}$ &          $0.99$ \\
   HV210205 &   $32 \pm 1$ &   $27^{+1}_{-1}$ &  $-2.38 \pm 0.15$ &  $-2.29 \pm 0.11$ &   $394.6^{+26.0}_{-20.9}$ &   $0.28^{+0.11}_{-0.09}$ &   $10.47^{+0.83}_{-0.83}$ &    $26^{+1}_{-1}$ &  $200^{+143}_{-61}$ &  $0.76^{+0.09}_{-0.08}$ &  $3.21^{+2.87}_{-1.06}$ &          $0.80$ \\
   HV205840 &   $27 \pm 1$ &   $22^{+1}_{-1}$ &  $-2.77 \pm 0.10$ &  $-2.38 \pm 0.09$ &   $393.8^{+16.7}_{-18.5}$ &   $0.19^{+0.08}_{-0.09}$ &    $8.58^{+0.70}_{-0.70}$ &    $22^{+1}_{-1}$ &   $140^{+56}_{-38}$ &  $0.73^{+0.06}_{-0.07}$ &  $2.09^{+0.97}_{-0.62}$ &          $1.00$ \\
\hline

\end{tabular}

\end{center}
\end{sidewaystable}

\begin{sidewaystable}\small
\caption{
Orbital parameters of the stars in our sample, integrated for 3\,Gyr using the adopted perturbed potential. 
}
\label{tab:orbits2}
\begin{center}

\begin{tabular}{|c|c|c|c|c|c|c|c|c|c|c|c|c|}

\hline

\multicolumn{1}{|c|}{ID} &
\multicolumn{1}{c|}{$d_{\rm H}$} &
\multicolumn{1}{c|}{$R$} &
\multicolumn{1}{c|}{$\mu_{\alpha^*}$} &
\multicolumn{1}{c|}{$\mu_\delta$} &
\multicolumn{1}{c|}{$V$} &
\multicolumn{1}{c|}{$E$} &
\multicolumn{1}{c|}{$L$} &
\multicolumn{1}{c|}{$r_{\rm peri}$} &
\multicolumn{1}{c|}{$r_{\rm apo}$} &
\multicolumn{1}{c|}{$e$} &
\multicolumn{1}{c|}{Orb. Period} &
\multicolumn{1}{c|}{Bound Likelihood}\\

     
     &   (kpc) &   (kpc) &   (mas\,yr$^{-1}$) &  (mas\,yr$^{-1}$) &  (km\,s$^{-1}$) &  ($10^{5}$\,km$^{2}$\,s$^{-2}$) &  ($10^{3}$\,kpc\,km\,s$^{-1}$)  &    (kpc) & (kpc) &   & (Gyr) &     \\     

\hline
 HiTS104009 &  $105 \pm 5$ &  $105^{+5}_{-5}$ &  $-0.53 \pm 1.09$ &  $-0.26 \pm 1.25$ &   $196.5^{+72.0}_{-45.6}$ &   $0.19^{+0.18}_{-0.05}$ &   $19.77^{+7.69}_{-3.40}$ &  $53^{+33}_{-17}$ &    $109^{+113}_{-5}$ &  $0.44^{+0.23}_{-0.14}$ &  $1.57^{+2.68}_{-0.38}$ &          $0.13$ \\
 HiTS112524 &   $86 \pm 4$ &   $86^{+4}_{-3}$ &  $-0.84 \pm 0.88$ &  $-0.81 \pm 0.63$ &   $255.1^{+34.1}_{-83.2}$ &   $0.26^{+0.11}_{-0.16}$ &   $19.59^{+4.36}_{-6.30}$ &  $48^{+16}_{-23}$ &    $155^{+86}_{-51}$ &  $0.57^{+0.09}_{-0.09}$ &  $2.42^{+1.98}_{-0.93}$ &          $0.34$ \\
    J050226 &   $48 \pm 2$ &   $51^{+3}_{-2}$ &   $0.39 \pm 0.19$ &  $-0.42 \pm 0.22$ &    $94.5^{+31.6}_{-26.8}$ &  $-0.33^{+0.05}_{-0.03}$ &    $4.13^{+1.98}_{-1.70}$ &     $9^{+7}_{-5}$ &      $54^{+10}_{-3}$ &  $0.72^{+0.17}_{-0.17}$ &  $0.61^{+0.19}_{-0.12}$ &          $1.00$ \\
 HiTS100956 &   $89 \pm 4$ &   $90^{+5}_{-2}$ &  $-0.24 \pm 0.73$ &   $0.38 \pm 0.67$ &  $235.1^{+68.9}_{-103.5}$ &   $0.25^{+0.17}_{-0.21}$ &  $21.12^{+5.58}_{-11.88}$ &  $68^{+10}_{-40}$ &   $170^{+209}_{-75}$ &  $0.52^{+0.17}_{-0.12}$ &  $2.91^{+4.50}_{-1.66}$ &          $0.32$ \\
    J051424 &   $47 \pm 2$ &   $47^{+2}_{-2}$ &   $1.42 \pm 0.27$ &   $1.20 \pm 0.28$ &   $332.5^{+53.7}_{-45.9}$ &   $0.03^{+0.19}_{-0.14}$ &   $15.23^{+2.37}_{-2.48}$ &   $43^{+5}_{-21}$ &  $200^{+126}_{-104}$ &  $0.67^{+0.15}_{-0.12}$ &  $3.38^{+2.11}_{-2.42}$ &          $0.88$ \\
    J023001 &   $30 \pm 2$ &   $35^{+2}_{-1}$ &  $-0.17 \pm 0.18$ &  $-1.57 \pm 0.14$ &   $225.3^{+18.1}_{-19.9}$ &  $-0.22^{+0.06}_{-0.05}$ &    $5.78^{+1.05}_{-1.01}$ &    $16^{+4}_{-4}$ &       $59^{+9}_{-5}$ &  $0.58^{+0.05}_{-0.03}$ &  $0.75^{+0.24}_{-0.21}$ &          $1.00$ \\
    J051213 &   $39 \pm 2$ &   $46^{+1}_{-2}$ &   $0.45 \pm 0.24$ &  $-0.53 \pm 0.22$ &   $101.7^{+21.2}_{-24.0}$ &  $-0.33^{+0.04}_{-0.03}$ &    $3.64^{+1.38}_{-1.69}$ &     $6^{+4}_{-5}$ &      $49^{+11}_{-3}$ &  $0.77^{+0.18}_{-0.09}$ &  $0.52^{+0.26}_{-0.08}$ &          $1.00$ \\
    J050902 &   $26 \pm 1$ &   $32^{+1}_{-1}$ &   $0.29 \pm 0.16$ &  $-1.08 \pm 0.13$ &   $101.0^{+19.4}_{-17.2}$ &  $-0.49^{+0.03}_{-0.03}$ &    $3.05^{+0.79}_{-0.55}$ &     $3^{+2}_{-2}$ &       $44^{+4}_{-8}$ &  $0.87^{+0.06}_{-0.09}$ &  $0.41^{+0.05}_{-0.04}$ &          $1.00$ \\
    J054653 &   $14 \pm 1$ &   $22^{+1}_{-1}$ &   $4.35 \pm 0.31$ &  $-0.81 \pm 0.27$ &   $241.8^{+16.1}_{-12.5}$ &  $-0.43^{+0.05}_{-0.04}$ &    $4.91^{+0.46}_{-0.41}$ &    $12^{+4}_{-4}$ &      $43^{+7}_{-12}$ &  $0.57^{+0.07}_{-0.13}$ &  $0.43^{+0.22}_{-0.13}$ &          $1.00$ \\
    J044339 &   $38 \pm 2$ &   $45^{+2}_{-2}$ &   $0.02 \pm 0.18$ &  $-1.38 \pm 0.17$ &   $196.7^{+25.2}_{-18.8}$ &  $-0.18^{+0.06}_{-0.04}$ &    $6.50^{+1.31}_{-0.96}$ &    $18^{+5}_{-4}$ &       $63^{+9}_{-5}$ &  $0.55^{+0.06}_{-0.06}$ &  $0.91^{+0.14}_{-0.16}$ &          $1.00$ \\
    J040422 &   $25 \pm 1$ &   $32^{+1}_{-1}$ &  $-0.01 \pm 0.17$ &  $-1.98 \pm 0.13$ &   $255.1^{+13.8}_{-16.3}$ &  $-0.21^{+0.05}_{-0.05}$ &    $4.52^{+0.76}_{-0.68}$ &    $11^{+2}_{-2}$ &       $61^{+8}_{-5}$ &  $0.70^{+0.03}_{-0.03}$ &  $0.79^{+0.08}_{-0.29}$ &          $1.00$ \\
    J034239 &   $27 \pm 1$ &   $33^{+1}_{-1}$ &   $0.24 \pm 0.17$ &  $-2.08 \pm 0.13$ &   $214.6^{+15.7}_{-15.2}$ &  $-0.28^{+0.04}_{-0.04}$ &    $4.31^{+0.58}_{-0.82}$ &    $11^{+2}_{-3}$ &       $53^{+4}_{-3}$ &  $0.66^{+0.07}_{-0.03}$ &  $0.58^{+0.13}_{-0.08}$ &          $1.00$ \\
 HiTS091050 &   $61 \pm 3$ &   $66^{+3}_{-3}$ &   $0.02 \pm 0.28$ &  $-0.54 \pm 0.25$ &   $155.7^{+33.7}_{-21.7}$ &  $-0.06^{+0.06}_{-0.04}$ &    $5.50^{+3.82}_{-2.79}$ &   $12^{+13}_{-9}$ &     $83^{+16}_{-10}$ &  $0.73^{+0.19}_{-0.19}$ &  $1.13^{+0.24}_{-0.48}$ &          $1.00$ \\
   HV204704 &   $35 \pm 1$ &   $30^{+1}_{-1}$ &  $-0.07 \pm 0.14$ &  $-1.80 \pm 0.11$ &   $145.4^{+12.0}_{-11.5}$ &  $-0.46^{+0.03}_{-0.03}$ &    $2.32^{+0.66}_{-0.84}$ &     $3^{+4}_{-2}$ &       $38^{+3}_{-1}$ &  $0.88^{+0.10}_{-0.18}$ &  $0.34^{+0.10}_{-0.07}$ &          $1.00$ \\
   HV210918 &   $62 \pm 2$ &   $57^{+2}_{-2}$ &   $0.19 \pm 0.39$ &  $-1.04 \pm 0.34$ &   $160.9^{+58.2}_{-43.8}$ &  $-0.12^{+0.12}_{-0.06}$ &    $8.18^{+4.14}_{-3.11}$ &    $14^{+6}_{-7}$ &     $89^{+72}_{-11}$ &  $0.76^{+0.10}_{-0.10}$ &  $1.15^{+1.29}_{-0.35}$ &          $0.99$ \\
   HV210205 &   $32 \pm 1$ &   $27^{+1}_{-1}$ &  $-2.38 \pm 0.15$ &  $-2.29 \pm 0.11$ &   $400.8^{+25.3}_{-22.8}$ &   $0.18^{+0.13}_{-0.09}$ &   $10.70^{+0.95}_{-0.88}$ &    $27^{+1}_{-1}$ &   $201^{+110}_{-25}$ &  $0.77^{+0.07}_{-0.03}$ &  $3.85^{+1.45}_{-0.88}$ &          $0.98$ \\
   HV205840 &   $27 \pm 1$ &   $22^{+1}_{-1}$ &  $-2.77 \pm 0.10$ &  $-2.38 \pm 0.09$ &   $392.3^{+15.0}_{-21.9}$ &   $0.06^{+0.07}_{-0.11}$ &    $8.52^{+0.52}_{-0.81}$ &    $22^{+1}_{-1}$ &    $147^{+54}_{-60}$ &  $0.74^{+0.06}_{-0.12}$ &  $2.64^{+1.14}_{-1.68}$ &          $1.00$ \\

\hline

\end{tabular}

\end{center}
\end{sidewaystable}

\bsp	
\label{lastpage}
\end{document}